\documentstyle[11pt,bezier]{article}
\font\frbig=eufm10 scaled\magstephalf
\font\frscr=eufm8
\font\frscrscr=eufm8
\newfam\frfam
\textfont\frfam=\frbig
\scriptfont\frfam=\frscr
\scriptscriptfont\frfam=\frscrscr
\def\fr{\fam\frfam}

\font\openbig=msbm10 scaled\magstephalf
\font\openscr=msbm7 
\font\openscrscr=msbm6
\newfam\openfam
\textfont\openfam=\openbig
\scriptfont\openfam=\openscr
\scriptscriptfont\openfam=\openscrscr
\def\open{\fam\openfam}

\font\ssfbig=cmss10 scaled\magstephalf
\font\ssfscr=cmss8 
\font\ssfscrscr=cmss8
\newfam\ssffam
\textfont\ssffam=\ssfbig
\scriptfont\ssffam=\ssfscr
\scriptscriptfont\ssffam=\ssfscrscr
\def\ssf{\fam\ssffam}

\def\mm{\cal}
\def\smm{\fr}
\def\auxm{\ssf}

\def\mC{{\mm C}}
\def\mF{{\mm F}}
\def\mM{{\mm M}}

\def\mP{{\mm P}}

\def\mR{{\mm R}}

\def\mU{{\mm U}}
\def\mV{{\mm V}}
\def\mW{{\mm W}}

\def\aM{{\auxm M}}
\def\aV{{\auxm V}}

\def\smM{{\smm M}}

\def\smV{{\smm V}}
\def\smR{{\smm R}}

\def\IIIpm#1{III${}_\pm^{#1}$}
\def\IIIpmim{III${}_\pm(1,{-})$}
\def\IIIpmip{III${}_\pm(1,{+})$}
\def\IIIpmzim{III${}_\pm^{0}(1,{-})$}
\def\IIIpmzip{III${}_\pm^{0}(1,{+})$}
\def\IIIpmziimm{III${}_\pm^{0}(2,{-}{-})$}
\def\IIIpmziipp{III${}_\pm^{0}(2,{+}{+})$}
\def\IIIpmziimp{III${}_\pm^{0}(2,{-}{+})$}
\def\IIIpmzziimp{III${}_\pm^{00}(2,{-}{+})$}
\def\IIIpmzziimm{III${}_\pm^{00}(2,{-}{-})$}
\def\IIIpmzziipp{III${}_\pm^{00}(2,{+}{+})$}
\def\Iiimm{I$(2,{-}{-})$}
\def\Iiipp{I$(2,{+}{+})$}
\def\Iiimp{I$(2,{-}{+})$}

\makeatletter
\newdimen\normalarrayskip
\newdimen\minarrayskip
\normalarrayskip\baselineskip
\minarrayskip\jot
\newif\ifold \oldtrue \def\new{\oldfalse}
\def\arraymode{\ifold\relax\else\displaystyle\fi}

\def\@arrayskip{\ifold\baselineskip\z@\lineskip\z@
  \else
  \baselineskip\minarrayskip\lineskip2\minarrayskip\fi}
\def\@arrayclassz{\ifcase \@lastchclass \@acolampacol \or
\@ampacol \or \or \or \@addamp \or
 \@acolampacol \or \@firstampfalse \@acol \fi
\edef\@preamble{\@preamble
 \ifcase \@chnum
  \hfil$\relax\arraymode\@sharp$\hfil
  \or $\relax\arraymode\@sharp$\hfil
  \or \hfil$\relax\arraymode\@sharp$\fi}}
\def\@array[#1]#2{\setbox\@arstrutbox=\hbox{\vrule
  height\arraystretch \ht\strutbox
  depth\arraystretch \dp\strutbox
  width\z@}\@mkpream{#2}\edef\@preamble{\halign \noexpand\@halignto
\bgroup \tabskip\z@ \@arstrut \@preamble \tabskip\z@ \cr}%
\let\@startpbox\@@startpbox \let\@endpbox\@@endpbox
 \if #1t\vtop \else \if#1b\vbox \else \vcenter \fi\fi
 \bgroup \let\par\relax
 \let\@sharp##\let\protect\relax
 \@arrayskip\@preamble}
\@addtoreset{equation}{section}
\makeatother

\def\tensor{\otimes}
\def\tilde{\widetilde}

\def\tp{{\widetilde p}}

\def\half{\frac{1}{2}}

\def\cE{{\cal E}}

\def\cG{{\cal G}}
\def\cH{{\cal H}}

\def\cL{{\cal L}}

\def\cO{{\cal O}}
\def\cP{{\cal P}}
\def\cQ{{\cal Q}}

\def\cT{{\cal T}}
\def\cU{{\cal U}}

\def\oC{{\open C}}
\def\oK{{\open K}}
\def\oN{{\open N}}
\def\oQ{{\open Q}}

\def\oZ{{\open Z}}

\def\Jplus{J^+}
\def\Jminus{J^-}
\def\Jnaught{J^0}

\def\jtop{{\ssf j}}
\def\jminus{{\ssf j}^-}

\def\C{A$\!$D$'$}

\def\tSL#1{{\widehat{s\ell}}(#1)}
\def\SL#1{s\ell(#1)}
\def\SSL#1#2{s\ell(#1|#2)}
\def\tSSL#1#2{\widehat{s\ell}(#1|#2)}

\def\ket#1{\mathchoice{%
    {\left|{#1}\right\rangle}}{|{#1}\rangle}{|{#1}\rangle}{|{#1}\rangle}}

\def\kettop#1{\mathchoice{{\left|{#1}\right\rangle}_{\rm top}}%
  {|{#1}\rangle_{\rm top}}{|{#1}\rangle_{\rm top}}{|{#1}\rangle_{\rm top}}}

\def\ketSL#1{\mathchoice{%
    {\left|{#1}\right\rangle}}{|{#1}\rangle}{|{#1}\rangle}{|{#1}\rangle}}

\def\frac#1#2{\mathchoice{%
{\textstyle{{#1}\over{#2}}}}{{#1\over#2}}{{#1\over#2}}{{#1\over#2}}}
\def\N#1{N\!=\!#1}

\def\d{\partial}
\def\bar{\overline}

\def\spsi{\psi^*}

\def\theell{{\ssf l}}
\def\ellch{{\ssf l}_{\rm ch}}

\def\ctop{{\ssf c}}
\def\Ctop{{\ssf C}}
\def\htop{{\ssf h}}

\def\hplus{{\ssf h}^+}
\def\hminus{{\ssf h}^-}
\def\jplus{{\ssf j}^+}
\def\jminus{{\ssf j}^-}
\def\jtop{{\ssf j}}

\def\req#1{(\ref{#1})}

\def\RV{relaxed}

\def\hw{highest-weight}

\def\Relaxed{\raisebox{.5pt}{\rule{5pt}{6pt}}}
\def\Verma{\Large$\bullet$}
\def\TVerma{{\Large $\circ$}}
\def\state{\ast}

\def\NPB{Nucl.\ Phys.\ B}
\def\PLB{Phys.\ Lett.\ B}
\def\MPLA{Mod.\ Phys.\ Lett.\ A}

\def\IJMPA{Int.\ J.\ Mod.\ Phys.\ A}

\newtheorem{lemma}{Lemma}[section]

\newtheorem{thm}[lemma]{Theorem}

\newtheorem{dfn}[lemma]{Definition}

\newlength\myleftmargin
\def\lvm{\leavevmode\hbox to\parindent{\hfill}}

\begin{document}
\hfuzz=1pt
\addtolength{\baselineskip}{2pt}

\begin{flushright}
  {\tt hep-th/9712102}
\end{flushright}
\thispagestyle{empty}

\bigskip

\begin{center}
  {\Large{\sc Embedding Diagrams of $\N2$ Verma Modules and\\[6pt]
      Relaxed $\tSL2$ Verma Modules}}\\[16pt]
  {\large A.~M.~Semikhatov and V.~A.~Sirota}\\[4pt]
  {\small\sl Tamm Theory Division, Lebedev Physics Institute, Russian
    Academy of Sciences}
\end{center}

\addtolength{\baselineskip}{-2pt}

\noindent\hbox to.05\hsize{\hfill}
\parbox{.9\hsize}{\footnotesize We classify and explicitly construct
  the embedding diagrams of Verma modules over the $\N2$
  supersymmetric extension of the Virasoro algebra. The essential
  ingredient of the solution consists in drawing the distinction
  between two different types of submodules appearing in $\N2$ Verma
  modules. The problem is simplified by associating to every $\N2$
  Verma module a {\it relaxed Verma\/} module over the affine
  algebra~$\tSL2$ with an isomorphic embedding diagram. We then make
  use of the mechanism according to which the structure of the
  $\N2$/relaxed-$\tSL2$ embedding diagrams can be found knowing
  the standard embedding diagrams of $\tSL2$ Verma modules. The
  resulting classification of the $\N2$/relaxed-$\tSL2$ embedding
  diagrams follows the I-II-III pattern extended by an additional
  indication of the number (0, 1, or 2) and the twists
  of the standard $\tSL2$ embedding diagrams contained in a given
  $\N2$/relaxed-$\tSL2$ embedding diagram.}

{\small \tableofcontents}

\addtolength{\baselineskip}{2pt}

\section{Introduction}\lvm
In this paper, we solve the long-standing problem of constructing
embedding diagrams of Verma modules over the $\N2$ superconformal
algebra in two dimensions (the $\N2$ extension of the Virasoro
algebra). This algebra appeared originally in the construction of
$\N2$ strings~\cite{[Ade]}, however it turned out later on that, in
addition to its role in the $\N2$ strings (\cite{[OV23],[Marcus]}, and
references therein), the $\N2$ superconformal symmetry in two
dimensions is important in the heterotic string
compactifications~\cite{[G]}, which has lead to investigating its
relations with Calabi--Yau manifolds and with the Landau--Ginzburg
theories and singularity theory~\cite{[Mar],[VW]}.  This algebra is in
fact the symmetry of topological conformal
theories~\cite{[W-top],[EY]} (see also~\cite{[Jose],[Get]}
and~\cite{[Get2]} for the discussion in terms of equivariant
cohomology) and is therefore related to topological field theories in
two dimensions.  Much attention has been given to investigating the
space of $\N2$ supersymmetric quantum field theories by perturbing
$\N2$ superconformal theories, see, e.g.,~\cite{[C],[CV]}.  An
important tool in constructing $\N2$ superconformal theories are the
Kazama--Suzuki models~\cite{[KS]} (see also~\cite{[OS]} for their
free-superfield realizations).  The $\N2$ algebra is actually realized
on the worldsheet of any non-critical string
theory~\cite{[GS2],[BLNW]}.  Therefore, deriving this algebra via the
Hamiltonian reduction of affine Lie
superalgebra~$\tSSL21$~\cite{[BO],[IK]} can be interpreted as deriving
the entire bosonic string via the Hamiltonian reduction; this approach
has been extended to other string theories~\cite{[BLLS],[RSS]}, with
the corresponding symmetry algebras containing the $\N2$ algebra as a
subalgebra. The $\N2$ algebra realized in the bosonic string
\hbox{actually underlies the construction of the physical
  states~\cite{[LZ]}.}  \vspace{-8pt}

\paragraph{`Non-affine' complications.}
{}From the representation-theoretic point of view (as well as in
several other respects, see, e.g.,~\cite{[EG]}), the $N=2$ algebra is
essentially different from the $N=1$ super-Virasoro algebra, which is
primarily due to the existence of the spectral flow
transform~\cite{[SS],[LVW]} (see also~\cite{[KL]} for the role of the
spectral flow in the (critical) $\N2$ strings).  In addition, certain
complications arising in the $\N2$ representation theory are due to
the fact that this algebra is {\it not\/} affine.  This affects, in
particular, finding all possible sequences of submodules of submodules
of\ldots, appearing in a given $\N2$ Verma module, that is, the {\it
  embedding diagrams}. An important related construction is the
appropriate version of the BGG-resolution~\cite{[BGG]}. \ For the
Verma modules over affine Lie algebras, the structure of the embedding
diagrams and of the BGG resolution is governed by the affine Weyl
group.  While the embedding diagrams of the $\tSL2$ as well as the
Virasoro Verma modules are well-known~\cite{[FF],[RCW],[Mal],[FFr]},
the problem of constructing the $\N2$ embedding diagrams, although it
has been around for some time~\cite{[Kir],[Dob],[M],[Doerr2]}, has not
been solved yet\footnote{That the earlier conjectures for the $\N2$
  embedding diagrams need being corrected was for the first time
  noticed in~\cite{[Doerr2]}, see also the remarks in~\cite{[EG]}.
  }.  No standard construction of an `affine' Weyl group applies here.
The problem is focused on the embedding diagrams, since the
BGG-resolution can be obtained by finding the embedding diagrams first
and then reconstructing the associated exact sequences (which, with
the underlying Weyl group representation unknown, appears to be the
only way to derive the resolution). The resolution should provide one
with a tool for systematically deriving the $\N2$ characters.
\vspace{-8pt}

\paragraph{Submodules and singular vectors.} It should be stressed
that the embedding diagrams describe embeddings of {\it modules\/},
i.e., the occurrence of submodules in a given Verma modules.
Submodules in a Verma module $\mU$ are often `identified' with
singular vectors, since, whenever one finds a state $\ket{S}\in\mU$
annihilated by the same elements of the algebra as the \hw{} vector
of~$\mU$, then there is a submodule $\mU'\subset\mU$ generated
from~$\ket{S}$. \ For the (affine) Lie algebras of rank
$\leq2$~\cite{[Mal]}, {\it any\/} submodule is a Verma module (or a
sum therof).  This is so, in particular for $\tSL2$, where the notions
of the submodule and of the singular vector can thus be identified.
However, the relation between singular vectors and submodules in Verma
modules is much more complicated in
general~\cite[Chap.~7]{[Dix]}.\nopagebreak

In the literature on representations of the $\N2$ superconformal
algebra, such an identification between submodules and singular
vectors appears to have resulted in replacing the analysis of
embeddings of $\N2$ Verma modules by the investigation of certain
algebraic constructions---some kind of singular vectors, or, more
precisely, singular vector operators. In particular, the conclusions
regarding the embeddings of submodules were arrived at according to
how these constructions compose.  This approach, however, runs into a
problem whenever, as it may happen, the composition of two $\N2$
singular vector operators vanishes. What happens to the corresponding
{\it submodules\/} then? Is there an {\it embedding\/}, and how a
composition of two embeddings (the mappings with {\it trivial
  kernels}) can possibly vanish?

The resolution of these problems requires, first of all, that one
analyses the {\it submodules\/} rather than singular vectors as such.
As realized in~\cite{[ST3]}, $\N2$ Verma modules contain submodules of
two different types, which we distinguish as the massive and the
(twisted) topological Verma modules (for example, the submodules
generated from the charged singular vectors~\cite{[BFK]} in a massive
Verma module are the twisted topological Verma modules).  A given
state $\ket{S}\in\mU$ satisfying the same annihilation conditions as
the \hw{} vector of a massive $\N2$ Verma module $\,\mU$ may give rise
to either a massive Verma module or a twisted topological Verma
module; in this respect, such singular vectors $\ket{S}$ are not very
useful as regards investigating the structure of {\it submodules\/}.
While the topological Verma modules can appear as submodules of the
massive ones, the converse is not true: what seems to be an embedding
of a massive Verma module \,$\mU$ into a topological Verma module
\,$\mV$ necessarily has a kernel given by a topological Verma
submodule in~\,$\mU$:
\begin{equation}
  0\longrightarrow\mV_1\longrightarrow\mU\longrightarrow
  \mV\longrightarrow0\,,
\end{equation}
hence the apparent vanishing of the `embedding'.  Moreover, once one
wishes to deal with the $\N2$ Verma-like modules of only one, massive,
type, one extends this sequence into
\begin{equation}
  \ldots\longrightarrow\mU_2\longrightarrow\mU_1\longrightarrow
  \mU\longrightarrow\mV\longrightarrow0\,,
\end{equation}
which is simply a resolution of a topological Verma module in terms of
the massive ones; the mappings in this {\it exact\/} sequence are by
no means embeddings then. If, further, the fact that \,$\mV$ is a
topological, not massive, Verma module is not recognized, the
interpretation of such a sequence becomes quite obscure.

Thus, in particular, one should clearly distinguish between the
embedding diagrams and the resolutions. Although the latter are of a
primary importance for constructing the characters, we concentrate in
this paper on the {\it embedding\/} diagrams, as a prerequisite for
constructing the resolutions. The $\N2$ embedding diagrams necessarily
involve modules of two types, the massive and the twisted topological
Verma modules.  A convenient way to differentiate between these
modules is to have them generated from the vectors satisfying the
\hw{} conditions that are correlated with the type of the
corresponding submodule. That this is possible for the $\N2$ algebra
is a lucky circumstance. One takes the singular vectors to satisfy
either the {\it twisted} (spectral-flow-transformed) topological \hw{}
conditions or the massive \hw{} conditions. These singular vectors,
constructed in~\cite{[ST3],[ST4]}, adequately describe the relations
between the corresponding submodules; at the same time, they generate
{\it maximal\/} submodules, therefore providing the description of all
\hbox{submodules in an $\N2$ Verma module without introducing
  subsingular vectors~\cite{[ST4]}.}  \vspace{-8pt}



\paragraph{Equivalence of categories.}
The problem of ``comparing'', in some sense, the {\it
  representations\/} of the $\N2$ algebra with those of the affine Lie
algebra~$\tSL2$ reappears in various contexts both in conformal field
theory and in representation theory. It was solved in~\cite{[FST]},
where a new type of $\tSL2$ modules was introduced and the {\it
  equivalence\/} was then shown to take place between certain
categories constructed out of the $\N2$ and $\tSL2$ representations
(because of the universal property of Verma modules in the~$\cO$-type
categories~\cite{[TheBook]}, one can start with the corresponding
Verma modules and then extend to the \hw-{\it type\/}
representations).  Namely, one considers an arbitrary complex level
$k\in\oC\setminus\{-2\}$ on the $\tSL2$ side; on the $\N2$ side, one
considers the `standard' representation category (the one implied in,
e.g.,~\cite{[BFK],[LVW]}), which includes the massive Verma modules
and their quotients, the (unitary and non-unitary) irreducible
representations,~etc., along with their images under the spectral flow
({\it twists\/}), with the central charge~$\ctop\in\oC\setminus\{3\}$.
{\it Modulo the spectral flow transform\/} (see~\cite{[FST]} for the
precise statement), this $\N2$ category is equivalent to the category
of {\it relaxed\/} \hw-type $\tSL2$ representations (and their
twists).\footnote{In this paper, `relaxed module'${}\equiv{}$`relaxed
  Verma module' of~\cite{[FST]}.} On the other hand, the standard
\hw-type $\tSL2$ representations turn out to be related to a narrower
category where \hbox{the `universal' Verma-like objects are the {\it
    twisted topological $\N2$ Verma modules}.}

A statement regarding the equivalence of two categories can often be
interpreted to the effect that there are two different languages
describing the same structure. In this way, any `structural' result
about $\N2$ Verma modules can in principle be seen in relaxed $\tSL2$
modules, and vice versa. As it may (and does) happen with equivalence
of categories, however, a number of facts that are fairly obvious on
one side translate into the statements which are not quite obvious on
the other side.  Thus, the $\tSL2$ representations that correspond to
the massive $\N2$ Verma modules are the relaxed (Verma) modules,
which, in general, have infinitely many \hw{} vectors.  Although this
may seem strange at a first glance, it is a good example of the
effects related to the equivalence of categories, since the
`proliferation' of the \hw{} vectors is a counterpart of the situation
well-known in the massive $\N2$ Verma modules. Further, singular
vectors in the relaxed-$\tSL2$ and massive $\N2$ Verma modules are
given by similar constructions (see~\cite{[FST]} and~\cite{[ST4]},
respectively), yet the $\tSL2$ one is somewhat more straightforward
algebraically.  On the other hand, proving that these constructions
give {\it all\/} singular vectors is easier on the $\N2$ side, where
the Ka\v c determinant has been known for some time~\cite{[BFK]}. \ An
important consequence of the equivalence is that the embedding
diagrams of massive $\N2$ Verma \hbox{modules are isomorphic to those
  of the relaxed $\tSL2$ modules}\footnote{The direct and the inverse
  functors of~\cite{[FST]} establish the equivalence between the {\it
    chains\/} of spectral-flow-transformed modules.  However, all
  degenerations occur simultaneously in every module in the chain and
  the embedding diagrams are isomorphic for all the
  spectral-flow-transformed modules.}.
\vspace{-8pt}

\paragraph{$\N2$ and relaxed-$\tSL2$ embedding diagrams.} Thus, the
$\N2$ embedding diagrams can be arrived at using the results of
either~\cite{[FST]} or~\cite{[ST4]}: \ Degenerations of massive $\N2$
Verma modules were directly analysed in~\cite{[ST4]}, where the
possible types of submodules were classified and the conditions were
given for the different combinations of submodules to appear in a
given massive Verma module.  Alternatively, the equivalence of
categories allows us to translate the $\N2$ problem into the
relaxed-$\tSL2$ context and then to apply the results of~\cite{[FST]}
on the classification of possible degenerations (reducibility
patterns) of the \RV{} $\tSL2$ modules.  In view of the equivalence of
categories, the distinction between two types of submodules applies on
the $\tSL2$ side as well. In the $\tSL2$ setting, however, things are
considerably simplified due to the absence of arbitrary twists.

In this paper, therefore, rather than directly analysing how the $\N2$
Verma modules can be embedded into one another, we take another root
to the $\N2$ embedding diagrams by making use of the fact that these
are isomorphic to the relaxed-$\tSL2$ embedding diagrams. The analysis
of the latter is easier also because the affine-Lie algebra
representation theory is available then and certain subdiagrams in the
relaxed embedding diagrams are literally the standard $\tSL2$
embedding diagrams~\cite{[RCW],[Mal]}.  Moreover, even though the
relaxed $\tSL2$ Verma modules are not a `classical' object in the
representation theory of affine Lie algebras, the problem of
enumerating submodules of \RV{} modules can be reduced, to a large
extent, to the classical problem of the standard~$\tSL2$-embedding
diagrams. Thus, for a given \RV{} module \,$\mR$, one can find an
``auxiliary'' usual-Verma module \,$\aM$ whose submodules are in a
$1:1$ (or, in some degenerate cases, essentially in a $2:1$)
correspondence with the \RV{} submodules in \,$\mR$. If, further,
\,$\mR$ contains no submodules generated from the {\it charged
  singular vectors\/}, then its embedding diagram is immediately read
off from the embedding diagram of the Verma module~\,$\aM$.  If, on
the other hand, there are charged singular vectors in \,$\mR$, the
picture is slightly more complicated because the entire embedding
diagram consisting of the usual-Verma modules is attached to the
embedding diagram of~\,$\mR$.

We would like to point out once again that all these effects can, of
course, be seen directly on the $\N2$ side as well; as we have
mentioned, the role of the usual-Verma $\tSL2$ modules is played there
by the topological Verma modules.  Analysing the structure of a given
massive $\N2$ Verma module \,$\mU$, again, is reduced~\cite{[ST4]} to
analysing an auxiliary twisted topological Verma module \,$\aV$.  In
particular, singular vectors (submodules) in \,$\aV$ translate, in a
certain way, into singular vectors in~\,$\mU$.  In addition, it may
happen that one or two of the auxiliary topological Verma modules
become actual submodules of the massive Verma module, in which case
the {\it charged\/} singular vectors occur in the latter module.
Accordingly, the embedding diagrams of massive Verma modules can be
reconstructed, with some work, from those of the topological Verma
modules\footnote{This is parallel and, apparently, not unrelated, to
  the well-known situation with $\N2$ theories, where the {\it
    topological\/} sector allows one to reconstruct a considerable
  amount of the information about the entire theory.}; the latter are
isomorphic, in their own turn, to embedding diagrams of the usual
$\tSL2$ Verma modules.  In the end of the day, one arrives at the same
embedding diagrams as for the relaxed $\tSL2$ modules, read with
somewhat different conventions.

In what follows, we first classify the relaxed-$\tSL2$ embedding
diagrams and then explain the legend that allows one to read the same
diagrams in the $\N2$ language. The classification of embedding
diagrams that we obtain agrees with the classification of degeneration
patterns of $\N2$ Verma modules given in~\cite{[ST4]}, more precisely,
the present construction is a (considerable) refinement of the
previous classification.

The $\N2$/relaxed-$\tSL2$ embedding diagrams which we construct are
somewhat more complicated than the Virasoro or the standard $\tSL2$
ones; in particular, a significant difference in the classification is
that there does not exist a `generic' case such that all the other
cases correspond to its degenerations\footnote{We thank F.~Malikov for
  stressing this point in relation with the search of a possible
  analogue of the affine Weyl group.}.  Yet, the embedding diagrams
follow the familiar I-II-III pattern~\cite{[FF]}, which has to be
extended by an additional indication of how many (0, 1, or 2) charged
singular vectors exist in the module.

From our results, one can reproduce the well-known embedding diagrams
of the Virasoro Verma modules~\cite{[FF]} by contracting---in the
conventions that we discuss in what follows---all the horizontal
arrows in the $\N2$/relaxed-$\tSL2$ embedding diagrams.  In addition,
there are two types of relations between the
relaxed-$\tSL2$/massive-$\N2$ and the standard $\tSL2$ Verma module
embedding diagrams~\cite{[RCW],[Mal]}: first, as we have noted,
different embedding patterns of the relaxed/$\N2$ Verma modules are
sensitive to the type of the embedding diagram of the auxiliary Verma
module.  Therefore, when enumerating all degenerations of \RV{}
modules, one has to use the classification of the standard embedding
diagrams of $\tSL2$ Verma modules (in fact, one has to be even more
detailed than in the standard case!).  Second, whenever a charged
singular vector appears in a \RV{} module \,$\mR$, the quotient of
\,$\mR$ over the corresponding (maximal) submodule is (up to a twist)
the usual Verma module, therefore the usual-Verma embedding diagrams
are also reproduced by taking the quotients of the relaxed/$\N2$
embedding diagrams with respect to the embedding diagrams of Verma
submodules.
\vspace{-8pt}

\paragraph{This paper is organized as follows.} We begin in
Sec.~\ref{sec:SL2} with recalling the $\tSL2$ algebra and
representations.  In Sec.~\ref{subsec:Verma}, we introduce the {\it
  twisted\/} $\tSL2$ Verma modules, the relaxed modules are defined in
Sec.~\ref{subsec:relaxed}, while in Sec.~\ref{subsec:r-singular} we
consider singular vectors in \RV{} modules and review the structural
results regarding the corresponding submodules.  Relaxed-$\tSL2$
embedding diagrams are considered in Sec.~\ref{sec:classifying}, where
we first classify the possible cases (Sec.~\ref{subsec:list}, with the
summary in the Table on p.~\pageref{table:table}) and then, in
Sec.~\ref{subsec:diagrams}, list the corresponding embedding diagrams.
The $\N2$ superconformal algebra is considered in Sec.~\ref{sec:N2}.
Here, we introduce the massive (Sec.~\ref{subsec:massive}) and
topological (Sec.~\ref{subsec:top}) Verma modules, and also recall, in
Sec.~\ref{subsec:anti-KS}, the construction that allows one to map the
$\N2$ \hw-type representations theory onto the relaxed-\hw-type
representations of the affine $\SL2$. The embedding diagrams of $\N2$
Verma modules are literally the same as those in
Sec.~\ref{subsec:diagrams}, however we give several comments in
Sec.~\ref{subsec:comments} as to the notations that apply in the $\N2$
case; we also repeat here, in the intrinsic $\N2$ language, the
classification of the embedding diagram patterns, so that these now
apply directly to the massive $\N2$ Verma modules. The case of the
$\N2$ central charge $\ctop=3$ has to be analysed separately, which is
done in Sec.~\ref{subsec:c3}. \ Section~\ref{sec:conclusion} contains
several concluding remarks.

\section{The $\tSL2$ side: \RV{} modules\label{sec:SL2}}
\subsection{Twisted Verma modules\label{subsec:Verma}}\lvm
The affine $\SL2$ algebra is defined as
\begin{equation}\new
  \begin{array}{rcl}
    {[}J^0_m,\,J^\pm_n]&=&{}\pm J^\pm_{m+n}\,,\qquad
    [J^0_m,\,J^0_n]~=~{}\frac{K}{2}\,m\,\delta_{m+n,0}\,,\\
    {[}J^+_m,\,J^-_n]&=&{}K\,m\,\delta_{m+n,0} + 2J^0_{m+n},
  \end{array}
  \label{sl2modes}
\end{equation}
with $K$ being the central element, whose eigenvalue will be denoted
by~$t-2$.  In what follows, we assume $t\neq0$.  The spectral flow
transform~\cite{[BH],[FST]} is the automorphism
\begin{equation}
  \cU_\theta:\quad
  J^+_n\mapsto J^+_{n+\theta}\,,\qquad
  J^-_n\mapsto J^-_{n-\theta}\,,\qquad
  J^0_n\mapsto J^0_n+\frac{t-2}{2}\theta\delta_{n,0}\,,\qquad\theta\in\oZ\,.
  \label{spectralsl2}
\end{equation}

A twisted Verma module $\smM_{j,t;\theta}$ is freely generated by
$J^+_{\leq\theta-1}$, $J^-_{\leq-\theta}$, and $J^0_{\leq-1}$ from a
twisted highest-weight vector $\ketSL{j,t;\theta}$ defined by the
conditions
\begin{equation}\new
  \begin{array}{l}
    J^+_{\geq\theta}\,\ketSL{j,t;\theta}=J^0_{\geq1}\,\ketSL{j,t;\theta}=
    J^-_{\geq-\theta+1}\,\ketSL{j,t;\theta}=0\,,\\
    \left(J^0_{0}+\frac{t-2}{2}\theta\right)\,\ketSL{j,t;\theta}=
    j\,\ketSL{j,t;\theta}\,.
  \end{array}
  \label{sl2higgeneral}
\end{equation}
In what follows, $j$ is referred to as the {\it spin} of the \hw{}
vector $\ketSL{j,t;\theta}$.  We define $\ketSL{j,t}=\ketSL{j,t;0}$.
We also denote $\mM_{j,t}=\smM_{j,t;0}$. \ The structure of $\tSL2$
Verma modules is conveniently encoded in the {\it extremal diagram},
which in the `untwisted' case $\theta=0$ reads as
\begin{equation}
  \unitlength=1pt
  \begin{picture}(250,80)
    \put(-35,62){\Huge $\ldots$}
    \put(0,60){$\state$}
    \put(15,65){${}^{J^-_0}$}
    \put(28,63){\vector(-1,0){22}}
    \put(30,60){$\state$}
    \put(45,65){${}^{J^-_0}$}
    \put(58,63){\vector(-1,0){22}}
    \put(60,60){$\state$}
    \put(75,65){${}^{J^-_0}$}
    \put(88,63){\vector(-1,0){22}}
    \put(90,60){$\bullet$}
    \put(103,64){${}_{J^+_{-1}}$}
    \put(97,60){\vector(2,-1){17}}
    \put(115,47){$\state$}
    \put(24,-13){%
      \put(103,64){${}_{J^+_{-1}}$}
      \put(97,60){\vector(2,-1){17}}
      \put(115,47){$\state$}
      }
    \put(48,-26){%
      \put(103,64){${}_{J^+_{-1}}$}
      \put(97,60){\vector(2,-1){17}}
      \put(115,47){$\state$}
      }
    \put(72,-39){%
      \put(103,50){\Huge $\cdot$}
      \put(109,47){\Huge $\cdot$}
      \put(115,44){\Huge $\cdot$}
      }
  \end{picture}
  \label{Vermaextr}
\end{equation}
This expresses the fact that $\Jplus_{-1}$ and $\Jminus_0$ are the
highest-level operators that do not yet annihilate the \hw{} state
$\bullet$.  The $\tSL2$ modules being graded with respect to the
charge\footnote{by the {\it charge\/}---more precisely, the
  $J^0_0$-charge--- we mean the eigenvalue of $J^0_0$; this is
  different from spin $j$ for the twisted modules, in accordance
  with~\req{sl2higgeneral}.} and the level, {\it extremal vectors\/}
are those with the boundary value of (charge,\,level). All the other
states of the module should be thought of as lying in the interior of
the angle in the above diagram.  In our conventions, increasing the
charge by 1 corresponds to shifting to the neighbouring site on the
right, while increasing the level corresponds to moving down.  In
these conventions, e.g., $\Jnaught_{-1}$ is represented by a downward
vertical arrow.

A singular vector exists in the Verma module $\mM_{j,t}$ over the
affine $\SL2$ algebra if and only if either $j=\jplus(r,s,t)$ or
$j=\jminus(r,s,t)$, where
\begin{equation}\new
  \begin{array}{l}
    \jplus(r,s,t)=\frac{r-1}{2}-t\frac{s-1}{2}\,,\\
    \jminus(r,s,t)=-\frac{r+1}{2}+t\frac{s}{2}\,,
  \end{array}
  \qquad r,s\in\oN\,,\quad t\in\oC\,.
  \label{sl2singcond}
\end{equation}
Singular vectors in the module $\mM_{\jtop^\pm(r,s,t),t}$ are
given by the explicit constructions~\cite{[MFF]}:
\begin{equation}
  \new\begin{array}{rcl}
    \ket{{\rm MFF}^+(r,s,t)}\kern-6pt&=&\kern-6pt
    (J^-_0)^{r+(s-1)t}(J^+_{-1})^{r+(s-2)t}
    \ldots
    (J^+_{-1})^{r-(s-2)t}
    (J^-_0)^{r-(s-1)t}\ket{\jplus(r,s,t),t}_{\SL2}\,,\\
    \ket{{\rm MFF}^-(r,s,t)}\kern-6pt&=&\kern-6pt
    (J^+_{-1})^{r+(s-1)t}(J^-_0)^{r+(s-2)t}
    \ldots
    (J^-_0)^{r-(s-2)t}
    (J^+_{-1})^{r-(s-1)t}\ket{\jminus(r,s,t),t}_{\SL2}\,.
  \end{array}
  \label{mff}
\end{equation}
As is well known, these formulae evaluate as Verma-module elements
using the following relations:
\begin{equation}\!\!\!\!\new
  \begin{array}{rcl}
    (\Jminus_0)^\alpha\,\Jplus_m &=&
    -\alpha (\alpha - 1) \Jminus_m(\Jminus_0)^{\alpha-2} -
    2 \alpha \Jnaught_m\,(\Jminus_0)^{\alpha-1} +
    \Jplus_m\,(\Jminus_0)^{\alpha}
    \,,\\
    (\Jminus_0)^{\alpha}\,\Jnaught_m &=&
    \alpha \Jminus_m(\Jminus_0)^{\alpha-1} +
    \Jnaught_m\,(\Jminus_0)^{\alpha}
    \,,\\
    (\Jplus_{-1})^\alpha\,\Jminus_m &=&
    -\alpha (\alpha - 1) \Jplus_{m-2}(\Jplus_{-1})^{\alpha-2} -
    k\,\alpha\,\delta_{m - 1, 0} (\Jplus_{-1})^{\alpha-1} +
    2 \alpha \Jnaught_{m-1}\,(\Jplus_{-1})^{\alpha-1} +
    \Jminus_m\, (\Jplus_{-1})^{\alpha}\,,\\
    (\Jplus_{-1})^{\alpha}\,\Jnaught_m &=&
    -\alpha \Jplus_{m-1}(\Jplus_{-1})^{\alpha-1} +
    \Jnaught_m\,(\Jplus_{-1})^{\alpha}\,,
  \end{array}
  \label{properties}
\end{equation}
which can be derived for $\alpha $ being a positive integer and then
continued to an arbitrary complex~$\alpha $.

Equations~\req{mff} produce a singular vector in $\mM_{j,t}$ for any
pair of positive integers $r$ and $s$ that solve $j=\jtop^\pm(r,s,t)$.
At most two of these singular vectors are {\it primitive}, i.e., are
not in a submodule determined by some other singular vector.  By
looking for primitive singular vectors in the submodules generated by
each of the primitive vector, etc., one obtains all of the singular
vectors in the Verma module. For generic rational $t$, one-half of
them are directly obtained from the \hw{} vector as described
in~\req{mff}.

\subsection{Relaxed modules\label{subsec:relaxed}}
\begin{dfn} For an {\it unordered\/} pair of
  complex numbers $\{\mu_1,\mu_2\}$, the following set of conditions
  define the relaxed \hw{} state $\ketSL{\mu_1,\mu_2,t}$:
  \begin{eqnarray}
    &J^+_{\geq1}\,\ketSL{\mu_1,\mu_2,t}=J^0_{\geq1}\,
    \ketSL{\mu_1,\mu_2,t}=
    J^-_{\geq 1}\,\ketSL{\mu_1,\mu_2,t}=0\,,\label{floorhw}\\
    &J^-_{0}J^+_0\,
    \ketSL{\mu_1,\mu_2,t}=-\mu_1\mu_2\,\ketSL{\mu_1,\mu_2,t}\,.
    \label{Lambda}\\
    &J^0_0 \, \ketSL{\mu_1,\mu_2,t}=-\half(1+\mu_1+\mu_2)\,
    \ketSL{\mu_1,\mu_2,t}\,,
    \label{jnaught0}\\
    &K\ketSL{\mu_1,\mu_2,t}=(t-2)\ketSL{\mu_1,\mu_2,t}\,.
  \end{eqnarray}
  These equations are called the relaxed \hw{} conditions.
\end{dfn}
\begin{dfn}
  The \RV{} module\footnote{We follow~\cite{[FST]}, however we have
    changed the parametrization of relaxed modules to the one that
    will be more convenient for our present purposes. We also call
    relaxed Verma modules simply as relaxed modules.}
  $\mR_{\mu_1,\mu_2,t}$ is generated from a relaxed \hw{} state
  $\ketSL{\mu_1,\mu_2,t}$ by the free action of operators
  $J^+_{\leq0}$, $J^-_{\leq0}$, and $J^0_{\leq-1}$ modulo the
  constraint~\req{Lambda}.
\end{dfn}

Thus, the \RV{} module can also be defined as the induced
representation from the $\SL2$ {\it Lie\/} algebra representation
without a highest- or a lowest-weight.  It is convenient to consider
the extremal diagram of the \RV{} module:
\begin{equation}
  \unitlength=1pt
  \begin{picture}(250,20)
    \put(-35,2){\Huge $\ldots$}
    \put(0,0){$\state$}
    \put(15,5){${}^{J^-_0}$}
    \put(28,3){\vector(-1,0){22}}
    \put(30,0){$\state$}
    \put(45,5){${}^{J^-_0}$}
    \put(58,3){\vector(-1,0){22}}
    \put(60,0){$\state$}
    \put(75,5){${}^{J^-_0}$}
    \put(88,3){\vector(-1,0){22}}
                                %
    \put(90,0){$\star$}
    \put(100,5){${}^{J^+_0}$}
    \put(97,3){\vector(1,0){22}}
    \put(120,0){$\state$}
    \put(130,5){${}^{J^+_0}$}
    \put(127,3){\vector(1,0){22}}
    \put(150,0){$\state$}
    \put(160,5){${}^{J^+_0}$}
    \put(157,3){\vector(1,0){22}}
    \put(180,0){$\state$}
    \put(193,2){\Huge $\ldots$}
  \end{picture}
  \label{floor}
\end{equation}
where $\star$ denotes the state $\ketSL{\mu_1,\mu_2,t}$. The {\it
  extremal states\/}
\begin{equation}
  (J^-_0)^{-i}\,\ketSL{\mu_1,\mu_2,t},~i<0,\quad {\rm and}\quad
  (J^+_0)^{i}\,\ketSL{\mu_1,\mu_2,t},~i>0,
  \label{extremal}
\end{equation}
constitute precisely the $\SL2$-representation, while the other states
of the $\tSL2$ module belong to the homogeneous (charge,\,level)
subspaces represented by a rectangular lattice lying {\it below\/} the
line of extremal states.

Whenever $\mu_\alpha=0$, the definition of the \RV{} module is such
that the state $\ket{0,\mu,t|1}\equiv J^+_0\ket{0,\mu,t}\neq0$
satisfies the condition $J^-_0\,\ket{0,\mu,t|1}=0$. Similarly,
$\mu_\alpha=-1$ means that the state $\ket{-1,\mu,t|{-1}}\equiv
J^-_0\ket{-1,\mu,t}\neq0$ satisfies the condition
$J^+_0\,\ket{-1,\mu,t|{-1}}=0$.

In general, the \RV{} module may also be generated from other extremal
states than $\ketSL{\mu_1,\mu_2,t}$ since each of the extremal
states~\req{extremal} satisfies the relaxed \hw{} conditions with
\begin{equation}
  \label{mushift}
  \mu_\alpha\mapsto\mu_\alpha^{(i)}=\mu_\alpha-i\,.
\end{equation}
Generically, all of these states generate the same module because each
is a descendant of the others:
\begin{equation}
  \unitlength=1pt
  \begin{picture}(250,30)
    \put(-35,12){\Huge $\ldots$}
    \put(0,10){$\state$}
    \put(15,17){${}^{J^-_0}$}
    \put(28,15){\vector(-1,0){22}}
    \put(7,11){\vector(1,0){22}}
    \put(15,3){${}_{J^+_0}$}
    \put(30,10){$\state$}
    \put(45,17){${}^{J^-_0}$}
    \put(58,15){\vector(-1,0){22}}
    \put(37,11){\vector(1,0){22}}
    \put(45,3){${}_{J^+_0}$}
    \put(60,10){$\state$}
    \put(75,17){${}^{J^-_0}$}
    \put(88,15){\vector(-1,0){22}}
    \put(67,11){\vector(1,0){22}}
    \put(75,3){${}_{J^+_0}$}
    \put(90,10){$\star$}
    \put(105,17){${}^{J^-_0}$}
    \put(118,15){\vector(-1,0){22}}
    \put(97,11){\vector(1,0){22}}
    \put(105,3){${}_{J^+_0}$}
    \put(120,10){$\state$}
    \put(135,17){${}^{J^-_0}$}
    \put(148,15){\vector(-1,0){22}}
    \put(127,11){\vector(1,0){22}}
    \put(135,3){${}_{J^+_0}$}
    \put(150,10){$\state$}
    \put(165,17){${}^{J^-_0}$}
    \put(178,15){\vector(-1,0){22}}
    \put(157,11){\vector(1,0){22}}
    \put(165,3){${}_{J^+_0}$}
    \put(180,10){$\state$}
    \put(193,12){\Huge $\ldots$}
  \end{picture}
  \label{bothways}
\end{equation}
where the composition of any pair of the outward and the return arrows
gives the operator of a multiplication with a number.  However, as
soon as this number is zero, the equivalence between the extremal
states breaks down and some of the states generate only a submodule of
the entire module. Those extremal states that do generate the same
module can be described as follows.

\begin{dfn}\label{def:orbit}
  By the {\sl orbit\/} of an (unordered) pair $\{\mu_1,\mu_2\}$ where
  $\mu_\alpha\neq0$ and $\neq-1$, we will mean the maximal set of
  unordered pairs
  $\left(\{\mu_1^{(i)},\mu_2^{(i)}\}\right)_{i\in[a,b]\subset\oZ}$,
  where $-\infty\leq a\leq-1$ and $1\leq b\leq+\infty$, such that
  $\mu^{(i)}_\alpha\neq0$ for $i\in[a,-1]$ and
  $\mu^{(i)}_\alpha\neq-1$ for $i\in[1,b]$. If $\mu_1=-1$,
  $\mu_2\neq0$, the orbit is the maximal set of unordered pairs
  $\left(\{\mu_1^{(i)},\mu_2^{(i)}\}\right)_{i\in[0,b]\subset\oZ}$,
  where $1\leq b\leq+\infty$, such that $\mu^{(i)}_2\neq-1$ for
  $i\in[1,b]$; if, further, $\mu_1=0$, $\mu_2\neq-1$, the orbit is the
  maximal set of unordered pairs
  $\left(\{\mu_1^{(i)},\mu_2^{(i)}\}\right)_{i\in[a,0]\subset\oZ}$,
  where $-\infty\leq a\leq-1$, such that $\mu^{(i)}_2\neq0$ for
  $i\in[a,-1]$; finally, the orbit of $\{\mu_1,\mu_2\}=\{-1,0\}$
  consists of only that point.
\end{dfn}
Then, as easy to see, {\it the \RV{} module $\mR_{\mu_1,\mu_2,t}$ is
  determined by any point of the orbit of $\{\mu_1,\mu_2\}$}.
If none of the $\mu_\alpha$ are integers, the orbit is infinite in
both directions and can be identified with the set of extremal states
in~\req{floor}.  On the other hand, whenever
$\mu_1=n\in-\oN$,\footnote{Here and henceforth, $\oN=1,2,\ldots$ and
  $\oN_0=0,1,2,\ldots\,$. We also use notations of the type of
  $\oQ_-=\{x\in\oQ\mid x<0\}$, etc.} we have
\begin{equation}
  J^+_0\cdot(J^-_0)^{-n}\ketSL{n,\mu_2,t}=0\,,
  \label{Vermaneg}
\end{equation}
in which case diagram \req{bothways} branches as
\begin{equation}
  \unitlength=1pt
  \begin{picture}(250,80)
    \put(-35,62){\Huge $\ldots$}
    \put(0,60){$\state$}
    \put(15,67){${}^{J^-_0}$}
    \put(28,65){\vector(-1,0){22}}
    \put(7,61){\vector(1,0){22}}
    \put(15,55){${}_{J^+_0}$}
    \put(30,60){$\state$}
    \put(45,67){${}^{J^-_0}$}
    \put(58,65){\vector(-1,0){22}}
    \put(37,61){\vector(1,0){22}}
    \put(45,55){${}_{J^+_0}$}
    \put(60,60){$\state$}
    \put(75,67){${}^{J^-_0}$}
    \put(88,65){\vector(-1,0){22}}
    \put(67,61){\vector(1,0){22}}
    \put(75,55){${}_{J^+_0}$}
    \put(90,60){$\bullet$}
    \put(105,67){${}^{J^-_0}$}
    \put(118,65){\vector(-1,0){22}}
    \put(120,60){$\state$}
    \put(135,67){${}^{J^-_0}$}
    \put(148,65){\vector(-1,0){22}}
    \put(127,61){\vector(1,0){22}}
    \put(137,55){${}_{J^+_0}$}
    \put(150,60){$\state$}
    \put(40,0){
      \put(150,60){$\state$}
      \put(165,67){${}^{J^-_0}$}
      \put(178,65){\vector(-1,0){22}}
      \put(157,61){\vector(1,0){22}}
      \put(165,55){${}_{J^+_0}$}
      \put(180,60){\Large$\star$}
      }
    \put(163,62){\Large$\ldots$}
    \put(233,62){\Large$\ldots$}
    \put(92,40){${}^{J^+_{-1}}$}
    \put(95,58){\vector(2,-1){17}}
    \put(114,53){\vector(-2,1){17}}
    \put(115,47){$\state$}
    \put(24,-13){%
      \put(92,40){${}^{J^+_{-1}}$}
      \put(95,58){\vector(2,-1){17}}
      \put(101,57.5){${}^{J^-_{1}}$}
      \put(114,53){\vector(-2,1){17}}
      \put(115,47){$\state$}
      }
    \put(48,-26){%
      \put(92,40){${}^{J^+_{-1}}$}
      \put(95,58){\vector(2,-1){17}}
      \put(106,55){${}^{J^-_{1}}$}
      \put(114,53){\vector(-2,1){17}}
      \put(115,47){$\state$}
      }
    \put(72,-39){%
      \put(103,50){\Huge $\cdot$}
      \put(109,47){\Huge $\cdot$}
      \put(115,44){\Huge $\cdot$}
      }
  \end{picture}
  \label{withVerma}
\end{equation}
where we recognize the ordinary Verma module~\req{Vermaextr} as a
submodule of the \RV{} module.  Similarly, whenever $\mu_1=n\in\oN_0$,
we have
\begin{equation}
  J^-_0\cdot(J^+_0)^{n+1}\ketSL{n,\mu_2,t}=0\,.
  \label{Vermapos}
\end{equation}
Somewhat more schematically than in \req{withVerma}, this can be
represented as follows:
\begin{equation}
  \label{rightVerma}
  \unitlength=1pt
  \begin{picture}(500,70)
    \put(0,-30){
      \put(60,80){\line(1,0){300}}
      \put(275,80){\vector(1,0){2}}
      \put(280,80){\line(-2,-1){120}}
      \put(200,77){\Large$\star$}
      \put(275,76){\Large$\circ$}
      {\linethickness{.1pt}
        \put(203,70){\line(0,1){30}}
        \put(278,70){\line(0,1){30}}
        \put(206,90){\vector(1,0){70}}
        \put(206,90){\vector(-1,0){2}}
        \put(236,91){${}^{n+1}$}
        }
      }
  \end{picture}
\end{equation}
In this case, the submodule is a Verma module twisted by the spectral
flow transform with~$\theta=1$; in particular, the state at the branch
point satisfies the annihilation conditions from~\req{sl2higgeneral}
with~$\theta=1$.

Whenever $\mu_\alpha$ are integers of different signs, the orbit
contains a finite number of points (and, thus, can be identified with
a finite-dimensional representation of the Lie algebra $\SL2$).  As
will be seen in what follows, this case is always special, being
considerably different even from the case where $\mu_1$ and $\mu_2$
are integers of the same sign.  Let us also note that, obviously,
$\mu_1 - \mu_2$ is invariant under the $\oZ$-action~\req{mushift}, as
is the fact that one or both of the $\mu_\alpha$ is an integer.
Further, the signs of each of the $\mu_\alpha$ are also preserved
along the orbit.

\bigskip

In contrast to \RV{} modules, {\it by {\sl Verma modules\/} over
  $\tSL2$, we will always mean the standard Verma modules}. The
twisted Verma modules that we encounter in this paper, are always with
the twist parameter $\theta=1$.



\bigskip

In what follows, we will also need the {\it twisted\/} \RV{} modules.
For a fixed $\theta\in\oN$, the module $\smR_{\mu_1,\mu_2,t;\theta}$
is generated from the state $\ketSL{\mu_1,\mu_2,t;\theta}$ that
satisfies the annihilation conditions
\begin{equation}
  J^+_{\geq1+\theta}\,\ketSL{\mu_1,\mu_2,t;\theta}=J^0_{\geq1}\,
  \ketSL{\mu_1,\mu_2,t;\theta}=
  J^-_{\geq 1-\theta}\,\ketSL{\mu_1,\mu_2,t;\theta}=0
  \label{floorhwtw}
\end{equation}
by a free action of the operators $J^+_{\leq\theta-1}$,
$J^-_{\leq-\theta-1}$, and $J^0_{\leq-1}$ and by the action of
$J^+_{\theta}$ and $J^-_{-\theta}$ subject to the constraint
\begin{equation}
  J^-_{-\theta}J^+_\theta\,
  \ketSL{\mu_1,\mu_2,t;\theta}=
  -\mu_1\mu_2\,\ketSL{\mu_1,\mu_2,t;\theta}\,.
\end{equation}
In addition, the \hw{} state $\ketSL{\mu_1,\mu_2,t;\theta}$ satisfies
\begin{equation}
  \left(J^0_0+\frac{t-2}{2}\theta\right)\,
  \ketSL{\mu_1,\mu_2,t;\theta}=-\half(1+\mu_1+\mu_2)\,
  \ketSL{\mu_1,\mu_2,t;\theta}\,.
  \label{jnaught0tw}
\end{equation}

\subsection{Submodules and singular vectors in \RV{}
  modules\label{subsec:r-singular}}\lvm The `Verma points' encountered
in the top floor of an extremal diagram of the \RV{} module can of
course be viewed as singular vectors. It is simply a reformulation of
the above observations that the states
\begin{equation}
  \ket{C(n,\mu,t)}=
  \left\{\kern-4pt\new
    \begin{array}{ll}
      (J^-_0)^{-n}\,\ketSL{n,\mu,t}\,,&n\in-\oN\,,\\
      (J^+_0)^{n+1}\,\ketSL{n,\mu,t}\,,&n\in\oN_0
    \end{array}
  \right.
  \label{chargedsl2}
\end{equation}
satisfy the Verma \hw{} conditions for $n\leq-1$ and the twisted Verma
\hw{} conditions with the twist parameter $\theta=1$ for $n\geq0$,
hence the corresponding submodule is the Verma module or the twisted
Verma module, respectively.  The states \req{chargedsl2} are called
the {\it charged singular vectors\/}\footnote{The reason being that
  they are in a $1:1$ correspondence with the $\N2$ singular vectors
  that are traditionally called charged~\cite{[BFK]} (although the
  submodules generated from charged $\N2$ singular vectors were
  described correctly only in~\cite{[ST3]}).}.

The (twisted) Verma submodules may also occur `inside' a given \RV{}
module \,$\mR$ rather than in its extremal diagram (such would be the
submodules in the (twisted) Verma module generated from a charged
singular vector; in fact, {\it every\/} Verma submodule in $\mR$ is
necessarily a submodule in some `charged' submodule').  Besides the
Verma or twisted Verma submodules, a \RV{} module $\mR$ may also
contain {\it \RV\/} submodules. Such submodules are generated from the
{\it relaxed singular vectors}. These states that satisfy the relaxed
\hw{} conditions~\req{floorhw} (with the eigenvalues of $J^0_0$ and
$J^-_0J^+_0$, obviously, labelled by different parameters than
in~\req{Lambda}--\req{jnaught0}) were defined in~\cite{[FST]} in such
a way as to guarantee that they generate the entire `floor' of the
extremal states. In what follows, we summarize the general
construction for the relaxed singular vectors in the way
that is more convenient for our present purposes.

Define the set
\begin{equation}
  \label{defoK}
  \oK(t)=\Bigl\{a - bt\Bigm|a,b\in\oZ,~a\cdot b>0\Bigr\}\,.
\end{equation}
and
\begin{equation}
  \bar\oK(t)=\left\{\new
    \begin{array}{ll}
      \oK(t)\setminus\{2\tp\}\,,&
      t=-\frac{\tp}{q}\,,~\tp\in\oN\,,~
      q\in\oN\,,\\
      \oK(t)&{\rm otherwise}\,.
    \end{array}
  \right.
\end{equation}
Then,
\begin{thm}\label{thm:3types}\mbox{}

  \begin{enumerate}\addtolength\parskip{-6pt}
  \item\label{general-types} Any submodule in a \RV{} module is (a sum
    of the submodules) of only the following three types: \RV{}
    modules, Verma modules, and the twisted Verma modules with the
    twist parameter~$\theta=1$.

  \item\label{general-charged} The charged singular vectors in the
    \RV{} module $\mR_{\mu_1,\mu_2,t}$ are in the $1:1$ correspondence
    with the distinct integers among $\{\mu_1,\mu_2\}$. Whenever
    $\mu_\alpha=n\in\oZ$, the corresponding charged singular vector is
    given by~\req{chargedsl2}. Further,
    \begin{enumerate}
    \item The submodule generated from $\ket{C(n,\mu,t)}$ is a Verma
      module if $n<0$ and a twisted Verma module if
      $n\geq0$.

    \item\label{same-side} If both $\mu_1$ and $\mu_2$ are integers
      and $(\mu_1+\half)(\mu_2+\half) >0$, then one of the charged
      singular vectors belongs to the submodule generated from the
      other charged singular vector.
    \end{enumerate}

  \item\label{general-relaxed} The \RV{} module $\mR_{\mu_1,\mu_2,t}$
    contains at least one \RV{} submodule if and only if
    \begin{itemize}\addtolength{\parskip}{-4pt}
    \item[(i)] either \ {\rm(}$\mu_1\notin\oZ$ or
      $\mu_2\notin\oZ${\rm)} and $\mu_1 - \mu_2 \in\oK(t)$,

    \item[(ii)] or {\rm(}$\mu_1\in\oZ$ and $\mu_2\in\oZ${\rm)} and
      $\mu_1 - \mu_2 \in\bar\oK(t)$
    \end{itemize}
  \end{enumerate}
\end{thm}
In terms of extremal diagrams, the condition in~\ref{same-side} means
that the two charged singular vectors are on the same side of the
\hw{} vector of the \RV{} module. That is, we write
$(\mu_1+\half)(\mu_2+\half)>0$ or $<0$ to say that the cases where
($\mu_1=0$, $\mu_2<0$) are considered along with those where
$\mu_1\mu_2<0$, while ($\mu_1=0$, $\mu_2>0$) is `unified' with
$\mu_1\mu_2>0$.

Part~\ref{general-types} follows from the analysis of extremal
diagrams and the fact that the modules under consideration are the
induced representations. Part~\ref{general-charged} is obvious from
the previous subsection and the explicit formulae~\req{chargedsl2}.
Apart from the `exception' in \ref{general-relaxed}(ii), `if'
statement of Part~\ref{general-relaxed} follows from the explicit
construction for singular vectors that we review shortly,
while the `only if' part relies on the evaluation of the Ka\v c
determinant. This has been done for the $\N2$ Verma
modules~\cite{[BFK]}, while the equivalence results of~\cite{[FST]}
show that the zeros of the determinant are given by~$\mu_\alpha\in\oZ$
(which we already know to correspond to the charged singular vectors)
and by condition $\mu_1 - \mu_2 \in\oK(t)$. The exceptional case
occurs for negative rational $t=-\frac{\tp}{q}$ when, in addition,
$\mu_1,\;\mu_2\in\oZ$ are such that $\mu_2-\mu_1=2\tp$. The
determinant formula counts this case along with all the other zeros,
however it follows from the analysis of~\cite{[FST]} that the
{\it\RV\/} submodule {\it does not exist\/} in this case -- instead,
there is a direct sum of a Verma and a twisted Verma submodule: on a
given level, there are precisely as many states satisfying the relaxed
\hw{} conditions as there would be if the \RV{} submodule existed,
however these states do not make up the extremal diagram of a \RV{}
module.

\medskip

Further, all the instances where (twisted) Verma modules appear in a
\RV{} module are described in the following Theorem, which, again, is
an immediate consequence of~\cite{[FST]}.

Given an embedding of a (twisted) Verma module $\mM$ into a \hw{}
module $\mP$, let $\widetilde{\ket{e}}$ be the image of the \hw{}
vector of $\mM$.  For brevity, we will say that $\mM$ is embedded onto
level $l$ if $\widetilde{\ket{e}}$ is at level~$l$ relative to the
\hw{} vector of~$\mP$. Similarly, we will say that a \RV{} module
$\mR'$ is at level~$l$ in another \RV{} module $\mR$ if the extremal
states of $\mR'$ are on level $l$ relative to the extremal states
of~\,$\mR$.  Now,
\begin{thm}\label{thm:Verma}Let \,$\mR$ be a \RV{} module.
  \begin{enumerate}\addtolength\parskip{-6pt}
  \item\label{in-Verma} Every Verma or twisted Verma submodule $\mC'$
    of $\mR$ is necessarily a submodule of a submodule $\mC\subset\mR$
    generated from a charged singular vector in~$\mR$.

  \item\label{intersec} If $\mR$ contains a \RV{} submodule~$\mR'$ and
    also a charged singular vector generating a (twisted) Verma
    module~$\mC$, then the intersection \,$\mC'\equiv\mR'\cap\mC$ is
    non-empty and is a (twisted) Verma submodule in~\,$\mR$. The \hw{}
    vector of~\,$\mC'$ is a charged singular vector in~\,$\mR'$.

  \item Let $\ket{e}$ be the \hw{} vector of a Verma (respectively, a
    twisted Verma) submodule in \,$\mR$ which is not a charged
    singular vector in~\,$\mR$. Let $j$ be the charge of $\ket{e}$:
    $J^0_0\ket{e}=j\ket{e}$. Then one of the following is true:
    \begin{enumerate}
    \item\label{is-relaxed} there exists a \RV{} submodule
      $\mR'\subset\mR$ on the same level in $\mR$ as the
      vector~$\ket{e}$;

    \item\label{no-relaxed} there exists exactly one twisted Verma
      (respectively, Verma) submodule in \,$\mR$ on the same level as
      $\ket{e}$ such that
      $J^0_0\ket{e'}=j'\ket{e'}$ with $j'=j+1$ (respectively, $j-1$).
      Then   
      there are no \RV{} submodules in \,$\mR$ on the same level as
      $\ket{e}$.
    \end{enumerate}
  \end{enumerate}
\end{thm}
We see from Part~\ref{in-Verma} that Verma submodules (respectively,
twisted Verma submodules) exist in $\mR$ if and only if there are
charged singular vectors~\req{chargedsl2} with $n<0$ (respectively,
$n\geq0$).  For our present purposes, the significance of this fact is
that, whenever a charged singular vector exists in $\mR$, much of the
embedding structure of $\mR$ can be deduced from the embedding diagram
of the Verma module \,$\mC$ generated from the charged singular vector.
\RV{} submodules in $\mR$ correspond to submodules in~\,$\mC$. Since
the enumeration of the latter is a classical result~\cite{[RCW],[Mal]},
we recover the embedding structure of \RV{} submodules in~\,$\mR$.

\medskip

Even more generally, irrespective of whether or not there are charged
singular vectors in the \RV{} module $\mR$, one can still establish a
correspondence between relaxed singular vectors in $\mR$ and singular
vectors in certain Verma modules.

We will say that condition~\C{} is satisfied for a Verma submodule
$\mM'\subset\mM_{j,t}$ if

\smallskip

\noindent
(\C{})\qquad\parbox[t]{.9\textwidth}{$\mM'\subset\mM_{j,t}$,
  $t=\frac{p}{q}<0$ is rational and negative, $(2j+1)/(2p)\in\oZ$, and
  the weight of $\mM'$ is antidominant.}

\begin{thm}\label{thm:relax} Let $\mR\equiv\mR_{\mu_1,\mu_2,t}$ be a
  \RV{} module 
  and let $\aM=\mM_{j,t}$ be the Verma module with spin $j =
  \half(\mu_1-\mu_2-1)$.
  \begin{enumerate}\addtolength{\parskip}{-6pt}

  \item\label{1:1} If $\mu_1-\mu_2 \notin \oZ$, then \RV{} submodules
    in \,$\mR$ are in a $1:1$ correspondence with submodules in
    \,$\aM$, the corresponding submodules being on the same
    level in $\aM$ and in \,$\mR$.
    Moreover, two submodules \,$\mM'\subset\aM$ and
    \,$\mM''\subset\aM$ are related by an embedding,
    \,$\mM''\subset\mM'$, and only if the respective \RV{} submodules
    in \,$\mR$ are related by the embedding
    \,$\mR''\subset\mR'$.

  \item\label{2:1} If $\mu_1-\mu_2\in\oZ$, then each submodule
    \,$\mM'\subset\aM$ that does not satisfy condition~\C{} corresponds
    to a \RV{} submodule in \,$\mR$. Two Verma modules
    \,$\mM'\subset\aM$ and \,$\mM''\subset\aM$ correspond in this way
    to the same submodule in \,$\mR$ if and only if \,$\mM'$ and
    \,$\mM''$ are on the same level in~\,$\aM$.

    Vice versa, to every \RV{} submodule $\mR'\subset\mR$ on level $l$
    there corresponds at least one Verma submodule $\mM'\subset\aM$ on
    level~$l$.
\end{enumerate}
\end{thm}

We now outline the proof of this Theorem in the case where
$\mu_1\notin\oZ$ and $\mu_2\notin\oZ$ (the case where the integers
among $\{\mu_1,\mu_2\}$ give rise to one charged singular vector or to
two charged singular vectors of the same twist is obtained by minor
modifications, while  the case with two charged singular
vectors of different twists is more complicated and we refer the
reader to Theorem~\ref{thm:special}).

Using Eqs.~\req{properties} (and similar relations involving
$(J^+_0)^\alpha$), it is easy to check that, even though $\mu_\alpha$
is complex, the states
\begin{equation}
  (\Jminus_0)^{-\mu_\alpha}\,\ket{\mu_1,\mu_2, t}
  \qquad {\rm and}\qquad
  (\Jplus_0)^{\mu_\alpha+1}\,\ket{\mu_1,\mu_2, t}
  \label{auxVerma}
\end{equation}
formally satisfy the Verma \hw{} conditions~\req{sl2higgeneral} with
$\theta=0$ or $1$ respectively. The {\it auxiliary Verma modules} are
defined as the Verma modules built on states~\req{auxVerma}. We denote
these modules as $\aM^-_{1,2}$ and $\aM^+_{1,2}$, where the subscript
corresponds to taking $\mu_\alpha=\mu_{1,2}$ and the superscript, to
the respective vectors in~\req{auxVerma}. We see that
\begin{equation}\new
  \begin{array}{rclcrcl}
    \aM^-_1&\approx&\mM_{\half(\mu_1 - \mu_2 - 1),t}\,,
    &~&\aM^+_1&\approx&\smM_{\half(t + \mu_1 - \mu_2 - 1),t;1}\,,\\
    \aM^-_2&\approx&\mM_{\half(\mu_2 - \mu_1 - 1),t}\,,
    & &\aM^+_2&\approx&\smM_{\half(t + \mu_2 - \mu_1 - 1),t;1}\,.
  \end{array}
\end{equation}
Now, the condition that there be singular vectors in the auxiliary
(twisted) Verma modules reads as follows:
\begin{equation}\new
  \begin{array}{rcl}
    \half(\pm(\mu_1-\mu_2)-1) = \jplus(r, s, t)
    &\Longleftrightarrow&
    \half(t \mp (\mu_1-\mu_2) - 1) = \jminus(r,s,t)\,,\\
    \half(\pm(\mu_1-\mu_2)-1) = \jminus(r, s, t)
    &\Longleftrightarrow&
    \half(t \mp (\mu_1-\mu_2) - 1) = \jplus(r,s,t)\,.
  \end{array}
  \label{conditions}
\end{equation}
We now see that singular vectors exist in (at least one of) the
auxiliary Verma modules if and only if $\mu_1 - \mu_2 \in\oK(t)$.  For
example, singular vectors in the auxiliary Verma module $\aM^-_1$ and
in the twisted auxiliary Verma module $\aM^+_2$ read as
\begin{equation}\label{work10}
  \kern-4pt\new
  \begin{array}{ll}
    {\cal MFF}^-(r, s, t)\,(\Jminus_0)^{-\mu_1}\,
    \ketSL{\mu_1,\mu_2,t}\,,&
    {\cal MFF}^+(r', s', t)\,(\Jminus_0)^{-\mu_1}\,
    \ketSL{\mu_1,\mu_2,t}\,,\\
    {\cal MFF}^{+, 1}(r, s, t)\,
    (\Jplus_0)^{\mu_2 + 1}\,
    \ketSL{\mu_1,\mu_2, t}\,,&
    {\cal MFF}^{-, 1}(r', s', t)\,
    (\Jplus_0)^{\mu_2 + 1}\,
    \ketSL{\mu_1,\mu_2, t}\,,
  \end{array}~
  \left\{\kern-4pt
    \begin{array}{l}
     \mu_1 - \mu_2 = -r+ts\,,\\
     \mu_1 - \mu_2 = r'-t(s'-1)\,,\\
      r,s,r',s'\in\oN\,,
    \end{array}
  \right.
\end{equation}
where ${\cal MFF}^-(r, s, t)$ is the singular vector {\it operator\/}
(read off by dropping the \hw{} state in~\req{mff}), and ${\cal
  MFF}^{+, \theta}(r, s, t)$ is the spectral flow transform
\req{spectralsl2} of the singular vector operator~${\cal MFF}^+(r, s,
t)$.  Similar expressions for the other two auxiliary vectors are
obtained by changing $\mu_1 \leftrightarrow \mu_2$.

Note that, because of equivalence \req{conditions}, the auxiliary
Verma module $\aM^-_1$ and the twisted module $\aM^+_2$ have the
embedding diagrams that are mirror-symmetric to each other with
respect to charge in the charge-level lattice: we see
from~\req{conditions} that a singular vector exists in $\aM^-_1$ at
level $l$ if and only if a singular vector exists in $\aM^+_2$ at the
same level~$l$. If the singular vector in \,$\aM^-_1$ is given by one
of the expressions in the first line of~\req{work10}, then the vector
in $\aM^+_2$ is given by the corresponding \hbox{expression in the
  second line of~\req{work10}.}

The construction of~\cite{[FST]}, which we review in a moment, maps
singular vectors~\req{work10} in the auxiliary modules into the states
in $\mR_{\mu_1,\mu_2,t}$ that satisfy the relaxed \hw{} conditions.
Then in all cases except those described in
Part~\ref{general-relaxed}(ii) of Theorem~\ref{thm:3types}, there
exists~\cite{[FST]} at least one extremal state that generates a \RV{}
submodule, therefore condition $\mu_1 - \mu_2 \in\bar\oK(t)$ is
sufficient in order that a relaxed singular vector exist in
$\mR_{\mu_1,\mu_2,t}$. That the condition $\mu_1 - \mu_2 \in\oK(t)$ is
necessary follows from the equivalence with the $\N2$ superconformal
representation theory that was proved independently in~\cite{[FST]}.
On the $\N2$ side, the expression of the Ka\v c determinant is known,
with this condition guaranteeing a zero (which is at the same time
{\it not\/} the one corresponding to a charged singular vector).

To map singular vectors \req{work10} back to $\mR_{\mu_1,\mu_2, t}$, we
use, again, complex powers of $J^\pm_0$. In order that no non-integral
powers of $J^\pm_0$ remain in the final expressions, we act on the
respective states in~\req{work10} with $(\Jminus_0)^{\mu_1 + N}$ and
$(\Jminus_0)^{-\mu_2 - 1 + N}$, where $N$ is an integer, and make use
of Eqs.~\req{properties}.  Analysing the relative $J^0_0$ charge and
level of singular vectors~\req{mff}, it is not difficult to see that,
in order to be left with only {\it positive\/} integral powers after
  the rearrangements, the integer $N$ has to be $\geq r+rs$.  We thus
define
\begin{equation}
  \ket{\Sigma^-_1(r,s,\mu_1+\mu_2,t)}=
  (\Jminus_0)^{\mu_1 + r + r s}\,
  {\cal MFF}^-(r, s, t)\,(\Jminus_0)^{-\mu_1}\,
  \ketSL{\mu_1,\mu_2, t},
  \quad \mu_1 - \mu_2 = -r + ts\,,\quad r,s\in\oN
  \label{sigmaminus}
\end{equation}
and
\begin{equation}
  \ket{\Sigma^+_2(r,s,\mu_1+\mu_2,t)}=
  (\Jplus_0)^{- \mu_2 - 1 + r + r s}
  {\cal MFF}^{+, 1}(r, s, t)\,
  (\Jplus_0)^{\mu_2 + 1}\,\ketSL{\mu_1,\mu_2, t},~
  \mu_1 - \mu_2 = -r + ts\,,~r,s\in\oN \,,
  \label{sigmaplus}
\end{equation}
and also $\ket{\Sigma^-_2(r,s,\mu_1+\mu_2,t)}$ and
$\ket{\Sigma^+_1(r,s,\mu_1+\mu_2,t)}$, which are obtained by replacing
$\mu_1 \leftrightarrow \mu_2$.  Note that we do not discuss the
vectors constructed by applying the ${\cal MFF}^{+}$ operators to
$(\Jminus_0)^{-\mu_\alpha}\cdot\ketSL{\mu_1,\mu_2, t}$ and the ${\cal
  MFF}^{-,1}$ operators, to $(\Jplus_0)^{\mu_\alpha
  +1}\,\ketSL{\mu_1,\mu_2, t}$.  The reason is that these differ from
the singular vectors in the auxiliary Verma (twisted Verma) module
corresponding to the complementary $\mu$ by a power of $\Jminus_0$
(resp., of $\Jplus_0$) and hence the construction of the type of
\req{sigmaminus} (resp.,~\req{sigmaplus}) produces the same vectors in
the \RV{} module. Also, there is no need to consider all the four
auxiliary Verma modules built on vectors~\req{auxVerma}, since singular
vectors in these modules are correlated in accordance
with~\req{conditions}.

Thus, \ $\aM^-_1\ni{\cal MFF}^+(r, s, t)\,(\Jminus_0)^{-\mu_1}\,
\ketSL{\mu_1,\mu_2,t}$ with $s\neq1$ if and only if $\aM^-_2\ni {\cal
MFF}^-(r, s-1, t)\cdot(\Jminus_0)^{-\mu_2}\,\ketSL{\mu_1,\mu_2,t}$;
similarly, $\aM^+_2\ni{\cal MFF}^{-, 1}(r, s, t)\, (\Jplus_0)^{\mu_2 +
1}\, \ketSL{\mu_1,\mu_2, t}$ with $s\neq1$ if and only if $\aM^+_1\ni
{\cal MFF}^{+, 1}(r, s-1, t)\, (\Jplus_0)^{\mu_1 + 1}\,
\ketSL{\mu_1,\mu_2, t}$ (where $\mu_1 - \mu_2 = -r+t(s-1)$,
$r,s\in\oN$).

The basic facts about singular vectors~\req{sigmaminus} and
\req{sigmaplus} are as follows:
\begin{thm}[\cite{[FST]}]\label{thm:vectors}\addtolength\parskip{-4pt}
  Let $\mR=\mR_{\mu_1,\mu_2,t}$ be a \RV{} module with
  $\mu_\alpha\notin\oZ$.
  \begin{enumerate}
  \item Expressions~\req{sigmaminus} and~\req{sigmaplus} evaluate as
    elements of the \RV{} module $\mR$ and satisfy the relaxed-\hw{}
    conditions
    \begin{equation}\kern-8pt\new
      \begin{array}{l}
        J^+_{\geq1}\,\ket{\Sigma^\pm_{1,2}(r,s,\mu_1+\mu_2,t)}=
        J^0_{\geq1}\,\ket{\Sigma^\pm_{1,2}(r,s,\mu_1+\mu_2,t)}=
        J^-_{\geq1}\,\ket{\Sigma^\pm_{1,2}(r,s,\mu_1+\mu_2,t)}=0\,,\\
        J^-_{0}J^+_{0}\,\ket{\Sigma^\pm_{1,2}(r,s,\mu_1+\mu_2,t)}=
        -\mu_1^\pm\mu_2^\pm\,\ket{\Sigma^\pm_{1,2}(r,s,\mu_1+\mu_2,t)}\,,\\
        J^0_{0}\,\ket{\Sigma^\pm_{1,2}(r,s,\mu_1+\mu_2,t)}=
        -\half(\mu_1^\pm + \mu_2^\pm + 1)\,
        \ket{\Sigma^\pm_{1,2}(r,s,\mu_1+\mu_2,t)}\,,
      \end{array}
      \label{hwsing}
    \end{equation}
    where $\mu_{1,2} - \mu_{2,1} = -r + ts$ (with the first or the
    second subscript taken depending on whether {\rm(}$\Sigma_1^-$ and
    $\Sigma_2^+${\rm)} or {\rm(}$\Sigma_2^-$ and $\Sigma_1^+${\rm)}
    are being considered) and
    \begin{equation}\new
      \begin{array}{rcl}
        \mu_1^\pm &=&\pm\frac{r}{2} \pm \frac{t}{2}s \mp rs +
        \frac{\mu_1+ \mu_2}{2}\,,\\
        \mu_2^\pm &=&\mp\frac{r}{2} \mp \frac{t}{2}s \mp rs +
        \frac{\mu_1 + \mu_2}{2}\,.
      \end{array}
      \label{Lambdaplusminus}
    \end{equation}

  \item Each of the vectors
    $\ket{\Sigma_{1,2}^\pm(r,s,\mu_1+\mu_2,t)}$ generates a \RV{}
    submodule. Moreover, the vectors
    $\ket{\Sigma^-_{1}(r,s,\mu_1+\mu_2,t)}$ and
    $\ket{\Sigma^+_{2}(r,s,\mu_1+\mu_2,t)}$ and, on the other hand,
    $\ket{\Sigma^-_{2}(r',s',\mu_1+\mu_2,t)}$ and
    $\ket{\Sigma^+_{1}(r',s',\mu_1+\mu_2,t)}$, are on the same level
    in~$\mR$ and, moreover, each vector from the respective pair is a
    $J^\pm_0$-descendant of the other:
    \begin{equation}\kern-8pt\new
      \begin{array}{rcl}
        \ket{\Sigma^-_{1,2}(r,s,\mu_1+\mu_2,t)}\kern-5pt&=&\kern-5pt
        c^-_{1,2}\,(J^-_0)^{2rs}\,
        \ket{\Sigma^+_{2,1}(r,s,\mu_1+\mu_2,t)}\,,  \\
        \ket{\Sigma^+_{1,2}(r,s,\mu_1+\mu_2,t)}\kern-5pt&=&\kern-5pt
        c^+_{1,2}\,(J^+_0)^{2rs}\,
        \ket{\Sigma^-_{2,1}(r,s,\mu_1+\mu_2,t)}\,,
      \end{array}
      \label{lercompare}
    \end{equation}
    where the numerical coefficients $c^{\pm}_{1,2}$ depend on $r$,
    $s$, $\mu_1+\mu_2$, and~$t$.
  \end{enumerate}
\end{thm}
This construction can be illustrated in the following extremal
diagram:
\begin{equation}
  \unitlength=1pt
  \begin{picture}(250,60)
    \put(30,0){
      {\linethickness{.7pt}
        \bezier{70}(87.5,45)(22.5,12)(-60,41)
        \bezier{30}(-78,26)(-22.5,21.5)(0,5)
        \bezier{70}(96.5,45)(163.5,12)(244,40)
        \bezier{30}(262,26)(206.5,19.5)(184,5)
        }
      \put(-60,41){\vector(-3,1){1}}
      \put(0,5){\vector(3,-2){1}}
      \put(244,40){\vector(3,1){1}}
      \put(186,6){\vector(-3,-2){1}}
      \put(-66.5,38.5){$\bullet$}
      \put(246.5,38.5){$\bullet$}
      {\linethickness{0.5pt}
        \put(251.5,41.5){\line(1,0){20}}
        \put(247.7,40.3){\line(-1,-1){15}}
        }
      \put(20,-14){\put(246.5,38.5){$\bullet$}
        {\linethickness{0.5pt}
          \put(251.5,41.5){\line(1,0){20}}
          \put(247.7,40.3){\line(-1,-1){10}}
          }}
      {\linethickness{0.5pt}
        \put(-86.5,41.5){\line(1,0){20}}
        \put(-61.7,40.3){\line(1,-1){15}}
        }
      \put(-20,-15){\put(-66.5,38.5){$\bullet$}
        {\linethickness{0.5pt}
          \put(-86.5,41.5){\line(1,0){20}}
          \put(-61.7,40.3){\line(1,-1){15}}
          } }
      \put(0,36){${}_{{}{(\Jminus_0)^{-\mu_1}}}$}
      \put(164,36){${}_{{}{(\Jplus_0)^{\mu_2+1}}}$}
      \put(-60,13){${}_{{}^{(\Jminus_0)^{\mu_1+r+rs}}}$}
      \put(208,13){${}_{{}^{(\Jplus_0)^{-\mu_2-1+r+rs}}}$}
      \put(0,45){
        \put(-35,2){\Large $\ldots$}
        \put(0,0){$\state$}
        \put(15,5){${}^{J^-_0}$}
        \put(28,3){\vector(-1,0){22}}
        \put(30,0){$\state$}
        \put(45,5){${}^{J^-_0}$}
        \put(58,3){\vector(-1,0){22}}
        \put(60,0){$\state$}
        \put(75,5){${}^{J^-_0}$}
        \put(88,3){\vector(-1,0){22}}
        \put(90,0){$\star$}
        \put(100,5){${}^{J^+_0}$}
        \put(97,3){\vector(1,0){22}}
        \put(120,0){$\state$}
        \put(130,5){${}^{J^+_0}$}
        \put(127,3){\vector(1,0){22}}
        \put(150,0){$\state$}
        \put(160,5){${}^{J^+_0}$}
        \put(157,3){\vector(1,0){22}}
        \put(180,0){$\state$}
        \put(193,2){\Large $\ldots$}
        }
      \put(-35,2){\Large $\ldots$}
      \put(0,0){$\state$}
      \put(15,5){${}^{J^-_0}$}
      \put(28,3){\vector(-1,0){22}}
      \put(30,0){$\state$}
      \put(45,5){${}^{J^-_0}$}
      \put(58,3){\vector(-1,0){22}}
      \put(60,0){$\state$}
      \put(75,5){${}^{J^-_0}$}
      \put(88,3){\vector(-1,0){22}}
      \put(90,0){$\star$}
      \put(100,5){${}^{J^+_0}$}
      \put(97,3){\vector(1,0){22}}
      \put(120,0){$\state$}
      \put(130,5){${}^{J^+_0}$}
      \put(127,3){\vector(1,0){22}}
      \put(150,0){$\state$}
      \put(160,5){${}^{J^+_0}$}
      \put(157,3){\vector(1,0){22}}
      \put(180,0){$\state$}
      \put(193,2){\Large $\ldots$}
      }
  \end{picture}
  \label{twofloors}
\end{equation}
The lower floor is the extremal diagram of the \RV{} {\it
  sub\/}module, every point in the lower floor satisfying the relaxed
\hw{} conditions~\req{floorhw}. The auxiliary Verma module (and its
submodule) on the right are `rotated' by the spectral flow transform
with~$\theta=1$.  The submodules arise in the auxiliary Verma modules
simultaneously, in accordance with the above statements.  The
auxiliary Verma modules are to be thought of as disconnected from the
extremal diagram of the \RV{} module~$\mR_{\mu_1,\mu_2,t}$, since they
do not belong to~$\mR_{\mu_1,\mu_2,t}$ as long as the mappings shown
in dotted lines are given by complex powers of the generators.

\medskip

To return to Theorem~\ref{thm:relax}, we see that (in the case where
with the relevant parameters do {\it not\/} become integers), the
structure of embeddings of the \RV{} submodules repeats the
Verma-module embedding diagram for each of the auxiliary Verma
modules, while all of the latter are identical up to mirror-symmetry.
Theorem~\ref{thm:vectors} ensures that the
respective pairs of singular vectors in $\aM_1^-$ and in $\aM_2^+$ give
rise to the same \RV{} submodule in~$\mR_{\mu_1,\mu_2,t}$. This shows
Part~\ref{1:1} of Theorem~\ref{thm:relax}.

On the other hand, in the case where $s=1$, which is possible when
$\Delta\equiv\mu_2-\mu_1\in\oZ$, one of the auxiliary Verma modules is
embedded into the other, and similarly for the twisted auxiliary
modules; if, for definiteness, $\Delta\geq0$, we have
\begin{equation}\kern-10pt\new
  \begin{array}{lclcl}
    (\Jminus_0)^{-\mu_1}\, \ketSL{\mu_1,\mu_2,t}\,&=&
  (\Jminus_0)^{\Delta}\, \left( (\Jminus_0)^{-\mu_2}\,
    \ketSL{\mu_1,\mu_2,t} \right) \,&=&
    {\cal MFF}^-(\Delta, 1, t)\,(\Jminus_0)^{-\mu_2}\,
    \ketSL{\mu_1,\mu_2,t}\,,\\
    (\Jplus_0)^{\mu_2 +1}\, \ketSL{\mu_1,\mu_2,t}\,&=&
  (\Jplus_0)^{\Delta}\, \left( (\Jplus_0)^{\mu_1 +1}\,
    \ketSL{\mu_1,\mu_2,t} \right) \,&=&
    {\cal MFF}^{+,1}(\Delta, 1, t)\,(\Jplus_0)^{\mu_1 +1}\,
    \ketSL{\mu_1,\mu_2,t}\,.\kern-20pt
  \end{array}
  \label{lerintdelta}
\end{equation}
Now, in the situation where $\Delta \in \oK(t)$---i.e., the auxiliary
modules contain submodules---the standard fact about Verma-module
embedding diagrams is that submodules in the auxiliary modules occur
in pairs, with the \hw{} vectors of the submodules from the same pair
embedded onto the same level. Moreover, one of these two \hw{} vectors
is a $\Jminus_0$-descendant of the other. We thus see
that the corresponding $\Sigma^-$ vectors generate the same \RV{}
submodule in~$\mR_{\mu_1,\mu_2,t}$. The same is true for the twisted
auxiliary Verma module.  We thus arrive at Part~\ref{2:1} of
Theorem~\ref{thm:relax} in the case where none of the $\mu_\alpha$ are
integers.

\medskip

Let us now discuss briefly the case where exactly one of the
$\mu_{\alpha}$ is an integer. Then one of the vectors \req{auxVerma}
becomes a charged singular vector, and the corresponding (twisted)
auxiliary module becomes a Verma submodule of~$\mR_{\mu_1,\mu_2,t}$,
while the \RV{} submodules in $\mR$ are still generated from the other
auxiliary Verma modules in accordance with the above procedure.  One
of the relations \req{lercompare} may not hold then if the
corresponding $\Sigma^\pm$ vector lies in the Verma submodule $\mC$
generated from a charged singular vector in \RV{} submodule $\mR'$.
Then, obviously, the corresponding vector $\Sigma^+$ (or $\Sigma^-$)
does not generate the entire \RV{} module $\mR'$. However, the other
vector $\Sigma^-$ (respectively, $\Sigma^+$) does generate the \RV{}
submodule.

Next, in the case where the \RV{} module~$\mR_{\mu_1,\mu_2,t}$
contains two charged singular vectors of the same twist, the \RV{}
submodules may be generated by the auxiliary Verma modules with the
complementary twist.  Then, we have $\mu_1-\mu_2 \in \oZ$, hence each
\RV{} submodule corresponds to a pair of singular vectors in the
auxiliary Verma module and contains two charged singular vectors.  A
minor additional problem is encountered in the case where $\mu_1$
equals $-1$ or $0$. 
Then one may be unable to arrive at \hw{} state of the auxiliary Verma
module using formulae~\req{auxVerma}. However, replacing the \hw{}
vector with a different point from the orbit (see
Definition~\ref{def:orbit}), the situation is reduced to the previous
one.

Further, in the case where $\mu_1=\mu_2 \in \oZ$, there is only one
charged singular vector, however if $0=\mu_1-\mu_2 \in \oK(t)$ (which
may be the case only when $t \in \oQ\,, \ t>0$), then there are \RV{}
submodules in~\,$\mR_{\mu_1,\mu_2,t}$, each of which contains two
charged singular vectors (obviously, with the same sign of~$n$
from~\req{chargedsl2}) and, therefore, each \RV{} submodule corresponds
to a pair of Verma-module singular vectors.

\medskip

Finally, the case where $\mu_1\in -\oN$ and $\mu_2\in\oN_0$, is more
involved. It will be described in more detail in
Sec.~\ref{subsec:diagrams}, paragraphs~{\bf\IIIpmziimp}
and~{\bf\IIIpmzziimp}, while now we only formulate the following
general result~\cite{[FST]}:
\begin{thm}\label{thm:special} Let $\mR\equiv\mR_{\mu_1,\mu_2,t}$ be
  a \RV{} module, and $\mu_1\in -\oN$ and $\mu_2\in\oN_0$. Let then
  $\mC_-$, and $\mC_+$ be the corresponding Verma module and the
  twisted Verma module, respectively, generated from the charged
  singular vectors.  Then
  \begin{enumerate}\addtolength{\parskip}{-6pt}


  \item whenever two Verma modules $\widetilde\mC''_-$ and $\mC''_-$
    are embedded onto the same level~$l$ in \,$\mR$, there exist two
    twisted Verma modules $\widetilde\mC''_+$ and $\mC''_+$ embedded
    onto the same level, and vice versa. In this case, there is a
    \RV{} submodule $\mR'$ whose \hw{} vector is on the same
    level~$l$. We label the modules so that
    $\widetilde\mC''_-\supset\mC''_-$ and
    $\widetilde\mC''_+\supset\mC''_+$. Then we have the {\sl
      embeddings\/}
    \begin{equation}
      \label{direct-sum}
      \begin{array}{c}
        \widetilde\mC''_-\oplus\widetilde\mC''_+\\
        \nearrow~~~~\Bigm\uparrow~~~~\nwarrow\\
        \mC''_-\longrightarrow\mR'\longleftarrow\mC''_+
      \end{array}
    \end{equation}

  \item if there is exactly one Verma submodule $\mC'_-$ embedded on
    a given level $l$ in $\mR$, then there also exists exactly one
    twisted Verma module $\mC'_+$ on the same level, and vice versa.
    In this case, one of the following is true:
    \begin{enumerate}
    \item $\mC'_-$ ($\mC'_+$) satisfies condition~\C{} in $\mC_-$
      (respectively, $\mC_+$); then there are no \RV{} submodules on
      any level~$\geq l$ in~$\mR$;

    \item\label{maximal} the Verma module $\mC'_-$ (hence, also
      $\mC'_+$) contains proper submodule(s); then there exists a
      \RV{} submodule $\mR'$ on level $l$ in $\mR$ and, moreover,
      $\mC'_-$ and $\mC'_+$ are generated from the charged singular
      vectors in $\mR'$. In this case, further, $\mR'$ is a maximal
      submodule in $\mR$ in the following sense: for a \RV{} module
      \,$\tilde\mR$, we have \,$\mR'\subset\tilde\mR \subset\mR$
      $\Longrightarrow$ \,$\tilde\mR=\mR'$ or \,$\tilde\mR=\mR$.
    \end{enumerate}
  \end{enumerate}
\end{thm}

\section{Classifying the embedding
  diagrams\label{sec:classifying}}\lvm As we have seen, singular
vectors in \RV{} modules are of two basic types: the charged ones,
which occur directly in the extremal diagram, and all the others,
i.e., those occurring `inside' the module, which can be either
(twisted) Verma modules or the \RV{} modules.  The crucial point is
that, in the general position, the \RV{} submodules are determined by
the usual singular vectors in the auxiliary Verma modules.
Accordingly, the embedding diagrams of \RV{} modules can be determined
knowing the standard $\tSL2$ Verma-module embedding diagrams and
analysing the possible existence of charged singular vectors. The
appearance of a charged singular vector can be interpreted as the case
where the entire embedding diagram of auxiliary Verma module is
attached to the `relaxed' embedding diagram.  When this happens with
two charged singular vectors, the rule according to which the
`relaxed' part of the diagram is determined by its Verma parts becomes
slightly more complicated, but it is still true that the `relaxed'
part can be reconstructed knowing how the {\it Verma\/} modules are
embedded into each other. One can also have `combined' degenerations,
where charged singular vectors appear simultaneously with further
degenerations occurring in the Verma embedding diagrams.  A useful
observation is that whenever there is at least one charged singular
vector, then, taking the {\it quotient\/} (with respect to the maximal
submodule if there are two charged singular vectors of the same
twist), we are left with the known embedding diagrams of the usual
Verma modules. This, in particular, gives another way to recover the
standard $\tSL2$ Verma-module embedding diagrams and serves as a good
check on the relaxed embedding diagrams given below.

\subsection{Classifying the degeneration
  patterns\label{subsec:list}}\lvm We now apply the above Theorems in
order to derive the classification of embedding diagrams of \RV{}
modules. The result consists in the embedding diagrams of the \RV{}
modules $\mR_{\mu_1, \mu_2, t}$ listed below. The different entries of
the following list (summarized in the Table
on~p.~\pageref{table:table}) are labelled using the familiar I-II-III
pattern~\cite{[FF],[FFr]}, with an additional indication of how many
(0, 1, or 2) of the $\mu_\alpha$ are integers and what their signs are.
Thus, numbers 0, 1, and 2 refer to the existence of submodules
generated from charged singular vectors, whereas the Roman numbers I,
II, and III indicate the existence of \RV{} submodules (0, 1, and
$\geq2$, respectively).  Whenever there are submodules generated from
charged singular vectors, we also use the signs $-$ or $+$ to indicate
whether these correspond to negative or non-negative $n$
in~\req{chargedsl2}, respectively. The signs can also be read as the
indication of the twists: the modules generated from charged singular
vectors with $n\leq-1$ have twist~0, while those generated from charged
singular vectors with $n\geq0$, twist~1.  Thus, for example,
III{}$(2,{-}{+})$ indicates two submodules generated from charged
singular vectors, one of which is untwisted and the other is twisted
(always by~$\theta=1$).  Note, however, that we use the notation
III$_+$ for the {\it positive\/} zone $t>0$, and III$_-$ for the
negative zone $t<0$, which is opposite to the notations used by some
other authors.  Further, whenever exactly one of $\mu_1$, $\mu_2$ is an
integer, we take this to be $\mu_1$; when both $\mu_1$ and $\mu_2$ are
integers, we choose $\mu_1\leq\mu_2$.  Finally, for a rational $t$ in
the negative zone, $t=-\tp/q$, we will be interested in the total
number of submodules in a given \RV{} module. Whenever there {\it is\/}
  at least one \RV{} submodule, we have $\mu_1-\mu_2\in\bar\oK(t)$
(hence, in particular, $\mu_1-\mu_2\in\oQ$); in fact, already for
$\mu_1-\mu_2\in\oK(t)$, we can write
\begin{equation}
  |\mu_1-\mu_2| = \tp\xi + \zeta + \eta \frac{\tp}{q}
  \label{L-condition}
\end{equation}
where, in general, $0\leq\zeta<\tp$, $0\leq\eta<q$, and $\xi\in\oN_0$,
with $\xi^2 + \zeta^2\neq0$, $\xi^2 + \eta^2\neq0$, and $\xi^2 +
\eta^2 + \zeta^2\neq1$.  Then $\xi$ is responsible for the number
of submodules, as we will see in each of the particular cases.

An important remark is in order regarding how we divide the set of
$(\mu_1,\mu_2,t)$ into different cases.  One may classify the
different embedding diagram patterns using, among others, the same
criterion as in~\cite{[FF]}, where the cases were singled out
according to the number of integral points on a certain line in the
$(\mu_1-\mu_2,t)$ plane. Instead, we have chosen a more `direct'
classification according to the values of $\mu_1$, $\mu_2$,~etc.
However, the price to be paid is that we have to explicitly point out
that some of the diagrams with negative rational $t$ --- namely, those
in which there are a `minimal' number of submodules --- are identical
to the irrational-$t$ diagrams, where the very fact that $t\notin\oQ$
implies that only the `minimal' number of submodules be possible. That
is, while the generic situation for the rational $t$ is that there
exists a chain of embedded submodules, this chain becomes finite in
the negative zone and, moreover, may contain at most one submodule of
each kind when $\xi=0,1,2,3$ in Eq.~\req{L-condition}. We believe that
this obvious fact does not confuse the classification of different
degeneration patterns, therefore we will characterize the I-II-III
cases by the `generic' conditions (e.g., $t\notin\oQ$).  To be
completely explicit, however, we \hbox{will also point out such
  `small-$\xi$ exceptions' explicitly.}

Now, the classification is as follows.
\setlength{\leftmargini}{18pt}
\begin{enumerate}

  \def\theenumi{\Roman{enumi}} \renewcommand\labelenumi{\theenumi:}
  \renewcommand\labelenumii{\theenumi\theenumii:}
  \renewcommand\labelenumiii{\theenumi\theenumii,\theenumiii):}
\item{}\label{l:I} $\mu_1-\mu_2\notin\oK(t)$.  The simple diagrams
  shown in~\req{d:I} and~\req{d:I(2)} exhaust the possibilities that
  can occur in this case. Namely, the different cases are as follows:
  \vspace{-6pt}
  \setlength{\leftmarginii}{27pt}
  \begin{enumerate}

    \addtocounter{enumii}{-1} \def\theenumii{(\arabic{enumii})}
  \item \label{l:I(0)} $\mu_1\notin\oZ$, $\mu_2\notin\oZ$. \ This is
    the trivial case, the embedding diagram consisting of the lonely
    \RV{} module.

  \item \label{l:I(1)} $\mu_1\in\oZ$, $\mu_2\notin\oZ$ or
    $\mu_1=\mu_2\in\oZ$. In this case we have a single Verma or a
    twisted Verma submodule: \def\theenumii{(\arabic{enumii}}

    \begin{enumerate}
      \addtolength{\itemindent}{18pt} \def\theenumiii{${-}$}
    \item a Verma module if $\mu_1\in-\oN$, or

      \def\theenumiii{${+}$}
    \item a twisted Verma module if $\mu_1\in\oN_0$,
    \end{enumerate}
    which is embedded via a charged singular vector, Eq.~\req{d:I}.

    \def\theenumii{(\arabic{enumii})}
  \item \label{l:I(2)} $\mu_1\in\oZ$, $\mu_2\in\oZ$, $\mu_1\neq\mu_2$.
    \ Whenever $\mu_1\cdot\mu_2 > 0$, we have one of the following two
    `mirror-symmetric' cases depending on whether $\mu_1$ and $\mu_2$
    are negative or positive:

    \def\theenumii{(\arabic{enumii}} 

    \setlength{\leftmarginiii}{48pt}
    \begin{enumerate}
      \def\theenumiii{${-}{-}$}
    \item $\mu_1,\mu_2\in-\oN$, the first diagram
      in~\req{d:I(2)}.

      \def\theenumiii{${+}{+}$}
    \item $\mu_1,\mu_2\in\oN_0$, the second diagram in~\req{d:I(2)}.
    \end{enumerate}
    If, on the other hand, $\mu_1$ and $\mu_2$ are of different signs,
    we have
    \begin{enumerate}

      \def\theenumiii{${-}{+}$}
    \item \label{l:II(2,mp)} $\mu_1\in-\oN$, $\mu_2\in\oN_0$, with
      the third diagram in~\req{d:I(2)}.

    \end{enumerate}

  \end{enumerate}
  \vspace{-6pt}

\item \label{l:II} $\mu_1 - \mu_2\in\oK(t)$, $t\notin\oQ$. The first
  condition guarantees that there exists at least one \RV{} submodule;
  on the other hand, $t\notin\oQ$ implies that there will be not more
  than one \RV{} submodule.
    \vspace{-4pt}
    \setlength{\leftmarginii}{32pt}
  \begin{enumerate}

    \addtocounter{enumii}{-1}
    \def\theenumii{(\arabic{enumii})}
  \item \label{l:II(0)} $\mu_1\notin\oZ$, $\mu_2\notin\oZ$. In this
    case we have the first of the diagrams~\req{d:II}. In addition,
    this diagram can be viewed as a particular case of some of the
    diagrams of case III (those with rational $t$), namely with
    $t=-\frac{\tp}{q}$ and $|\mu_1-\mu_2|$ as in~\req{move2II0}.

  \item \label{l:II(1)} $\mu_1\in\oZ$, $\mu_2\notin\oZ$. In this case,
    we have the respective diagrams~\req{d:II}, depending on the sign
    of~$\mu_1$:
    \def\theenumii{(\arabic{enumii}} 
    \setlength{\leftmarginiii}{43pt}
    \begin{enumerate}

      \def\theenumiii{${-}$}
    \item $\mu_1\in-\oN$,

      \def\theenumiii{${+}$}
    \item $\mu_1\in\oN_0$.
    \end{enumerate}
    These two diagrams are mirror-symmetric in the obvious sense.

    In addition, the same diagram can be viewed as a particular case
    of some of the embedding diagrams of the rational-$t$ case III,
    namely those where $t=-\frac{\tp}{q}$ and $|\mu_1-\mu_2|$ is as
    in~\req{move2II1}.
  \end{enumerate}
  \vspace{-6pt}

\item{} $\mu_1 - \mu_2\in\oK(t)$, $t\in\oQ$. \ This case comprises all
  the most interesting embedding diagrams.
  \vspace{-4pt}

      \setlength{\leftmarginii}{28pt}
  \begin{enumerate}

    \def\theenumii{$_\pm$}
  \item{}$\mu_1 - \mu_2\notin\oZ$, $(\mu_1 - \mu_2)/t\notin\oZ$. \ We
    then have the following cases:

    \setlength{\leftmarginiii}{42pt}
    \begin{enumerate}
      \addtocounter{enumiii}{-1}
      \def\theenumiii{(\arabic{enumiii})}
      \renewcommand\labelenumiii{\theenumi\theenumii\theenumiii:}

    \item{} \label{l:IIIpm(0)} $\mu_1\notin\oZ$, $\mu_2\notin\oZ$.
      This corresponds to the double-chain~\req{d:IIIpm(0)} of \RV{}
      modules. The chain is finite or infinite depending on whether
      $t$ is negative/positive. In either case, the structure of the
      embedding diagram of the \RV{} module repeats~\cite{[FST]} the
      structure of the embedding diagram of the auxiliary Verma
      module.

    \item{} \label{l:IIIpm(1)} $\mu_1\in\oZ$, $\mu_2\notin\oZ$.
      Depending on the sign of $\mu_1$, we have one of the following
      `mirror-symmetric' cases:

      \setlength{\leftmarginiv}{54pt}
      \begin{enumerate}
        \def\theenumiii{(\arabic{enumiii}}
        \renewcommand\labelenumiv{\theenumi\theenumii\theenumiii,\theenumiv):}

        \def\theenumiv{${-}$}
      \item \label{l:IIIpm(1,m)} $\mu_1\in-\oN$, diagram
        \req{d:IIIpm(1)}. Embeddings via charged singular vectors (the
        horizontal arrows) repeat for every \RV{} module in the
        diagram. Again, the Verma dots are placed on the left of the
        boxes representing \RV{} modules, in accordance with how the
        Verma modules are embedded via charged singular vectors into
        the \RV{} module, see~\req{withVerma}.

        \def\theenumiv{${+}$}
      \item \label{l:IIIpm(1,p)} $\mu_1\in\oN_0$. We have diagram
        \req{d:RTV} with the open dots representing twisted Verma
        modules. In view of how the twisted Verma submodules appear in
        \RV{} modules, see~\req{rightVerma}, the twisted-Verma modules
        are drawn on the right of the relaxed ones.
      \end{enumerate}

    \end{enumerate}

    \def\theenumii{$_\pm^0$}
  \item{} {\it Either\/} ($\mu_1 - \mu_2\in\oZ$, $(\mu_1 -
    \mu_2)/t\notin\oZ$) {\it or\/} ($\mu_1 - \mu_2\notin\oZ$, $(\mu_1
    - \mu_2)/t\in\oZ$).

    \begin{enumerate}{}
      \addtocounter{enumiii}{-1}
      \def\theenumiii{(\arabic{enumiii})}
      \renewcommand\labelenumiii{\theenumi\theenumii\theenumiii:}

    \item{} \label{l:IIIpm0(0)} $\mu_1\notin\oZ$, $\mu_2\notin\oZ$,
      with the embedding diagram~\req{d:IIIpm0(0)}. One of the
      following two things happens in the auxiliary Verma module, with
      the same effect on the embedding diagram of \RV{} modules: \
      (a)~the auxiliary Verma-module embedding diagram degenerates
      into a single-chain (whenever $(\mu_1 - \mu_2)/t\in\oZ$); \
      (b)~it acquires embeddings via ${\rm MFF}^+(r, 1, t)$ singular
      vectors (whenever $\mu_1-\mu_2\in\oZ$, see Part~\ref{2:1} of
      Theorem~\ref{thm:relax}).  In either case, the `relaxed'
      double-chain~\req{d:IIIpm(0)} collapses to a single
      chain~\req{d:IIIpm0(0)}.

    \item{} \label{l:IIIpm0(1)} $\mu_1\in\oZ$, $\mu_2\notin\oZ$.
      This means that $\mu_1-\mu_2\notin\oZ$, hence we should have
      $(\mu_1 - \mu_2)/t\in\oZ$. Then there are two possibilities:

      \def\theenumiii{(\arabic{enumiii}}
      \renewcommand\labelenumiv{\theenumi\theenumii\theenumiii,\theenumiv):}
      \setlength{\leftmarginiv}{54pt}
      \begin{enumerate}

        \def\theenumiv{${-}$}
      \item \label{l:IIIpm0(1m)} $\mu_1\in-\oN$,
        diagram~\req{d:IIIpm0(1)};

        \def\theenumiv{${+}$}
      \item \label{l:IIIpm0(1p)} $\mu_1\in\oN_0$, the embedding
        diagram being the vertical mirror of~\req{d:IIIpm0(1)}, with
        the replacement {\Verma}${}\leadsto{}${\TVerma}.
      \end{enumerate}
      As can be seen, the single-chain of \RV{} modules is in this
      case due to the fact that the Verma embedding diagram is also a
      single chain (because of $(\mu_1 - \mu_2)/t\in\oZ$); as usual,
      $\mu_1\in\oZ$ implies that the Verma and the `relaxed' diagrams
      are attached to each other by embeddings via charged singular
      vectors.

      \def\theenumiii{(\arabic{enumiii})}
    \item{} \label{l:IIIpm0(2)}$\mu_1\in\oZ$, $\mu_2\in\oZ$. \ Since
      $\mu_1-\mu_2\in\oZ$, we should have $(\mu_1 -
      \mu_2)/t\notin\oZ$. Then, there are the following cases
      depending on the signs of $\mu_1$ and $\mu_2$:

      \setlength{\leftmarginiv}{62pt}
      \begin{enumerate}
        \def\theenumiii{(\arabic{enumiii}}

        \def\theenumiv{${-}{-}$}
      \item \label{l:IIIpm0(2,mm)} $\mu_1\in-\oN$, $\mu_2\in-\oN$,
        with the diagram \req{d:IIIpm0(2)}, in which each pair of
        Verma modules on the same level (i.e., with one embedded into
        the other by an ${\rm MFF}^+(r, 1, t)$ singular vector)
        correspond to one \RV{} module.

        \def\theenumiv{${+}{+}$}
      \item \label{l:IIIpm0(2,pp)} $\mu_1\in\oN_0$, $\mu_2\in\oN_0$. \
        We have a vertical mirror of~\req{d:IIIpm0(2)}, with the Verma
        modules replaced by spectral-flow-transformed Verma modules
        with $\theta=1$.

        \def\theenumiv{${-}{+}$}
      \item{} \label{l:IIIpm0(2,mp)} $\mu_1\in-\oN$, $\mu_2\in\oN_0$,
        diagram~\req{d:IIIpm0(2,mp)}. This case is considerably
        different from the previous ones, since two charged singular
        vectors in the \RV{} module appear on different sides of the
        \hw{} vector.  Thus, both Verma and twisted-Verma submodules
        are present simultaneously.\pagebreak[3]

      \end{enumerate}

    \end{enumerate}

    \def\theenumii{$_\pm^{00}$}
  \item $\mu_1 - \mu_2\in\oZ$, $(\mu_1 - \mu_2)/t\in\oZ$. Setting
    $t=\frac{p}{q}$, we have $(\mu_1 - \mu_2)/t\in\oZ$ if and only if
    $(\mu_1 - \mu_2)/p\in\oZ$.

    \setlength{\leftmarginiv}{24pt}
    \begin{enumerate}
      \addtocounter{enumiii}{-1} \def\theenumiii{(\arabic{enumiii})}
      \def\labelenumiii{\theenumi\theenumii\theenumiii:}

    \item \label{l:IIIpm00(0)} $\mu_1\notin\oZ$, $\mu_2\notin\oZ$.  The
      embedding diagram is a single-chain as in~\req{d:IIIpm0(0)}.
      However, {\it in the negative zone}, the diagram is half that
      long as~\req{d:IIIpm0(0)}, which is in accordance with the
      `double' degeneration taking place in the auxiliary Verma module,
      where one has a single-chain of Verma modules, with {\it two\/}
      modules embedded on each level.

      \addtocounter{enumiii}{1}
      \def\labelenumiv{\theenumi\theenumii\theenumiii,\theenumiv):}
    \item $\mu_1\in\oZ$, $\mu_2\in\oZ$. \ Whenever
      $\mu_1\cdot\mu_2>0$, we have one of the following two
      possibilities:

      \setlength{\leftmarginiv}{64pt}
      \begin{enumerate}
        \def\theenumiii{(\arabic{enumiii}}
        \def\theenumiv{${-}{-}$}

      \item \label{l:IIIpm00(2,mm)} $\mu_1\in-\oN$, $\mu_2\in-\oN$,
        diagram \req{d:IIIpm00(2,mm)}, where the Verma dots make up
        the `doubly-degenerate' embedding diagram as the one mentioned
        in~\IIIpm{00}(0). {\it In the negative zone\/} ($t<0$), where
        the embedding diagram is finite, we can further distinguish
        the following two cases depending on how the modules arrange
        near the bottom of the embedding diagram:

        \begin{itemize}
          \addtolength{\itemindent}{4pt}

        \item[i)] $(\mu_1 - \mu_2)/p$ is odd, in which case the
          embeddings terminate as in the first diagram
          in~\req{d:terminate1}.

        \item[ii)] $(\mu_1 - \mu_2)/p$ is even, in which case the
          embeddings terminate as in the second diagram
          in~\req{d:terminate1}.
        \end{itemize}
        In the case where $\mu_1=\mu_2$ (which is possible only when
        $t>0$), a special subcase is described by
        diagram~\req{d:IIIpm00(2,mm)-corr}.

        \def\theenumiv{${+}{+}$}
      \item \label{l:IIIpm00(2,pp)} $\mu_1\in\oN_0$, $\mu_2\in\oN_0$.
        The embedding diagram is the mirror of~\req{d:IIIpm00(2,mm)},
        with Verma modules replaced by twisted Verma modules. {\it In
          the negative zone}, in complete similarity
        with~\IIIpmzziimm, we can distinguish two cases,
        \begin{itemize}
          \addtolength{\itemindent}{4pt}

        \item[i)] $(\mu_1 - \mu_2)/p$ odd,

        \item[ii)] $(\mu_1 - \mu_2)/p$ even,
        \end{itemize}
        which, again, are the mirror of~\req{d:terminate1}, and
        similarly with the special case where $\mu_1=\mu_2$.
      \end{enumerate}
      If, on the other hand, $\mu_1\cdot\mu_2<0$, we have the
      following case:
      \setlength{\leftmarginiv}{62pt}
      \begin{enumerate}
        \def\theenumiii{(\arabic{enumiii}}
        \def\labelenumiv{\theenumi\theenumii\theenumiii,\theenumiv):}
        \def\theenumiv{${-}{+}$}
        
      \item \label{l:IIIpm00(2,mp)} $\mu_1\in-\oN$, $\mu_2\in\oN_0$.
        We have the embedding diagram~\req{d:IIIpm00(2,mp)}, where the
        pairs of Verma modules are at the same level as the
        corresponding pair of twisted Verma modules. {\it In the
          negative zone}, the diagram is finite and we have two
        possibilities of its structure near the bottom, shown
        in~\req{last-bottom}:
        \begin{itemize}
          \addtolength{\itemindent}{4pt}

        \item[i)] $(\mu_1 - \mu_2)/p$ is odd,
          
        \item[ii)] $(\mu_1 - \mu_2)/p$ is even. Then the Verma modules
          with the antidominant weights are not embedded by charged
          singular vectors into any \RV{} module---{\it there is no
            \RV{} module\/} at the bottom level in the diagram.
          
          In the particular case where, in addition to the
          requirements specified above, $\mu_2 - \mu_1=2\tp$, with
          $t=-\frac{\tp}{q}$ for $\tp,q\in\oN$, we have the
          `exceptional' diagram~\req{exceptional} with no \RV{}
          submodules.
        \end{itemize}

      \end{enumerate}

    \end{enumerate}

  \end{enumerate}

\end{enumerate}


The above cases are summarized in the Table, where we include
references to the corresponding diagrams.  We now list the
corresponding embedding diagrams and comment on their structure.

\begin{table}[tb]
\newlength{\mytableheight}
\setlength{\mytableheight}{15pt}
\begin{center}
  \renewcommand{\arraystretch}{0}
  \begin{tabular}{|p{2.5cm}||p{2.9cm}|p{2.9cm}|p{3.2cm}|p{3.2cm}|}
    \hline    \label{table:table}
    \mbox{}& $\mu_1,\mu_2\notin\oZ$
    &$\mu_1\in\oZ$, $\mu_2\notin\oZ$
    &\multicolumn{2}{c|}{$\mu_1,\mu_2\in\oZ$}\\
    \cline{4-5}
    {}&{}&{}& $\mu_1\cdot\mu_2>0$ & $\mu_1\cdot\mu_2<0$\\
    \hline
    \multicolumn{5}{|c|}{\rule{0pt}{1pt}}\\
    \hline
    $\mu_1 - \mu_2\notin\oK(t)$,\rule{0pt}{\mytableheight}\hfill\break
    \hfill
    &{}\hfill\break\ref{l:I(0)}, Eq.~\req{d:I}
    &{}\hfill\break\ref{l:I(1)}, Eq.~\req{d:I}
    &{}\Iiimm{} and\hfill\break \Iiipp, Eq.~\req{d:I(2)}
    &{}\hfill\break \Iiimp,~Eq.~\req{d:I(2)}
    \\
    \hline
    $\mu_1 - \mu_2\in\oK(t)$,\rule{0pt}{\mytableheight}\hfill\break
    $t\notin\oQ$\hfill
    &{}\hfill\break\ref{l:II(0)}, Eq.~\req{d:II}
    &{}\hfill\break \ref{l:II(1)}, Eq.~\req{d:II}
    &\hfill\break---
    &\hfill\break---
    \\
    \hline
    \multicolumn{5}{|c|}{\rule{0pt}{1pt}}\\
    \hline
    $\mu_1 - \mu_2\in\oK(t)$,\rule{0pt}{\mytableheight}\hfill\break
    $t\in\oQ$,\hfill\break
    $\mu_1 - \mu_2\notin\oZ$,\hfill\break
    $(\mu_1 - \mu_2)/t\notin\oZ$
    &{}\hfill\break\IIIpm{}(0), Eq.~\req{d:IIIpm(0)}
    &{}\hfill\break\IIIpm{}(1), Eq.~\req{d:IIIpm(1)} and
    \req{d:RTV}
    &{}\hfill\break ---
    &{}\hfill\break --- \\
    \hline
    $\mu_1 - \mu_2\in\oK(t)$,\rule{0pt}{\mytableheight}\hfill\break
    $t\in\oQ$,\hfill\break
    $\mu_1 - \mu_2\in\oZ$,\hfill\break
    $(\mu_1 - \mu_2)/t\notin\oZ$
    &{}\mbox{}\hfill\break
    \mbox{}\hfill\break
    \mbox{}\hfill\break
    \IIIpm0(0),
    &{}\hfill\break ---
    &{}\hfill\break\IIIpmziimm, Eq.~\req{d:IIIpm0(2)},\hfill\break
    and \IIIpmziipp\hfill
    &{}\hfill\break \IIIpmziimp, Eq.~\req{d:IIIpm0(2,mp)}\hfill\\
    \cline{1-1}\cline{3-5}
    $\mu_1 - \mu_2\in\oK(t)$,\rule{0pt}{\mytableheight}\hfill\break
    $t\in\oQ$,\hfill\break
    $\mu_1 - \mu_2\notin\oZ$,\hfill\break
    $(\mu_1 - \mu_2)/t\in\oZ$
    &{}Eq.~\req{d:IIIpm0(0)}
    &{}\hfill\break\IIIpm0(1), Eq.~\req{d:IIIpm0(1)}
    &{}\hfill\break ---
    &{}\hfill\break --- \\
    \hline
    $\mu_1 - \mu_2\in\oK(t)$,\rule{0pt}{\mytableheight}\hfill\break
    $t\in\oQ$,\hfill\break
    $\mu_1 - \mu_2\in\oZ$,\hfill\break
    $(\mu_1 - \mu_2)/t\in\oZ$
    &{}\hfill\break\IIIpm{00}(0)
    &{}\hfill\break ---
    &{}\hfill\break
    \IIIpmzziimm, Eq.~\req{d:IIIpm00(2,mm)},\hfill\break
    and \IIIpmzziipp{}\hfill
    &{}\hfill\break \IIIpmzziimp, Eq.~\req{d:IIIpm00(2,mp)}\\
    \hline
  \end{tabular}
\end{center}
\caption{{\sl Classification of the embedding diagram patterns}.
  For brevity, the notation $\mu_1\mu_2<0$ is used for the column
  that also includes the cases where $\mu_1=0$
  and $\mu_2<0$, and similarly with $\mu_1\mu_2>0$,
  where it may be the case that $\mu_1=0$ and $\mu_2>0$.
}
\end{table}

\subsection{The diagrams\label{subsec:diagrams}}

\paragraph{Notations.} {\it The arrows are always drawn in the
  direction from the parent module to the child (embedded) module}.
The filled dots {\Verma} denote the usual Verma modules, the open dots
{\TVerma} are the twisted Verma modules with the twist parameter
$\theta=1$, and {\Relaxed} are the \RV{} modules.  Embeddings of Verma
modules and of twisted Verma modules into \RV{} modules associated
with charged singular vectors are shown with horizontal arrows,
because the \hw{} vector of the Verma submodule (see~\req{withVerma})
and that of the twisted Verma submodule (see~\req{rightVerma}) are on
the same level as the relaxed \hw{} vector. Then, we also have to use
horizontal arrows to represent the embeddings of Verma modules into
each other performed by the ${\rm MFF}^+(r, 1, t)$ singular vectors
and, similarly, for the embeddings of twisted Verma modules performed
by the ${\rm MFF}^-(r, 1, t)$ singular vectors. Moreover, in
accordance with how the Verma- and twisted Verma submodules appear in
\RV{} modules (see~\req{withVerma} and~\req{rightVerma}), we place the
Verma modules on the left, and the twisted Verma modules, on the right
of the symbol designating the \RV{} module into which they are
embedded via a charged singular vector.  Then the $x$-coordinate in
the embedding diagrams can qualitatively be associated with the
$J^0_0$-charge of \hw{} vectors\footnote{Although we do not indicate
  them explicitly, the (charge, level) coordinates are not difficult
  to specify for each module in every embedding diagram, for instance
  by taking the relevant parameters from the standard {\it Verma\/}
  embedding diagrams and then applying the above Theorems in order to
  translate these into the $\mu_1$ and~$\mu_2$ parameters of each of
  the \RV{} modules.}.

\paragraph{\ref{l:I}.} $\mu_1 - \mu_2\notin\oK(t)$. \
In the cases where $(\mu_1\notin\oZ,\mu_2\notin\oZ)$,
$(\mu_1\in-\oN,\mu_2\notin\oZ)$, and $(\mu_1\in\oN_0,\mu_2\notin\oZ)$,
we have the following simple diagrams, respectively:
\begin{equation}
  \label{d:I}
  \unitlength=1pt
  \begin{picture}(500,10)
    \put(100,-15){
      \put(30,20){\Relaxed}
      \put(25,0){I(0)}
      \put(120,20){\Verma}
      \put(160,20){\Relaxed}
      \put(157,23){\vector(-1,0){30}}
      \put(130,0){I(1,${-}$)}
      \put(220,0){I(1,${+}$)}
      \put(90,0){
        \put(120,20){\Relaxed}
        \put(160,20){\TVerma}
        \put(128,23){\vector(1,0){30}}
        }
      }
  \end{picture}
\end{equation}
The second and the third diagram also describe the case of
$\mu_1=\mu_2\in\oZ$.  Note that, in the I(1,$\pm$) cases, taking the
quotient with respect to the Verma submodule embedded via the charged
singular vector, we are left with an irreducible (twisted) Verma
module, in accordance with the fact that the Verma-module spin $j$ is
such that $2j+1=\mu_1-\mu_2\notin\oK(t)$.

Whenever both $\mu_1$ and $\mu_2$ are (distinct) integers, the
diagrams are as follows:
\begin{equation}
  \label{d:I(2)}
  \unitlength=1pt
  \begin{picture}(500,25)
\put(0,-10){
    \put(-50, 0){
    \put(100,25){\Verma}
    \put(138,28){\vector(-1,0){31}}
    \put(140,25){\Verma}
    \put(178,28){\vector(-1,0){31}}
    \put(180,25){\Relaxed}
    \put(120,0){I(2,${-}{-}$)}
    \put(120,0){
    \put(100,25){\Relaxed}
    \put(108,28){\vector(1,0){31}}
    \put(140,25){\TVerma}
    \put(148,28){\vector(1,0){31}}
    \put(180,25){\TVerma}
    \put(120,0){I(2,${+}{+}$)}
    }
}
      \put(240,0){
        \put(120,25){\Relaxed}
        \put(160,25){\TVerma}
        \put(80,25){\Verma}
        \put(128,28){\vector(1,0){30}}
        \put(117,28){\vector(-1,0){30}}
        \put(100,0){I(2,${-}{+}$)}
        }
}
  \end{picture}
\end{equation}
for $(\mu_1,\mu_2\in-\oN)$, $(\mu_1,\mu_2\in\oN_0)$, and
$(\mu_1\in-\oN,\mu_2\in\oN_0)$, respectively.

As regards the cases shown in \req{d:I(2)}, it may be useful to
explicitly solve the conditions ($\mu_1-\mu_2\notin\oK(t)$,
$\mu_1-\mu_2\in\oZ$). We thus find that either $t\notin\oQ$ \ or \
$t\in\oQ_-$, $t=-\frac{\tp}{q}$, $|\mu_1-\mu_2|\leq\tp$.

\paragraph{\ref{l:II}.} $\mu_1 - \mu_2\in\oK(t)$, $t\notin\oQ$. \ When
none or precisely one of $\mu_1$ and $\mu_2$ is an integer, we have
the diagrams (in the cases where $(\mu_1\notin\oZ,\mu_2\notin\oZ)$,
$(\mu_1\in-\oN,\mu_2\notin\oZ)$, and $(\mu_1\in\oN_0,\mu_2\notin\oZ)$,
respectively):
\begin{equation}
  \label{d:II}
  \unitlength=1pt
  \begin{picture}(500,55)
    \put(0,-33){
      \put(50,80){\Relaxed}
      \put(50,40){\Relaxed}
      \put(53,78){\vector(0,-1){30}}
      \put(45,20){II(0)}
      \put(-20,0){
        \put(190,80){\Verma}
        \put(190,40){\Verma}
        \put(193,78){\vector(0,-1){30}}
        \put(230,80){\Relaxed}
        \put(227,83){\vector(-1,0){30}}
        \put(230,40){\Relaxed}
        \put(233,78){\vector(0,-1){30}}
        \put(227,43){\vector(-1,0){30}}
        \put(200,20){II(1,${-}$)}
        \put(130,0){
          \put(190,80){\Relaxed}
          \put(190,40){\Relaxed}
          \put(193,78){\vector(0,-1){30}}
          \put(230,80){\TVerma}
          \put(198,83){\vector(1,0){30}}
          \put(230,40){\TVerma}
          \put(233,78){\vector(0,-1){30}}
          \put(198,43){\vector(1,0){30}}
          \put(195,20){II(1,${+}$)}
          }
        }
      }
  \end{picture}
\end{equation}
Diagrams II(1,${-}$) and II(1,${+}$) are mirror-symmetric in the
obvious sense, which is in fact the general relation between the
$(1,{-})$ and $(1,{+})$ embedding diagrams. In the diagrams, we place
the submodules embedded via charged singular vectors with
$\mu_1\in-\oN$ on the left of the \RV{} module in accordance with how
such a charged singular vector actually appears in the \RV{} module,
Eq~\req{withVerma}. Similarly (see~\req{rightVerma}), the twisted
Verma submodules embedded via charged singular vectors with
$\mu_1\in\oN_0$ are shown on the right of the \RV{} module.

The same extremal diagrams occur for the rational negative
$t=-\frac{\tp}{q}$ and for $\mu_\alpha\notin\oZ$ such that
\begin{equation}\label{move2II0}
  \pm(\mu_1-\mu_2)=\left\{                                          
    \begin{array}{ll}
      \tp + \zeta\,,&0<\zeta<\tp\,,\\
      \omega+\eta\frac{\tp}{q}\,,&0<\omega\leq\tp,~0<\eta<q\,,\\
      2\tp\,,\\
      3\tp
    \end{array}\right.
  \quad
  \Longrightarrow{\rm diagram}\ {\rm II}(0)
\end{equation}\label{move2II1}
and for $\mu_1\in\oZ$, $\mu_2\notin\oZ$ such that
\begin{equation}
  |\mu_1-\mu_2|=
      \omega + \eta\frac{\tp}{q}\,,~~0<\omega\leq\tp\,,~0<\eta<q\,,
  \quad
  \Longrightarrow{\rm diagrams}\ {\rm II}(1)\,.
\end{equation}

Again, taking the quotient of, e.g., the II(1,$+$) diagram with
respect to the charged singular vector, we are left with a {\it
  Verma\/} module $\mM_{-\half(\mu_2-\mu_1+1),t}$, which has precisely
one singular vector.

\paragraph{\IIIpm{}(0).} $\mu_1 - \mu_2\in\oK(t)$, $t\in\oQ$,
$\mu_1 - \mu_2 \notin\oZ$, $(\mu_1 - \mu_2)/t\notin\oZ$,
$\mu_1\notin\oZ$, $\mu_2\notin\oZ$. \ In accordance with
Part~\ref{1:1} of Theorem~\ref{thm:relax}, we have the double-chain of
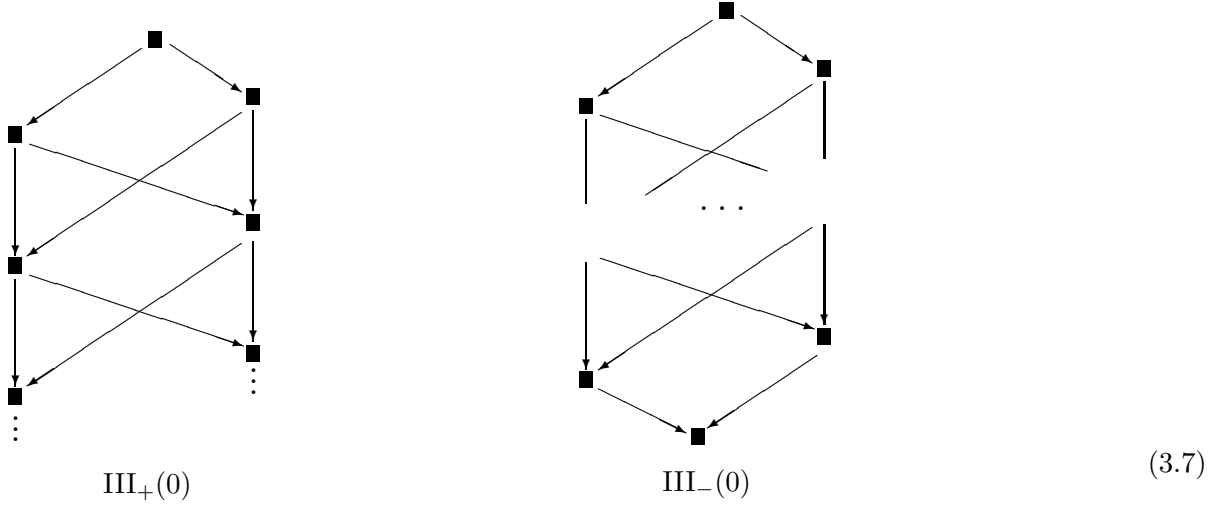
\RV{} modules
\begin{equation}
  \label{d:IIIpm(0)}
  \unitlength=.9pt
  \begin{picture}(500,198)
    \put(40,48){
      \put(59,130){\Relaxed}
      \put(56,130){\vector(-3,-2){48}}\put(68,132){\vector(3,-2){30}}
      \put(00,90){\Relaxed}\put(100,106){\Relaxed}
      \put(09,90){\vector(3,-1){90}}
      \put(98,103){\vector(-3,-2){90}}
      \put(03,88){\vector(0,-1){45}}\put(103,104){\vector(0,-1){42}}
      \put(0,-55){
        \put(0,90){\Relaxed}\put(100,108){\Relaxed}
        \put(9,90){\vector(3,-1){90}}
        \put(98,103){\vector(-3,-2){90}}
        \put(3,88){\vector(0,-1){45}}\put(103,104){\vector(0,-1){42}}
        }
      \put(0,-110){
        \put(0,90){\Relaxed}\put(100,108){\Relaxed}
        }
      \put(1,-35){\Large$\vdots$}\put(101,-15){\Large$\vdots$}
      \put(40,-57){III$_+(0)$}
      }
    \put(280,60){
      \put(59,130){\Relaxed}
      \put(56,130){\vector(-3,-2){48}}\put(68,132){\vector(3,-2){30}}
      \put(00,90){\Relaxed}\put(100,106){\Relaxed}
      \put(09,90){\line(3,-1){70}}
      \put(98,103){\line(-3,-2){70}}
      \put(03,88){\line(0,-1){35}}
      \put(103,104){\line(0,-1){32}}
      \put(50,50){\LARGE\ldots}
      \put(0,-60){
        \put(9,90){\vector(3,-1){90}}
        \put(98,103){\vector(-3,-2){90}}
        \put(3,88){\vector(0,-1){45}}\put(103,104){\vector(0,-1){42}}
        }
      \put(0,-115){
        \put(0,90){\Relaxed}\put(100,108){\Relaxed}
        \put(8,90){\vector(2,-1){37}}
        \put(100,104){\vector(-3,-2){46}}
        \put(47,66){\Relaxed}
        }
      \put(35,-68){{III{}$_-(0)$}}
      }
  \end{picture}
\end{equation}
which, therefore, looks identical to the standard embedding diagrams
of the ordinary Verma modules.

In the negative zone $t\in\oQ_-$, the chain is finite. Since we
have required $\mu_1 - \mu_2 \notin\oZ$ and $(\mu_1 -
\mu_2)/t\notin\oZ$, we have
\begin{equation}
  0<\zeta<\tp,\qquad
  0<\eta<q,\qquad
  \xi\in\oN_0
\end{equation}
in~\req{L-condition}. Then the total number of {\it embedded\/}
submodules in the III$_-(0)$ case is $2\xi + 1$ ($\xi$ submodules in
each branch, plus the bottom one). Thus, the case where $\xi=0$ may be
placed into~II(0), since the embedding diagrams are then identical to
those in~\req{d:II}.  We leave it to the reader's choice to either
place the set of the corresponding parameters
$(\mu_1,\mu_2,t=-\frac{\tp}{q})$ with $|\mu_1-\mu_2|=
\zeta+\eta\frac{\tp}{q}$, $0<\zeta<\tp$, $0<\eta<q$, into case~II(0),
at the same time excluding it from the III$_-(0)$ case, or to be
content with the understanding that, as the number of \RV{} submodules
diminishes to~1, the embedding diagram becomes identical to the
one for irrational~$t$ (where~$1$ is the largest possible number of
\RV{} submodules).

\medskip

Further degenerations can occur along two lines. First, whenever
charged singular vectors appear in the \RV{} module, the entire {\it
  Verma\/}-module embedding diagram joins the above
diagram~\req{d:IIIpm(0)}, the embeddings being given by charged
singular vectors.  In the case where the integer $\mu_1$ is positive,
as we have seen, the Verma module and hence all of the modules
embedded into it are twisted by the spectral flow transform
with~$\theta=1$.  Second, the Verma-module embedding diagram may
acquire a special form, which would also affect the `relaxed'
embedding diagram, making it into a single chain.  In addition, these
possibilities may occur simultaneously.

\paragraph{\IIIpmim.} 
$t\in\oQ$, $(\mu_1 - \mu_2)/t\notin\oZ$, $\mu_1\in-\oN$,
$\mu_2\notin\oZ$. \ Then the correspondence between \RV{} submodules in
$\mR_{\mu_1,\mu_2,t}$ and the Verma submodules in the module generated
from the charged singular vector is described in Parts~\ref{intersec}
and~\ref{is-relaxed} of Theorem~\ref{thm:Verma}. We have the following
embedding diagram:
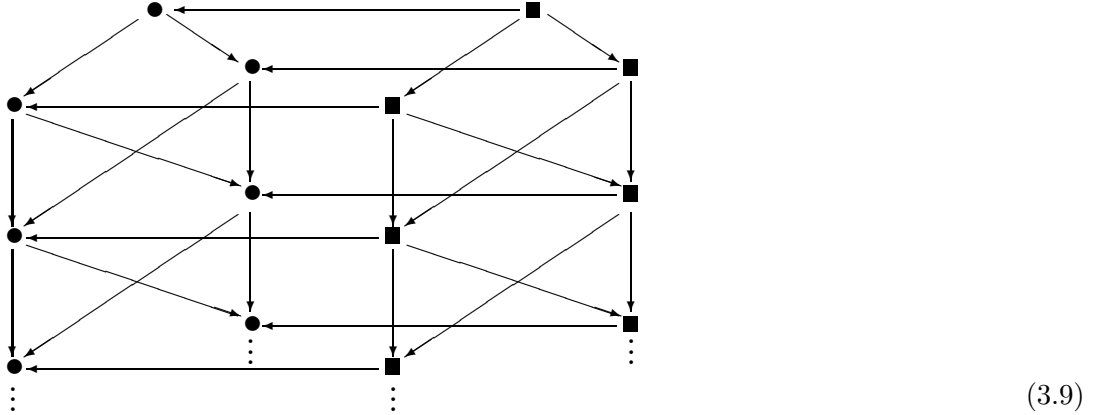
\begin{equation}
  \label{d:IIIpm(1)}
  \unitlength=0.9pt
  \begin{picture}(500,175)
    \put(-110,32){
      \put(259,130){\Verma}
      \put(256,130){\vector(-3,-2){48}}\put(268,132){\vector(3,-2){30}}
      \put(200,90){\Verma}\put(300,106){\Verma}
      \put(209,90){\vector(3,-1){90}}
      \put(298,103){\vector(-3,-2){90}}
      \put(203,88){\vector(0,-1){45}}\put(303,104){\vector(0,-1){42}}
      \put(0,-55){
        \put(200,90){\Verma}\put(300,108){\Verma}
        \put(209,90){\vector(3,-1){90}}
        \put(298,103){\vector(-3,-2){90}}
        \put(203,88){\vector(0,-1){45}}\put(303,104){\vector(0,-1){42}}
        }
      \put(0,-110){
        \put(200,90){\Verma}\put(300,108){\Verma}
        }
      \put(201,-35){\Large$\vdots$}\put(301,-15){\Large$\vdots$}
      }
    \put(50,32){
      \put(259,130){\Relaxed}
      \put(256,134){\vector(-1,0){145}}
      \put(256,130){\vector(-3,-2){48}}\put(268,132){\vector(3,-2){30}}
      \put(200,90){\Relaxed}\put(300,106){\Relaxed}
      \put(197,93){\vector(-1,0){148}}\put(297,109){\vector(-1,0){149}}
      \put(209,90){\vector(3,-1){90}}
      \put(298,103){\vector(-3,-2){90}}
      \put(203,88){\vector(0,-1){45}}\put(303,104){\vector(0,-1){42}}
      \put(0,-55){
        \put(200,90){\Relaxed}\put(300,108){\Relaxed}
      \put(197,93){\vector(-1,0){148}}\put(297,111){\vector(-1,0){149}}
        \put(209,90){\vector(3,-1){90}}
        \put(298,103){\vector(-3,-2){90}}
        \put(203,88){\vector(0,-1){45}}\put(303,104){\vector(0,-1){42}}
        }
      \put(0,-110){
        \put(200,90){\Relaxed}\put(300,108){\Relaxed}
      \put(197,93){\vector(-1,0){148}}\put(297,111){\vector(-1,0){149}}
        }
      \put(201,-35){\Large$\vdots$}\put(301,-15){\Large$\vdots$}
      }
  \end{picture}
\end{equation}
which is finite or infinite depending on whether $t<0$ (III{}$_-$) or
$t>0$~(III{}$_+$), respectively.  The top Verma module is the submodule
in the \RV{} module associated with a charged singular vector. Each of
the subsequent Verma modules is embedded via a charged singular vector
into the corresponding \RV{} module.  In the negative zone, the
\RV-module part of the diagram ends in the same way as in the
corresponding case in~\req{d:IIIpm(0)}, and similarly with the
Verma-module part of the diagram, the Verma module with the
antidominant weight is then embedded into the bottom \RV{} \hbox{module
  via a charged singular vector.}

As in the above, in the special case where $\xi=0$ in
\req{L-condition}, the III{}${}_-(1,{-})$ diagram becomes that of
case~II{}$(1,{-})$, Eq.~\req{d:II}. Thus, one could explicitly exclude
from case III{}${}_-(1)$ the parameters $(\mu_1\in\oZ,\mu_2\notin\oZ,
t=-\frac{\tp}{q})$ with $|\mu_1-\mu_2|=\zeta + \eta\frac{\tp}{q}$,
$0<\zeta<\tp$, $0<\eta<q$, placing these in the II(1) case.

\paragraph{\IIIpmip.} $\mu_1 - \mu_2\in\oK(t)$, $t\in\oQ$,
$(\mu_1 - \mu_2)/t\notin\oZ$, $\mu_1\in\oN_0$, $\mu_2\notin\oZ$. \
This case, too, is described by Parts~\ref{intersec}
and~\ref{is-relaxed} of Theorem~\ref{thm:Verma}.  We have a diagram
similar to the above, with Verma modules {\Verma} replaced by
twisted-Verma modules {\TVerma} (with the spectral parameter
$\theta=1$).
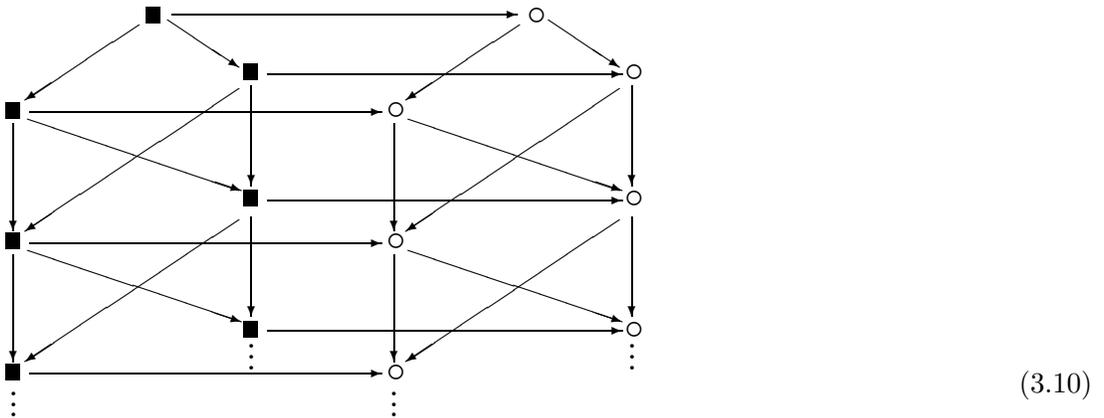
\begin{equation}
  \label{d:RTV}
  \unitlength=0.9pt
  \begin{picture}(500,165)
    \put(-110,25){
      \put(259,130){\Relaxed}
      \put(256,130){\vector(-3,-2){48}}\put(268,132){\vector(3,-2){30}}
      \put(200,90){\Relaxed}\put(300,106){\Relaxed}
      \put(209,90){\vector(3,-1){90}}
      \put(298,103){\vector(-3,-2){90}}
      \put(203,88){\vector(0,-1){45}}\put(303,104){\vector(0,-1){42}}
      \put(0,-55){
        \put(200,90){\Relaxed}\put(300,108){\Relaxed}
        \put(209,90){\vector(3,-1){90}}
        \put(298,103){\vector(-3,-2){90}}
        \put(203,88){\vector(0,-1){45}}\put(303,104){\vector(0,-1){42}}
        }
      \put(0,-110){
        \put(200,90){\Relaxed}\put(300,108){\Relaxed}
        }
      \put(201,-35){\Large$\vdots$}\put(301,-15){\Large$\vdots$}
      }
    \put(50,25){
      \put(259,130){\TVerma}
      \put(110,134){\vector(1,0){145}}
      \put(256,130){\vector(-3,-2){48}}\put(268,132){\vector(3,-2){30}}
      \put(200,90){\TVerma}\put(300,106){\TVerma}
      \put(50,93){\vector(1,0){148}}\put(150,109){\vector(1,0){149}}
      \put(209,90){\vector(3,-1){90}}
      \put(298,103){\vector(-3,-2){90}}
      \put(203,88){\vector(0,-1){45}}\put(303,104){\vector(0,-1){42}}
      \put(0,-55){
        \put(200,90){\TVerma}\put(300,108){\TVerma}
      \put(50,93){\vector(1,0){148}}\put(150,111){\vector(1,0){149}}
        \put(209,90){\vector(3,-1){90}}
        \put(298,103){\vector(-3,-2){90}}
        \put(203,88){\vector(0,-1){45}}\put(303,104){\vector(0,-1){42}}
        }
      \put(0,-110){
        \put(200,90){\TVerma}\put(300,108){\TVerma}
      \put(50,93){\vector(1,0){148}}\put(150,111){\vector(1,0){149}}
        }
      \put(201,-35){\Large$\vdots$}\put(301,-15){\Large$\vdots$}
      }
  \end{picture}
\end{equation}
This diagram, too, is finite in the III$_-(1,{+})$ case, and infinite,
in the III$_+(1,{+})$ case (i.e., for $t<0$ and $t>0$ respectively).
Obviously, the case of $\xi=0$ in~\req{L-condition} leaves us with the
diagram~II{}$(1,{+})$, Eq.~\req{d:II}.\nopagebreak

As before, taking the quotient of $\mR_{\mu_1,\mu_2,t}$ with respect
to the charged singular vector amounts to removing the {\TVerma} dots
and replacing {\Relaxed}${}\to{}${\Verma}. In this way, the remaining
part of the embedding diagram becomes that of the Verma module
$\mM_{-\half(\mu_2-\mu_1+1),t}$, which, of course, reproduces the
standard result.\pagebreak[3]

\paragraph{\IIIpm0(0).} $\mu_1 - \mu_2\in\oK(t)$, $t\in\oQ$,
$\mu_1\notin\oZ$, $\mu_2\notin\oZ$, and, in addition, one {\it but not
  both\/} of the following conditions satisfied: either $\mu_1 - \mu_2
\in\oZ$ or $(\mu_1 - \mu_2)/t\in\oZ$. \ We then have a single-chain of
\RV{} modules,
\begin{equation}
  \label{d:IIIpm0(0)}
  \unitlength=0.8pt
  \begin{picture}(500,200)
    \put(-50,100){
      \put(200,90){\Relaxed}
      \put(203,88){\vector(0,-1){45}}
      \put(0,-55){
        \put(200,90){\Relaxed}
        \put(203,88){\vector(0,-1){45}}
        }
      \put(0,-110){
        \put(200,90){\Relaxed}
        \put(203,88){\vector(0,-1){45}}
        }
      \put(0,-165){
        \put(200,90){\Relaxed}
        }
      \put(201,-90){\Large$\vdots$}
      \put(195,-110){III$_+^0(0)$}
      }
    \put(100,100){
      \put(200,90){\Relaxed}
      \put(203,88){\vector(0,-1){45}}
      \put(0,-55){
        \put(200,90){\Relaxed}
        \put(201,58){\Large$\vdots$}
        }
      \put(0,-110){
        \put(200,90){\Relaxed}
        \put(203,88){\vector(0,-1){45}}
        }
      \put(0,-165){
        \put(200,90){\Relaxed}
        }
      \put(195,-110){III$_-^0(0)$}
      }
  \end{picture}
\end{equation}
That the single chain takes the place of the double chain can be
explained in terms of the auxiliary Verma module~$\aM$.  There, either
the Verma-module embedding diagram becomes a single chain (in the case
where $(\mu_1-\mu_2)/t\in\oZ$), in which case we still have the
correspondence described in Part~\ref{1:1} of Theorem~\ref{thm:relax},
or
the embedding diagram of the auxiliary Verma module contains pairs of
Verma modules embedded onto the same level; every such pair
corresponds to one \RV{} module, as described in Part~\ref{2:1} of
Theorem~\ref{thm:relax}.

In the negative zone, we use~\req{L-condition}, where now $\eta=0$,
$0<\zeta<\tp$ in the case where $\mu_1-\mu_2\in\oZ$, and $\zeta=0$,
$0<\eta<q$ in the case where $(\mu_1-\mu_2)/t\in\oZ$. In either case,
we have $\xi=\left[\frac{|\mu_1-\mu_2|}{\tp}\right]\in\oN$ {\it
  embedded\/} \RV{} submodules. In the case where $\xi=1$, the diagram
`degenerates' to that of II(0), Eq.~\req{d:II}, therefore one may wish
to explicitly exclude from III${}_-^0(0)$ the parameters
$(\mu_1,\mu_2,t=-\frac{\tp}{q})$ such that either
$|\mu_1-\mu_2|=\tp+\zeta$, $0<\zeta<\tp$, or
$|\mu_1-\mu_2|=\tp+\eta\frac{\tp}{q}$, $0<\eta<q$ (and with
$\mu_\alpha\notin\oZ$ in both cases, obviously).

\paragraph{\IIIpmzim.} $\mu_1 - \mu_2\in\oK(t)$, $t\in\oQ$,
$\mu_1\in-\oN$, $\mu_2\notin\oZ$, $(\mu_1 - \mu_2)/t\in\oZ$. \ This is
a particular case described in Part~\ref{1:1} of
Theorem~\ref{thm:relax}.  The embedding diagram can be written in one
of the following ways:
\begin{equation}
  \label{d:IIIpm0(1)}
  \unitlength=0.85pt
  \begin{picture}(500,185)
    \put(-200,40){
      \put(209,135){\Verma}
      \put(367,139){\vector(-1,0){149}}\put(370,135){\Relaxed}
      \put(218,137){\vector(3,-1){80}}
      \put(300,106){\Verma}
      \put(367,109){\vector(-1,0){59}}\put(370,106){\Relaxed}
      \put(372.5,133.5){\vector(0,-1){20}}
      \put(298,103){\vector(-3,-2){90}}
      \put(0,-55){
        \put(200,90){\Verma}
        \put(367,93){\vector(-1,0){157}}\put(370,90){\Relaxed}
        \put(372.5,158.5){\vector(0,-1){60}}
        \put(209,90){\vector(3,-1){90}}
        }
      \put(0,-110){
        \put(300,108){\Verma}
        \put(367,112){\vector(-1,0){58}}\put(370,108){\Relaxed}
        \put(372.5,143){\vector(0,-1){27}}
        \put(298,106){\line(-3,-2){40}}
        }
      \put(0,-165){
        \put(372.5,159){\line(0,-1){20}}
        }
      \put(280,-50){$\mu_1-\mu_2<0$}
      }
    \put(140,40){
      \put(209,135){\Verma}
      \put(297,139){\vector(-1,0){79}}\put(300,135){\Relaxed}
      \put(209,137){\vector(-3,-1){80}}
      \put(122,106){\Verma}
      \put(297,109){\vector(-1,0){166}}\put(300,106){\Relaxed}
               \put(302.5,133.5){\vector(0,-1){20}}
      \put(129,104){\vector(3,-2){90}}
      \put(0,-55){
        \put(219,90){\Verma}
        \put(297,93){\vector(-1,0){72}}\put(300,90){\Relaxed}
               \put(302.5,158.5){\vector(0,-1){60}}
        \put(217,90){\vector(-3,-1){90}}
        }
      \put(0,-110){
        \put(120,108){\Verma}
        \put(297,112){\vector(-1,0){168}}\put(300,108){\Relaxed}
               \put(302.5,143){\vector(0,-1){27}}
      \put(125,106){\line(3,-2){40}}
      }
    \put(0,-165){
      \put(302.5,159){\line(0,-1){20}}
      }
    \put(200,-50){$\mu_1-\mu_2>0$}
      }
  \end{picture}
\end{equation}
These diagrams are of course isomorphic as embedding diagrams, however
it may be helpful to associate the $x$-coordinate in the {\it
  Verma\/}-part with the $J^0_0$-eigenvalue of the \hw{} vector in the
corresponding modules; then, depending on whether $\mu_1-\mu_2$ is
positive or negative, the first Verma-module embedding is performed by
an ${\rm MFF}^-(r,s,t)$ or an ${\rm MFF}^+(r,s,t)$ vector, we have the
first or the second of the above diagrams, respectively.  In fact,
specifying precise $x\!\equiv\!{\rm charge}$ coordinates for the \hw{}
vector of every module requires drawing the {\Verma}-part of the
diagrams as growing either wider or narrower as one moves down,
according to whether $t>0$ or $t<0$, respectively (which we do not do
here).

The diagrams are finite in the negative zone (III$^0_-(1,{-})$) and
infinite, in the positive zone (III$^0_+(1,{-})$). The structure of
the III$_-^0(1,-)$-diagram at the bottom is obvious. As in the
corresponding case in~\IIIpm0(0){}, we have only one embedded \RV{}
submodule whenever~$\xi=1$ in the negative zone and, thus, the diagram
becomes that of~II$(1,{-})$, Eq.~\req{d:II}.  Again, whenever
$|\mu_1-\mu_2|=\tp+\eta\frac{\tp}{q}$ with $0<\eta<q$, the
parameters $(\mu_1\in\oZ,\mu_2\notin\oZ,t=-\frac{\tp}{q})$ may be
removed from case~III${}^0_-(1)$.

After taking the quotient over the charged singular vector and
replacing {\Relaxed}${}\to{}${\TVerma} and then `untwisting', we
obtain the embedding diagram of the quotient module
$\mM_{-\half(\mu_2-\mu_1+1),t}$, which is a single-chain in accordance
with the fact that $(\mu_1-\mu_2)/t\in\oZ$.

\paragraph{\IIIpmzip.} $\mu_1 - \mu_2\in\oK(t)$, $t\in\oQ$,
$\mu_1\in\oN_0$, $\mu_2\notin\oZ$, $(\mu_1 - \mu_2)/t\in\oZ$. \ In
complete similarity with the previous case, this one is also described
in Part~\ref{1:1} of Theorem~\ref{thm:relax}. The embedding
diagram is a mirror of the above with the Verma modules {\Verma}
replaced by twisted-Verma modules~{\TVerma}.  The diagrams are finite
in the negative zone (III$^0_-(1,{+})$) and infinite, in the positive
zone (III$^0_+(1,{+})$). Clearly, $\xi=1$ leaves with the
diagram~II$(1,{+})$.

\paragraph{\IIIpmziimm.} $\mu_1 - \mu_2\in\oK(t)$, $t\in\oQ$,
$(\mu_1 - \mu_2)/t\notin\oZ$, $\mu_1\in-\oN$, $\mu_2\in-\oN$. \ This
is described in Part~\ref{2:1} of Theorem~\ref{thm:relax}.  With
$\mu_1$ and $\mu_2$ being integers of the same sign, there are two
charged singular vectors in the \RV{} module on one side of the \hw{}
vector.  In the present case, with $\mu_1$ and $\mu_2$ both negative,
the embedding diagram takes the following form, where the Verma part
is nothing but the standard double-chain embedding diagram with a part
of the arrows drawn horizontal:
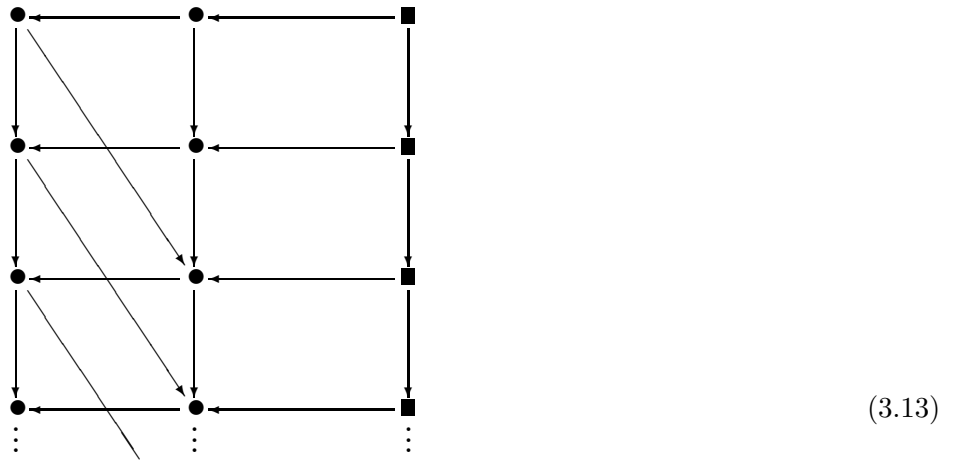
\begin{equation}
  \label{d:IIIpm0(2)}
  \unitlength=0.9pt
  \begin{picture}(500,174)
    \put(120,75){
      \put(200,90){\Relaxed}
      \put(197,93){\vector(-1,0){78}}
      \put(203,88){\vector(0,-1){45}}
      \put(0,-55){
        \put(200,90){\Relaxed}
      \put(197,93){\vector(-1,0){78}}
        \put(203,88){\vector(0,-1){45}}
        }
      \put(0,-110){
        \put(200,90){\Relaxed}
      \put(197,93){\vector(-1,0){78}}
        \put(203,88){\vector(0,-1){45}}
        }
      \put(0,-165){
        \put(200,90){\Relaxed}
      \put(197,93){\vector(-1,0){78}}
        }
      \put(201,-90){\Large$\vdots$}
      }
    \put(30,75){
      \put(200,90){\Verma}
      \put(197,93){\vector(-1,0){63}}
      \put(203,88){\vector(0,-1){45}}
      \put(0,-55){
        \put(200,90){\Verma}
      \put(197,93){\vector(-1,0){63}}
        \put(203,88){\vector(0,-1){45}}
        }
      \put(0,-110){
        \put(200,90){\Verma}
      \put(197,93){\vector(-1,0){63}}
        \put(203,88){\vector(0,-1){45}}
        }
      \put(0,-165){
        \put(200,90){\Verma}
      \put(197,93){\vector(-1,0){63}}
        }
      \put(201,-90){\Large$\vdots$}
      }
    \put(-45,75){
      \put(200,90){\Verma}
      \put(208,88){\vector(2,-3){66}}
      \put(203,88){\vector(0,-1){45}}
      \put(0,-55){
        \put(200,90){\Verma}
      \put(208,88){\vector(2,-3){66}}
        \put(203,88){\vector(0,-1){45}}
        }
      \put(0,-110){
        \put(200,90){\Verma}
      \put(208,88){\line(2,-3){47}}
        \put(203,88){\vector(0,-1){45}}
        }
      \put(0,-165){
        \put(200,90){\Verma}
        }
      \put(201,-90){\Large$\vdots$}
      }
  \end{picture}
\end{equation}

\bigskip

\noindent
As usual, the diagram is finite or infinite depending on whether
$t<0$ (III{}$^0_-(2,{-}{-})$) or $t>0$ (III{}$^0_+(2,{-}{-})$)
respectively. In the negative zone, the structure of the embedding
diagram near the bottom is as follows:
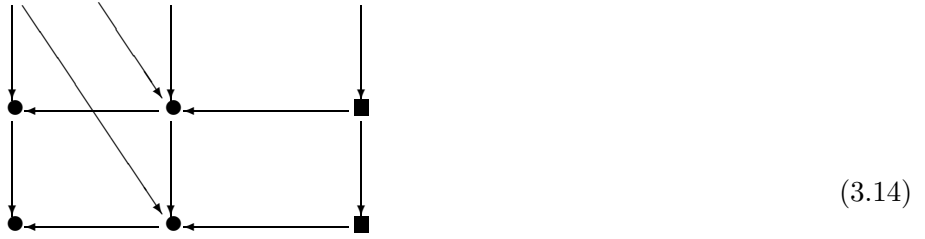
\begin{equation}
  \label{bottom-1}
  \unitlength=0.8pt
  \begin{picture}(500,95)
    \put(120,60){
      \put(0,-55){
        \put(203,88){\vector(0,-1){45}}
        }
      \put(0,-110){
        \put(200,90){\Relaxed}
        \put(197,93){\vector(-1,0){78}}
        \put(203,88){\vector(0,-1){45}}
        }
      \put(0,-165){
        \put(200,90){\Relaxed}
        \put(197,93){\vector(-1,0){78}}
        }
      }
    \put(30,60){
      \put(0,-55){
        \put(203,88){\vector(0,-1){45}}
        }
      \put(0,-110){
        \put(200,90){\Verma}
        \put(197,93){\vector(-1,0){63}}
        \put(203,88){\vector(0,-1){45}}
        }
      \put(0,-165){
        \put(200,90){\Verma}
        \put(197,93){\vector(-1,0){63}}
        }
      }
    \put(-45,60){
      \put(244,34){\vector(2,-3){30}}
      \put(0,-55){
        \put(208,88){\vector(2,-3){66}}
        \put(203,88){\vector(0,-1){45}}
        }
      \put(0,-110){
        \put(200,90){\Verma}
        \put(203,88){\vector(0,-1){45}}
        }
      \put(0,-165){
        \put(200,90){\Verma}
        }
      }
  \end{picture}
\end{equation}

\smallskip

\paragraph{\IIIpmziipp.} $\mu_1 - \mu_2\in\oK(t)$, $t\in\oQ$,
$(\mu_1 - \mu_2)/t\notin\oZ$, $\mu_1\in\oN_0$, $\mu_2\in\oN_0$. \ This
time, $\mu_1$ and $\mu_2$ are both positive, and the embedding diagram
is obtained from~\req{d:IIIpm0(2)}--\req{bottom-1} by replacing the
Verma modules {\Verma} with twisted-Verma modules {\TVerma} and
placing them on the right of the \RV{} modules (cf.~\req{d:RTV}).
Thus, the diagram is a vertical mirror of \req{d:IIIpm0(2)} with
{\Verma}${}\leadsto{}${\TVerma}.

\paragraph{\IIIpmziimp.} $\mu_1 - \mu_2\in\oK(t)$, $t\in\oQ$,
$(\mu_1 - \mu_2)/t\notin\oZ$, $\mu_1\in-\oN$, $\mu_2\in\oN$. \ With
$\mu_1$ and $\mu_2$ being of different signs, there are charged
singular vectors on different sides of the \hw{} vector of the \RV{}
module. One of the charged singular vectors comes with the embedding
diagram of Verma modules, while the other contributes a similar (in
fact, mirror-symmetric with respect to charge in the (charge, level)
coordinates) diagram of twisted-Verma modules. Superposition of these
diagrams has a nontrivial effect which we describe in two steps: We
first draw a diagram showing all the relevant modules, but {\it not\/}
all of the embeddings. Then we describe the subtleties related to the
embeddings as those in~\req{direct-sum}, after which we draw the final
embedding diagram (Eq.~\req{d:IIIpm0(2,mp)}), which may indeed look
rather complicated.

The superposition of the Verma and twisted-Verma embedding diagrams
looks as follows:
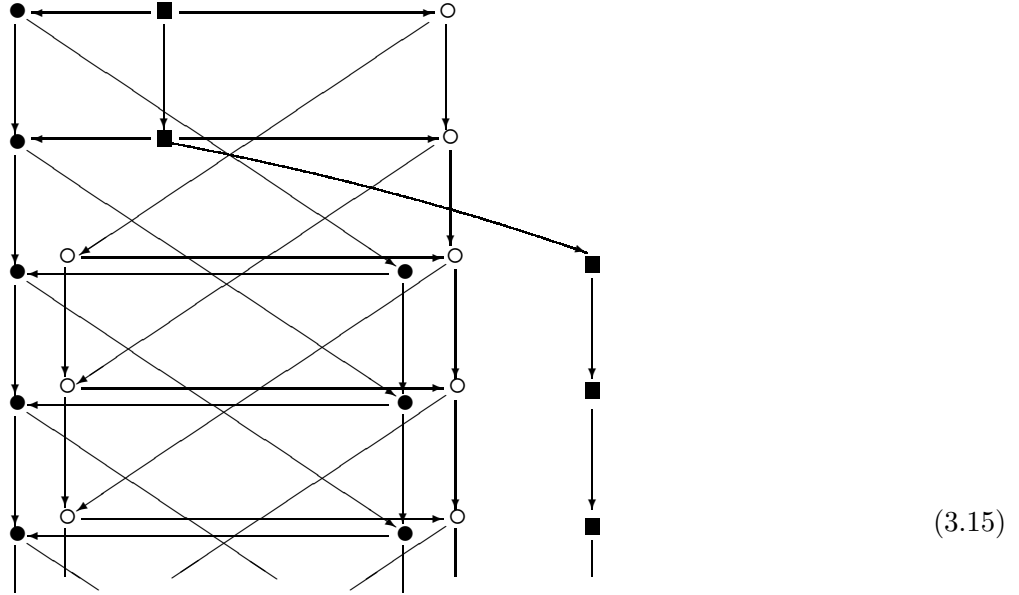
\begin{equation}
  \unitlength=0.9pt
  \label{d:IIIpm0(2,mp)-first}
  \begin{picture}(500,225)
\put(80,0){
    \bezier{400}(116,163)(210,145)(290,117)
    \put(288,118){\vector(3,-1){2}}
    \put(292.5,106){\vector(0,-1){42}}
    \put(292.5,51){\vector(0,-1){42}}
    \put(292.5,-4){\line(0,-1){15}}
    \put(210,125){
      \put(-163,90){\Verma}
      \put(-155,90){\vector(3,-2){155}}
      \put(-160,88){\vector(0,-1){46}}
      \put(0,-55){
        \put(-163,90){\Verma}
        \put(-155,90){\vector(3,-2){155}}
        \put(-160,88){\vector(0,-1){46}}
        }
      \put(0,-110){
        \put(-163,90){\Verma}\put(-3,93){\vector(-1,0){152}}
        \put(0,90){\Verma}
        \put(3,89){\vector(0,-1){46}}
        \put(-155,90){\vector(3,-2){155}}
        \put(-160,88){\vector(0,-1){46}}
        }
      \put(0,-165){
        \put(-163,90){\Verma}\put(-3,93){\vector(-1,0){152}}
        \put(0,90){\Verma}
        \put(3,89){\vector(0,-1){46}}
        \put(-155,90){\line(3,-2){105}}
        \put(-160,88){\vector(0,-1){46}}
        }
      \put(0,-220){
        \put(-163,90){\Verma}\put(-3,93){\vector(-1,0){152}}
        \put(0,90){\Verma}
        \put(3,89){\line(0,-1){20}}
        \put(-155,90){\line(3,-2){30}}
        \put(-160,89){\line(0,-1){20}}
        }
      }
    \put(210,125){
      \put(-100,90){\Relaxed}
      \put(-97.5,88){\vector(0,-1){44}}
      \put(-103,93){\vector(-1,0){50}} \put(-91,93){\vector(1,0){107}}
      \put(-100,36){\Relaxed}
      \put(-103,40){\vector(-1,0){50}} \put(-91,40){\vector(1,0){109}}
      }
    \put(75,132){
      \put(153,83){\TVerma}
      \put(149,82){\vector(-3,-2){147}}
      \put(156,81){\vector(0,-1){44}}
      \put(0,-55){
        \put(154,85){\TVerma}
        \put(151,85){\vector(-3,-2){150}}
        \put(158,83){\vector(0,-1){40}}
        }
      \put(0,-110){
        \put(156,90){\TVerma}\put(3,93){\vector(1,0){152}}
        \put(-7,90){\TVerma}
        \put(-4,89){\vector(0,-1){46}}
        \put(156,90){\vector(-3,-2){155}}
        \put(160,88){\vector(0,-1){46}}
        }
      \put(0,-165){
        \put(157,90){\TVerma}\put(3,93){\vector(1,0){152}}
        \put(-7,90){\TVerma}
        \put(-4,89){\vector(0,-1){46}}
        \put(156,90){\line(-3,-2){115}}
        \put(160,88){\vector(0,-1){46}}
        }
      \put(0,-220){
        \put(157,90){\TVerma}\put(3,93){\vector(1,0){152}}
        \put(-7,90){\TVerma}
        \put(-4,89){\line(0,-1){20}}
        \put(156,90){\line(-3,-2){40}}
        \put(160,89){\line(0,-1){20}}
        }
      }
   \put(290, 108){\Relaxed}
    \put(290, 55){\Relaxed}
    \put(290, -2){\Relaxed}
}
  \end{picture}
\end{equation}

\bigskip


\noindent
{\it where not all of the embeddings are yet shown}.  The three
modules at the second level (assuming the top level to be the first)
are described by Part~\ref{maximal} of Theorem~\ref{thm:special}. \ We
denote by $\mC_-'$, $\mR'$, and $\mC_+'$ the Verma module, the \RV{}
module, and the twisted Verma module at that level, respectively.
Further down, there are two Verma and two twisted-Verma modules
embedded onto the same level (in the diagram, these are shown slightly
displaced so as not to confuse the different arrows\footnote{and, for
  the same reason, the \RV{} modules are drawn outside the
  Verma/twisted-Verma part of the diagram.}). We introduce the
notations $\tilde\mC''_-$ and $\mC''_-$ for two Verma modules related
by the embedding $\mC''_-\subset\tilde\mC''_-$, and also
$\tilde\mC''_+\supset\mC''_+$ for the twisted Verma modules at a given
level.  Let also $\mR''$ be the \RV{} module at that level. If the
$J^0_0$-charge of the \hw{} vector of $\mC_-''$ is $j$, then the
charge of the \hw{} vector of $\tilde\mC''_+$ is~$j+1$.  Further, the
$J^0_0$-charge of the `parent' Verma module $\tilde\mC_-''$ is $j'>j$
and that of \hbox{the `child' {\it twisted\/} Verma module $\mC_+''$
is~$j'+1$.}

Let us first describe the embedding of \,$\mC_-'$, $\mR'$, and
\,$\mC_+'$ into \,$\mR\equiv\mR_{\mu_1,\mu_2,t}$.  The following
analysis is a direct consequence of~\cite{[FST]}.\footnote{The
  difference of our present approach from that of~\cite{[FST]} is that
  we are now interested in describing {\it all\/} submodules in a
  given \RV{} module and in finding the sequence in which submodules
  are embedded into one another, whereas the (simpler) problem
  addressed in~\cite{[FST]} was to give the configuration of
  submodules that occur `{\it somewhere\/}' inside the \RV{} module as
  soon as the parameters (in our present conventions) $\mu_1$,
  $\mu_2$, and $t$ {\it admit\/} a certain representation involving
  several integers.}  \ In the positive zone $t=\frac{p}{q}>0$, the
module \,$\mC'_-$ is embedded into \,$\mC_-$ by singular vector
$\ket{{\rm MFF}^+(r,s,t)}$ with $r<p$; in the negative zone, this is
$\ket{{\rm MFF}^-(r,s,t)}$ with $r<|p|$; then \,$\mC'_+$ is embedded
into \,$\mC_+$ by $\ket{{\rm MFF}^{-,1}(r,s,t)}$ (in the negative
zone, by $\ket{{\rm MFF}^{+,1}(r,s,t)}$). The modules \,$\mC_-'$ and
\,$\mC_+'$ are embedded into $\mR'$ via charged singular vectors in
accordance with Theorem~\ref{thm:Verma} (Parts~\ref{is-relaxed}
and~\ref{intersec}).  We, thus, have the extremal diagram\label{begin}
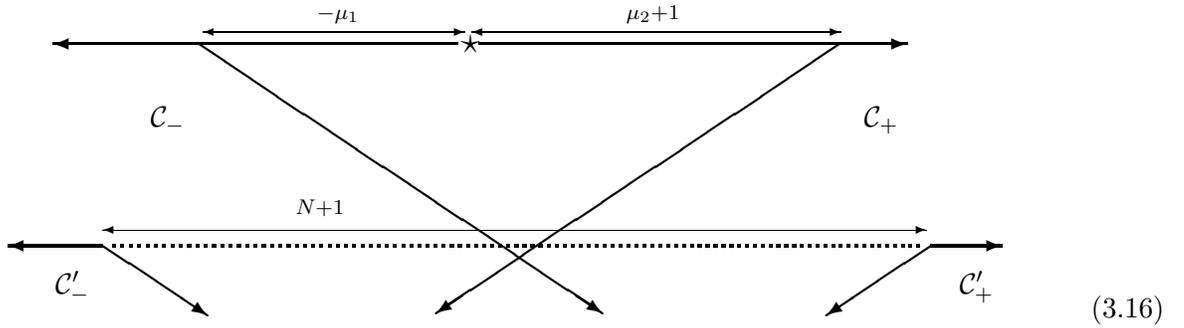
\begin{equation}
  \unitlength=.9pt
  \begin{picture}(250,130)
    \put(-20,30){
      {\thicklines
        \put(138,85){\vector(1,0){180}}
        \put(129,85){\vector(-1,0){170}}
        \put(20,85){\vector(3,-2){170}}  
        \put(290,85){\vector(-3,-2){170}}  
        \put(-20,0){\vector(-1,0){40}}
        \put(-20,0){\vector(3,-2){44}} 
        \put(328,0){\vector(-3,-2){44}} 
        \put(328,0){\vector(1,0){30}} 
        }
      \put(0, 50){$\mC_-$}
      \put(300, 50){$\mC_+$}
      \put(130.5,81){\Large$\star$}
      {\linethickness{1.2pt}\bezier{100}(-16,0)(160,0)(322,0)}
      \linethickness{.2pt}
      \put(-18,6.5){\vector(1,0){344}}
      \put(-18,6.5){\vector(-1,0){2}}
      \put(61,9){${}^{N+1}$} 
      \put(24,90){\vector(1,0){108}}     
      \put(24,90){\vector(-1,0){2}}
      \put(70,91){${}^{-\mu_1}$}
      \put(137,90){\vector(1,0){153}}     
      \put(137,90){\vector(-1,0){2}}
      \put(200,91){${}^{\mu_2+1}$}
      \put(-40, -20){$\mC'_-$}
      \put(340, -20){$\mC'_+$}
      }
  \end{picture}
  \label{1st-level}
\end{equation}

\medskip

\noindent
Here, the \RV{} module $\mR'$ can be generated from any of $N=\mu_2 -
\mu_1 + 2r$ (or, in the negative zone, $N=\mu_2 - \mu_1 - 2r$)
extremal states
\begin{equation}
  \ket{g'(i)}\,,\qquad i = 1,\ldots, N. 
  \label{g-states}
\end{equation}
{\it between\/} the two charged singular vectors.  We can choose
$\ket{g'(i+1)}=J^+_0\,\ket{g'(i)}$ for $1\leq i\leq N$. \ Let also
$\ket{g'(0)}$ be the \hw{} state of the $\mC'_-$ submodule, and
$\ket{g'(N + 1)}$, the \hw{} state of $\mC'_+$. We know
from~\cite{[FST]} that relation
\begin{equation}
  \label{solve}
  J^-_0\,\ket{g'(1)}=\ket{g'(0)}
\end{equation}
can be inverted as
\begin{equation}
  \ket{g'(1)}=(J^-_0)^{-1}\ket{g'(0)}\,,
\end{equation}
where the action of $(J^-_0)^{-1}$ is defined in the spirit
of~\req{properties} and $\ket{g'(1)}$ satisfies relaxed \hw{}
conditions. Such a $\ket{g'(1)}$ is unique.  A similar construction
starting with the \hw{} vector of \,$\mC'_+$ allows us to define the
extremal states
\begin{equation}
  \ket{g'_+(N)}=
  (J^+_0)^{-1}\ket{g'_+(N + 1)}\,,
  \qquad
  \ket{g'_+(i)}=
  (J^-_0)^{N - i}
  \ket{g'_+(N)}\,.
\end{equation}
Then it follows from the results of~\cite{[FST]} that
\begin{equation}
  \ket{g'(i)} = a'(i)\ket{g'_+(i)}\,,\quad
  1\leq i\leq N\,,\qquad a'(i)\neq0\,,
  \label{a(i)}
\end{equation}
where $a'(i)$ depends also on $t$, $\mu_1$, $\mu_2$, and $r$.

Let now $\mR''$ be the \RV{} module that is the third from the top in
\req{d:IIIpm0(2,mp)-first}. The embedding
$\iota'':\mR''\hookrightarrow\mR'$ is described precisely as the
embedding $\iota':\mR'\hookrightarrow\mR$, the only difference being
that $\mu_1$, $\mu_2$, and $r$ have to be replaced with $\mu'_1$,
$\mu'_2$, and $r'$ such that $\mu'_2 - \mu'_1 = N$, $r'=|p|-r$, and
$N'=N+2r'$ (or $N'=N-2r'$ in the negative zone), where
$t=\frac{p}{q}$. \ Among four (twisted) Verma modules
$\mC''_-\subset\tilde\mC''_-$ and $\tilde\mC''_+\supset\mC''_+$ that
are embedded onto the same level as $\mR''$, \ $\mC''_-$ and $\mC''_+$
are embedded into $\mR''$ via charged singular vectors. Thus, we
simply rewrite the above extremal diagram in the modified notations:
\begin{equation}
  \unitlength=.8pt
  \begin{picture}(250,140)
    \put(-20,30){
      {\thicklines
        \put(138,85){\vector(1,0){180}}
        \put(129,85){\vector(-1,0){170}}
        \put(20,85){\vector(3,-2){170}}  
        \put(290,85){\vector(-3,-2){170}}  
        \put(-20,0){\vector(-1,0){40}}
        \put(-20,0){\vector(3,-2){44}} 
        \put(328,0){\vector(-3,-2){44}} 
        \put(328,0){\vector(1,0){30}} 
        }
      \put(0, 50){$\mC'_-$}
      \put(300, 50){$\mC'_+$}
      \put(130.5,81){\Large$\star$}
      {\linethickness{1.2pt}\bezier{100}(-16,0)(160,0)(322,0)}
      \linethickness{.2pt}
      \put(-18,6.5){\vector(1,0){344}}
      \put(-18,6.5){\vector(-1,0){2}}
      \put(50,9){${}^{N'+1}$}
      \put(24,90){\vector(1,0){108}}     
      \put(24,90){\vector(-1,0){2}}
      \put(70,91){${}^{-\mu'_1}$}
      \put(137,90){\vector(1,0){153}}     
      \put(137,90){\vector(-1,0){2}}
      \put(200,91){${}^{\mu'_2+1}$}
      \put(-40, -20){$\mC''_-$}
      \put(340, -20){$\mC''_+$}
      }
  \end{picture}
  \label{wide-gap}
\end{equation}
We now have, in complete similarity with the above,
\begin{eqnarray}
  &\ket{g''(1)}=(J^-_0)^{-1}\ket{g''(0)}\,,~~~~\quad
  \ket{g''(i)}=(J^+_0)^{i-1}\ket{g''(1)}\,,\\
  &\ket{g''_+(N')}=(J^+_0)^{-1}\ket{g''_+(N'+1)},\quad
  \ket{g''_+(i)}=
  (J^-_0)^{N' - i}
  \ket{g''_+(N')},\\
  &\ket{g''(i)} = a''(i)\ket{g''_+(i)}\,.\label{a''(i)}
\end{eqnarray}

Let now $\ket{G(i)}=\iota\,\ket{g''(i)}$ be the image of
$\ket{g''(i)}$ under the embedding \ 
$\iota=\iota'\circ\iota'':\mR''\hookrightarrow\mR$; in particular,
$\ket{G(0)}$ and $\ket{G(N' + 1)}$ are the \hw{} vectors of the
embedding of $\mC''_-$ and $\mC''_+$, respectively.

{}From the embedding diagrams of the Verma module $\mC_-$
(respectively, $\mC_+$) we see that there actually exists the
submodule $\tilde\mC''_-$ (resp., $\tilde\mC''_+$) at the same level
as $\mC''_-$ (resp., $\mC''_+$) such that
$\mC''_-\subset\tilde\mC''_-$ (resp., $\mC''_+\subset\tilde\mC''_+$).
Moreover, these are embedded into $\mR$ as follows (with the \hw{}
vector of each module shown, for convenience, by the same symbol as
the corresponding {\it module\/} in the {\it embedding\/} diagram)
\begin{equation}
  \unitlength=1pt
  \begin{picture}(250,170)
    \put(-20,30){
      \put(100,131){\Verma}
      \put(203,131){\TVerma}
      \put(78,91){\Verma}
      \put(225,91){\TVerma}
      {\thicklines
        \put(130.5,131){\Relaxed}
        \put(138,135){\vector(1,0){180}}
        \put(129,135){\vector(-1,0){170}}
        \put(102,135){\vector(3,-2){44}}
        \put(208,135){\vector(-3,-2){44}}
        \put(70, 120){$\mC_-$}
        \put(220, 120){$\mC_+$}
        }
      \put(0,10){
        \put(60, 50){$\mC'_-$}
        \put(240, 50){$\mC'_+$}
        \put(130.5,81){\Relaxed}
        {\linethickness{.2pt}
          \put(84,90){\vector(1,0){48}}     
          \put(84,90){\vector(-1,0){2}}
          \put(90,91){${}^{-\mu_1}$}
          \put(137,90){\vector(1,0){93}}     
          \put(137,90){\vector(-1,0){2}}
          \put(175,91){${}^{\mu_2+1}$}
          }
        \thicklines
        \put(138,85){\vector(1,0){180}}
        \put(129,85){\vector(-1,0){170}}
        \put(80,85){\vector(3,-2){70}}  
        \put(230,85){\vector(-3,-2){70}}  
        }
      \put(0,-10){
      \put(227,-4){\Verma}
      \put(243,-0.5){\TVerma}
      \put(74,-0.5){\TVerma}
      \put(58,-4){\Verma}
        \thicklines
          \put(230,0){\vector(-1,0){290}} 
          \put(60,0){\vector(3,-2){44}}
          \put(230,0){\vector(3,-2){44}}
          \put(37,-13){$\mC''_-$}
          \put(86,-14){$\tilde\mC''_+$}
          {\linethickness{.2pt}
            \put(60,-4){\line(0,1){20}}
            \put(77,-4){\line(0,1){20}}
            \put(65,12){\vector(1,0){12}}
            \put(63,12){\vector(-1,0){3}}
            \put(66,13){${}^1$}
            }
          \put(250,3){\linethickness{.2pt}
            \put(-20,-4){\line(0,1){20}}
            \put(-3,-4){\line(0,1){20}}
            \put(-15,12){\vector(1,0){12}}
            \put(-17,12){\vector(-1,0){3}}
            \put(-14,13){${}^1$}
            }
          \put(0,3){
            \put(248,0){\vector(-3,-2){44}}
            \put(78,0){\vector(-3,-2){44}}
            \put(78,0){\vector(1,0){258}}
            \put(251,-13){$\mC''_+$}
            \put(205,-16){$\tilde\mC''_-$}
            }
        }
      }
  \end{picture}
  \label{wide-gap-embedded2}
\end{equation}

\bigskip

\noindent
where distance $1$ is in the units of the $J^0_0$-charge. This has the
following drastic effect on the solutions to the analogue of
Eq.~\req{solve}, $J^-_0\,\ket{X}=\ket{G(0)}$ (where $\ket{X}$ is to
satisfy relaxed \hw{} conditions). The solution is not unique, since
one can add a vector proportional to the \hw{} vector of
$\tilde\mC''_+$; conversely, any two solutions (satisfying, in
addition, the relaxed \hw{} conditions) differ by a vector
proportional to the \hw{} state of~\,$\tilde\mC''_+$. \ The appearance
of submodule $\tilde\mC''_-$ has a similar effect on representing
$\ket{G(N'+1)}$ (which is the \hw{} vector of $\mC''_+$) as $J^+_0$
acting on another vector satisfying relaxed \hw{} conditions.

We can {\it define\/} $(J^-_0)^{-1}\,\ket{G(0)}$ to be a vector inside
the \,$\tilde\mC''_-$ Verma module, and similarly for the vector
$\ket{G(N')}$: 
\begin{equation}
  \ket{G_-(1)}=(J^-_0)^{-1}\ket{G_-(0)}\in\tilde\mC''_-\,,\quad
  \ket{G_+(N')}=
  (J^+_0)^{-1}\ket{G_+(N'+1)}\in\tilde\mC''_+
\end{equation}
Setting now
\begin{equation}
  \ket{G_-(i)}=(J^+_0)^{i-1}\ket{G_-(1)}\in\tilde\mC''_-\,,\quad
  \ket{G_+(i)}=
  (J^-_0)^{N' - i}\ket{G_+(N')}
  \in\tilde\mC''_+\,,
\end{equation}
we have
\begin{equation}
  \iota\ket{g''(i)} \equiv
  \ket{G(i)}=\ket{G_-(i)} + a''(i) \ket{G_+(i)}\,,
  \qquad i=1,\ldots,N'\,,
 \label{end}
\end{equation}
where $a''(i)$ are as in~\req{a''(i)}.
This defines the embedding
\begin{equation}
  \label{into-direct-sum}
  \mR''\hookrightarrow \tilde\mC''_-\oplus\tilde\mC''_+
\end{equation}
on the extremal states and, hence, on any vector from~\,$\mR''$. At
the same time,
\begin{equation}
  \mR''\cap\tilde\mC''_-=\mC''_-\,,\qquad
  \mR''\cap\tilde\mC''_+=\mC''_+\,.
\end{equation}

This situation repeats at every subsequent level
in~\req{d:IIIpm0(2,mp)-first}. Thus, in addition to the embeddings of
$\mC''_-$ and $\mC''_+$ into $\mR''$ performed by charged singular
vectors, there is embedding~\req{into-direct-sum} of $\mR''$ into the
direct sum.  Taking all this into account, we finally
complete~\req{d:IIIpm0(2,mp)-first} to the following embedding
diagram:
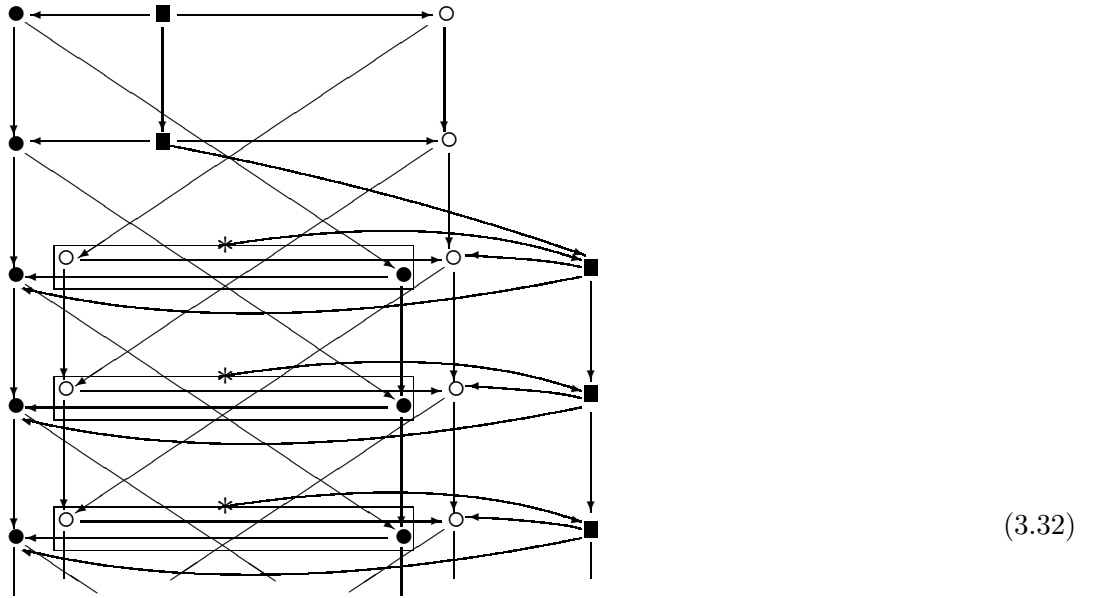
\begin{equation}
  \label{d:IIIpm0(2,mp)}
  \unitlength=0.9pt
  \begin{picture}(500,225)
    \put(50,0){
    \bezier{400}(116,163)(210,145)(290,117)
    \put(288,118){\vector(3,-1){2}}
    \put(292.5,106){\vector(0,-1){42}}
    \put(135,117.6){\Large$\ast$}
    \bezier{300}(136,121)(230,136)(288,115)
    \put(287,115.5){\vector(4,-1){2}}
    \bezier{500}(288,108)(150,80)(54,103)
    \put(55,102.5){\vector(-4,1){2}}
    \bezier{100}(288,112)(278,115)(240,117)
    \put(243,117){\vector(-1,0){2}}
    \put(0,-55){
    \put(292.5,106){\vector(0,-1){42}}
      \put(135,117.6){\Large$\ast$}
      \bezier{300}(136,121)(230,136)(288,115)
      \put(287,115.5){\vector(4,-1){2}}
      \bezier{500}(288,108)(150,80)(54,103)
      \put(55,102.5){\vector(-4,1){2}}
      \bezier{100}(288,112)(278,115)(240,117)
      \put(243,117){\vector(-1,0){2}}
      }
    \put(0,-110){
    \put(292.5,106){\line(0,-1){15}}
      \put(135,117.6){\Large$\ast$}
      \bezier{300}(136,121)(230,136)(288,115)
      \put(287,115.5){\vector(4,-1){2}}
      \bezier{500}(288,108)(150,80)(54,103)
      \put(55,102.5){\vector(-4,1){2}}
      \bezier{100}(288,112)(278,115)(240,117)
      \put(243,117){\vector(-1,0){2}}
      }
    {\linethickness{.2pt}
      \put(67.00,103.00){\framebox(151.00,18.00)[cc]{\mbox{}}}
      }
    \put(0,-55){
      {\linethickness{.2pt}
        \put(67.00,103.00){\framebox(151.00,18.00)[cc]{\mbox{}}}
        }
      }
    \put(0,-110){
      {\linethickness{.2pt}
        \put(67.00,103.00){\framebox(151.00,18.00)[cc]{\mbox{}}}
        }
      }
    \put(210,125){
      \put(-163,90){\Verma}
      \put(-155,90){\vector(3,-2){155}}
      \put(-160,88){\vector(0,-1){46}}
      \put(0,-55){
        \put(-163,90){\Verma}
        \put(-155,90){\vector(3,-2){155}}
        \put(-160,88){\vector(0,-1){46}}
        }
      \put(0,-110){
        \put(-163,90){\Verma}\put(-3,93){\vector(-1,0){152}}
        \put(0,90){\Verma}
        \put(3,89){\vector(0,-1){46}}
        \put(-155,90){\vector(3,-2){155}}
        \put(-160,88){\vector(0,-1){46}}
        }
      \put(0,-165){
        \put(-163,90){\Verma}\put(-3,93){\vector(-1,0){152}}
        \put(0,90){\Verma}
        \put(3,89){\vector(0,-1){46}}
        \put(-155,90){\line(3,-2){105}}
        \put(-160,88){\vector(0,-1){46}}
        }
      \put(0,-220){
        \put(-163,90){\Verma}\put(-3,93){\vector(-1,0){152}}
        \put(0,90){\Verma}
        \put(3,89){\line(0,-1){20}}
        \put(-155,90){\line(3,-2){30}}
        \put(-160,89){\line(0,-1){20}}
        }
      }
    \put(210,125){
      \put(-100,90){\Relaxed}
      \put(-97.5,88){\vector(0,-1){44}}
      \put(-103,93){\vector(-1,0){50}} \put(-91,93){\vector(1,0){107}}
      \put(-100,36){\Relaxed}
      \put(-103,40){\vector(-1,0){50}} \put(-91,40){\vector(1,0){109}}
      }
    \put(75,132){
      \put(153,83){\TVerma}
      \put(149,82){\vector(-3,-2){147}}
      \put(156,81){\vector(0,-1){44}}
      \put(0,-55){
        \put(154,85){\TVerma}
        \put(151,85){\vector(-3,-2){150}}
        \put(158,83){\vector(0,-1){40}}
        }
      \put(0,-110){
        \put(156,90){\TVerma}\put(3,93){\vector(1,0){152}}
        \put(-7,90){\TVerma}
        \put(-4,89){\vector(0,-1){46}}
        \put(156,90){\vector(-3,-2){155}}
        \put(160,88){\vector(0,-1){46}}
        }
      \put(0,-165){
        \put(157,90){\TVerma}\put(3,93){\vector(1,0){152}}
        \put(-7,90){\TVerma}
        \put(-4,89){\vector(0,-1){46}}
        \put(156,90){\line(-3,-2){115}}
        \put(160,88){\vector(0,-1){46}}
        }
      \put(0,-220){
        \put(157,90){\TVerma}\put(3,93){\vector(1,0){152}}
        \put(-7,90){\TVerma}
        \put(-4,89){\line(0,-1){20}}
        \put(156,90){\line(-3,-2){40}}
        \put(160,89){\line(0,-1){20}}
        }
      }
   \put(290, 108){\Relaxed}
    \put(290, 55){\Relaxed}
    \put(290, -2){\Relaxed}
}
  \end{picture}
\end{equation}

\bigskip

\medskip

\noindent
Here, the frame around the respective Verma and twisted-Verma modules
represents the direct sum of these modules. Accordingly, an arrow
drawn from that frame (symbolized by a $\ast$) to the corresponding
\RV{} module indicates that the \RV{} module is embedded into the
direct sum. On the other hand, the \RV{} module has two submodules
associated with charged singular vectors, as shown by the arrows drawn
{\it from\/} the \RV{} module.

The diagram is finite for III$^0_-(2,{-}{+})$ and infinite, for
III$^0_+(2,{-}{+})$. In the negative zone, where $t=-\tp/q$ with
$\tp,\,q\in\oN$, Eq.~\req{L-condition} now takes the form
\begin{equation}
  \mu_2-\mu_1=\xi\tp + \zeta\,,\qquad \xi\in\oN\,,\quad
  1\leq\zeta<\tp\,.
\end{equation}
As in all of the III{}${}_-^{0}$ cases, the number of the embedding
levels ($\equiv{}$the number of \RV{} modules except the top one) is
$\xi=\left[\frac{|\mu_1-\mu_2|}{\tp}\right]$.  In the negative zone,
further, the structure of the {\it Verma\/} and {\it twisted-Verma\/}
parts of the embedding diagram near the bottom is as follows:
\begin{equation}
  \unitlength=0.8pt
  \begin{picture}(400,150)
    \put(150,0){
      \put(18,121){\vector(2,-3){76}}
      \put(55,138){\vector(2,-3){40}}
      \put(8,127){\vector(0,-1){50}}\put(98,127){\vector(0,-1){50}}
      \put(5,70){\Verma}\put(13,73){\vector(1,0){80}}\put(95,70){\Verma}
      \put(8,67){\vector(0,-1){60}}\put(98,67){\vector(0,-1){60}}
      \put(5,0){\Verma}\put(13,3){\vector(1,0){80}}\put(95,0){\Verma}
      \put(20,10){
        \put(88,121){\vector(-2,-3){76}}
        \put(51,137){\vector(-2,-3){40}}
        \put(8,127){\vector(0,-1){50}}\put(98,127){\vector(0,-1){50}}
        \put(5,70){\TVerma}\put(93,73){\vector(-1,0){80}}
        \put(95,70){\TVerma}
        \put(8,67){\vector(0,-1){60}}\put(98,67){\vector(0,-1){60}}
        \put(5,0){\TVerma}\put(93,3){\vector(-1,0){80}}
        \put(95,0){\TVerma}}
      }
  \end{picture}
  \label{bottom}
\end{equation}
In the units of the $J^0_0$-charge, the distance between the two
bottom {\Verma}-modules in~\req{bottom} is $\zeta$ for $\xi$ even and
$(\tp-\zeta)$ for $\xi$ odd; the distance between the adjacent
{\Verma} and {\TVerma} modules is always~1.  Every quadruple of
(twisted) Verma modules is related to the \RV{} module at the same
level as explained above.

\paragraph{\IIIpm{00}(0).} $\mu_1 - \mu_2\in\oK(t)$, $t\in\oQ$,
$\mu_1 - \mu_2\in\oZ$, $\mu_1\notin\oZ$, $(\mu_1 - \mu_2)/t\in\oZ$. \
This case is covered by Part~\ref{2:1} of Theorem~\ref{thm:relax}.  We
have the diagram of the same form as in~\req{d:IIIpm0(0)}. However, in
the negative zone, where the diagrams are finite, the
\IIIpm{00}(0)-diagram is half that long as the \IIIpm0(0)-diagram.  In
terms of Eq.~\req{L-condition}, we now have $\zeta=\eta=0$, therefore
$|\mu_1-\mu_2|=\tp\xi$, $\xi\geq2$, the number of {\it embedded\/}
modules being $\left[\frac{\xi}{2}\right]$. This time, we have
small-$\xi$ `exceptions' at $\xi=2$ and $\xi=3$. Thus, the triples
$(\mu_1\notin\oZ,\mu_2\notin\oZ,t=-\frac{\tp}{q})$ with
$|\mu_1-\mu_2|$ equal to either~$2\tp$ or~$3\tp$ may be excluded from
the III${}_-^{00}(0)$ case and placed into~II${}_-(0)$.

\paragraph{\IIIpmzziimm.} $t\in\oQ$, $\mu_1\in-\oN$, $\mu_2\in-\oN$,
$(\mu_1 - \mu_2)/t\in\oZ$.  \ We are again in the situation described
by part~\ref{2:1} of Theorem~\ref{thm:relax}.  In this case, we have
the embedding diagram
\begin{equation}
  \label{d:IIIpm00(2,mm)}
  \unitlength=0.8pt
  \begin{picture}(500,145)
    \put(100,45){
      \put(200,90){\Relaxed}
      \put(197,93){\vector(-1,0){78}}
      \put(203,88){\vector(0,-1){45}}
      \put(0,-55){
        \put(200,90){\Relaxed}
      \put(197,93){\vector(-1,0){78}}
        \put(203,88){\vector(0,-1){45}}
        }
      \put(0,-110){
        \put(200,90){\Relaxed}
      \put(197,93){\vector(-1,0){78}}
        \put(203,88){\line(0,-1){30}}
        }
      }
    \put(10,45){
      \put(200,90){\Verma}
      \put(197,93){\vector(-1,0){46}}
            \put(0,-55){
        \put(200,90){\Verma}
      \put(197,93){\vector(-1,0){46}}
              }
      \put(0,-110){
        \put(200,90){\Verma}
      \put(197,93){\vector(-1,0){46}}
              }
      }
    \put(-48,45){
      \put(200,90){\Verma}
      \put(209,90){\vector(1,-1){48}}
            \put(0,-55){
        \put(200,90){\Verma}
      \put(209,90){\vector(1,-1){48}}
              }
      \put(0,-110){
        \put(200,90){\Verma}
      \put(209,90){\line(1,-1){30}}
              }
      }
  \end{picture}
\end{equation}

\bigskip

\bigskip

\noindent
which can be viewed as a degeneration of diagrams~\req{d:IIIpm0(2)}
and/or \req{d:IIIpm0(1)}.  It is finite or infinite depending on
whether $t$ is negative or positive respectively.  In the case where
$\mu_1=\mu_2$, $t>0$, the upper floor is somewhat changed:
\begin{equation}
  \label{d:IIIpm00(2,mm)-corr}
  \unitlength=0.7pt
  \begin{picture}(500,84)
    \put(0,-80){
    \put(100,65){
      \put(200,90){\Relaxed}
      \put(197,93){\vector(-1,0){107}}
      \put(203,88){\vector(0,-1){45}}
      \put(0,-55){
        \put(200,90){\Relaxed}
      \put(197,93){\vector(-1,0){78}}
        \put(203,88){\vector(0,-1){30}}
        }
      }
    \put(10,65){
            \put(0,-55){
        \put(200,90){\Verma}
      \put(197,93){\vector(-1,0){46}}
              }
      }
    \put(-48,65){
      \put(227,90){\Verma}
      \put(234,90){\vector(1,-2){24}}
            \put(0,-55){
        \put(200,90){\Verma}
      \put(209,90){\vector(1,-2){15}}
              }
      }
    }
  \end{picture}
\end{equation}

In the negative zone, the structure of diagram~\req{d:IIIpm00(2,mm)}
near the bottom depends on whether $\xi\equiv(\mu_1-\mu_2)/\tp$ is odd
or even:
\begin{equation}
  \label{d:terminate1}
  \unitlength=0.7pt
  \begin{picture}(500,80)
    \put(20,80){
      \put(163,-3){\vector(0,-1){68}}
      \put(10,-3){\vector(1,-1){68}}
      }
    \put(20,0){
      \put(0,0){\Verma}\put(77,3){\vector(-1,0){69}}
      \put(80,0){\Verma}\put(157,3){\vector(-1,0){69}}
      \put(160,0){\Relaxed}
      }
    \put(60,-20){$(\mu_1-\mu_2)/p~{\rm odd}$}
    \put(240,80){
      \put(163,-3){\vector(0,-1){68}}
      \put(10,-3){\vector(1,-1){68}}
      }
    \put(240,0){
      \put(80,0){\Verma}\put(157,3){\vector(-1,0){69}}
      \put(160,0){\Relaxed}
      }
    \put(300,-20){$(\mu_1-\mu_2)/p~{\rm even}$}
  \end{picture}
\end{equation}

\bigskip

\noindent
The distance (in the units of the $J^0_0$-charge) between two
{\Verma}-modules related by the charged singular vectors in the bottom
floor of the first of these diagrams is $\tp$.

\paragraph{\IIIpmzziipp.} $t\in\oQ$, $\mu_1\in\oN_0$, $\mu_2\in\oN_0$,
$(\mu_1 - \mu_2)/t\in\oZ$. \ In this case, the embedding diagrams are
the mirror-transform of the \IIIpmzziimm{} ones, with the Verma
modules replaced by twisted-Verma modules.

\paragraph{\IIIpmzziimp.} $t\in\oQ$, $\mu_1\in-\oN$, $\mu_2\in\oN_0$,
$(\mu_1 - \mu_2)/t\in\oZ$. \ This case can be considered as a further
degeneration of diagram \req{d:IIIpm0(2,mp)}. First, the charged
singular vectors give rise to the Verma and twisted-Verma embedding
diagrams which superpose as follows:
\begin{equation}
  \label{d:IIIpm00(2,mp)-first}
  \unitlength=0.9pt
  \begin{picture}(500,170)
    \put(300,125){
      \put(0,-55){
        \put(-160,90){\Verma}
        \put(-151,90){\vector(3,-1){150}}
        }
      \put(0,-110){
        \put(-160,90){\Verma}\put(-3,93){\vector(-1,0){148}}
        \put(0,90){\Verma}
        \put(-151,90){\vector(3,-1){150}}
        }
      \put(0,-165){
        \put(-160,90){\Verma}\put(-3,93){\vector(-1,0){148}}
        \put(0,90){\Verma}
        \put(-151,90){\vector(3,-1){150}}
        }
      \put(0,-220){
        \put(-160,90){\Verma}\put(-3,93){\vector(-1,0){148}}
        \put(0,90){\Verma}
        \put(-151,90){\line(3,-1){50}}
        }
      }
    \put(300,125){
      \put(-100,36){\Relaxed}
      \put(-103,40){\vector(-1,0){47}}\put(-91,40){\vector(1,0){106}}
      }
    \put(170,130){
      \put(0,-55){
        \put(150,86){\TVerma}
        \put(148,88){\vector(-3,-1){144}}
        }
      \put(0,-110){
        \put(154,90){\TVerma}\put(3,93){\vector(1,0){148}}
        \put(-7,90){\TVerma}
        \put(151,90){\vector(-3,-1){150}}
        }
      \put(0,-165){
        \put(153,90){\TVerma}\put(3,93){\vector(1,0){148}}
        \put(-7,90){\TVerma}
        \put(151,90){\vector(-3,-1){150}}
        }
      \put(0,-220){
        \put(153,90){\TVerma}\put(3,93){\vector(1,0){148}}
        \put(-7,90){\TVerma}
        \put(151,90){\line(-3,-1){50}}
        }
      }
  \end{picture}
\end{equation}

\bigskip

\noindent
{\it which does not show the \RV{} modules and some of the embeddings
  yet\/}! As in case~\IIIpmziimp{}, there are two Verma and two
twisted-Verma modules {\it all at the same level}, which are shown
somewhat displaced vertically in order to distinguish between the
different arrows.  The \hw{} vectors of the corresponding pairs of
Verma and twisted Verma modules are separated by the distance of
$J^0_0$-charge~1. In what follows, we denote by $\mC_-$ and $\mC_+$
the Verma and the twisted Verma module, respectively, generated from
  the corresponding charged singular vector.\pagebreak[3]

Now, let us describe the \RV{} modules that are not yet shown in the
above diagram.  Since $\mu_2 - \mu_1$ is now a multiple of $p$ (where
$t=\frac{p}{q}$), we see from the standard Verma embedding diagram
that the first two submodules in Verma module $\mC_-$ are
$\tilde\mC_-''$ and $\mC_-'$ such that they  are related by
$\mC_-'\subset\tilde\mC_-''$ and are embedded on the same level
in~\,$\mC_-$ (hence, by Theorem~\ref{thm:Verma}, on the same level
in~\,$\mR$).  We have the extremal diagram
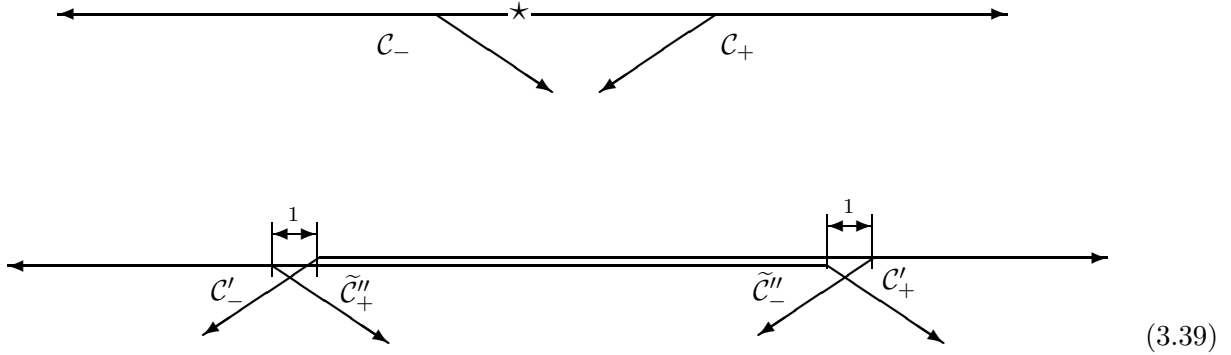
\begin{equation}
  \unitlength=1pt
  \begin{picture}(250,130)
    \put(-20,30){\thicklines
      \put(0,-40){
        \put(130,133){\Large$\star$}
        \put(138,135){\vector(1,0){180}}
        \put(129,135){\vector(-1,0){170}}
        \put(102,135){\vector(3,-2){44}}
        \put(208,135){\vector(-3,-2){44}}
        \put(80, 120){$\mC_-$}
        \put(210, 120){$\mC_+$}
        }
        \put(250,0){\vector(-1,0){310}}
        \put(40,0){\vector(3,-2){44}}
        \put(250,0){\vector(3,-2){44}}
        \put(17,-13){$\mC'_-$}
        \put(66,-14){$\tilde\mC''_+$}
        {\linethickness{.2pt}
          \put(40,-4){\line(0,1){20}}
          \put(57,-4){\line(0,1){20}}
          \put(45,12){\vector(1,0){12}}
          \put(43,12){\vector(-1,0){3}}
          \put(46,13){${}^1$}
          }
        \put(270,3){\linethickness{.2pt}
          \put(-20,-4){\line(0,1){20}}
          \put(-3,-4){\line(0,1){20}}
          \put(-15,12){\vector(1,0){12}}
          \put(-17,12){\vector(-1,0){3}}
          \put(-14,13){${}^1$}
          }
        \put(0,3){
          \put(268,0){\vector(-3,-2){44}}
          \put(58,0){\vector(-3,-2){44}}
          \put(58,0){\vector(1,0){298}}
          \put(271,-12){$\mC'_+$}
          \put(222,-16){$\tilde\mC''_-$}
          }
      }
  \end{picture}
  \label{wide-gap-embedded00}
\end{equation}
This picture---the grouping of (twisted) Verma submodules into
quadruples---is reproduced for every level except, possibly, the
bottom one in the negative zone (see below). As in case~\IIIpmziimp{},
the \RV{} module $\mR''$ at every level is embedded into the direct
sum of the corresponding $\tilde\mC''_-$ and $\tilde\mC''_+$, with
$\mR''\cap\,\tilde\mC''_-=\mC'_-$ and $\mR''\cap\tilde\mC''_+=\mC'_+$.
Thus, the embedding diagram reads as
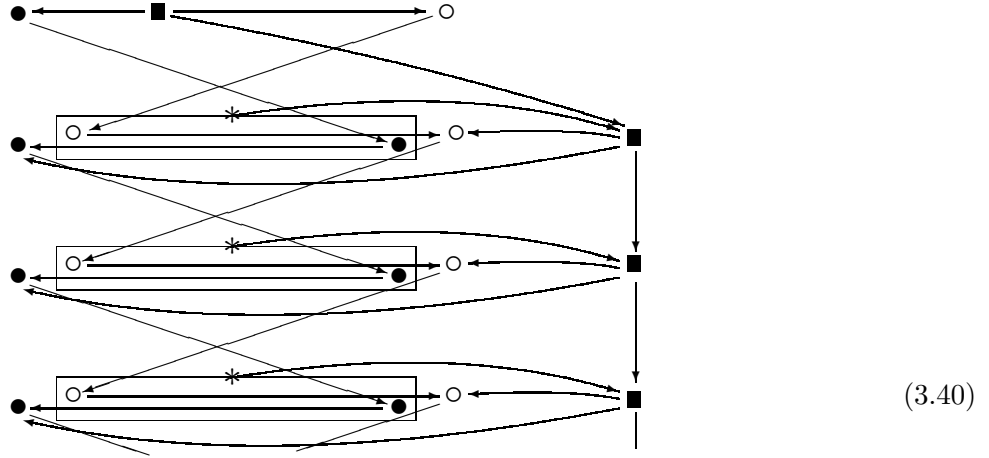
\begin{equation}
  \label{d:IIIpm00(2,mp)}
  \unitlength=0.9pt
  \begin{picture}(500,170)
    \put(403.5,106){\vector(0,-1){42}}
    \put(400, 108){\Relaxed}
    \put(400, 55){\Relaxed}
    \put(400, -2){\Relaxed}
    \bezier{400}(208,163)(310,145)(398,117)
    \put(396,118){\vector(3,-1){2}}
    \put(230,117.6){\Large$\ast$}
    \bezier{300}(233,121)(330,136)(396,115)
    \put(394,116){\vector(4,-1){2}}
    \bezier{500}(396,108)(250,80)(150,102)
    \put(148,102.5){\vector(-4,1){2}}
    \bezier{100}(396,112)(378,116)(333,114)
    \put(335,114){\vector(-1,0){2}}
    {\linethickness{.2pt}
      \put(160.00,103.00){\framebox(151.00,18.00)[cc]{\mbox{}}}
      }
    \put(0, -55){
      \put(403.5,106){\vector(0,-1){42}}
      \bezier{300}(233,121)(330,136)(396,115)
      \put(394,116){\vector(4,-1){2}}
      \bezier{500}(396,108)(250,80)(150,102)
      \put(148,102.5){\vector(-4,1){2}}
      \bezier{100}(396,112)(378,116)(333,114)
      \put(335,114){\vector(-1,0){2}}
      \linethickness{.2pt}
      \put(230,117.6){\Large$\ast$}
      \put(160.00,103.00){\framebox(151.00,18.00)[cc]{\mbox{}}}
      }
    \put(0, -110){
      \put(403.5,106){\line(0,-1){15}}
      \bezier{300}(233,121)(330,136)(396,115)
      \put(394,116){\vector(4,-1){2}}
      \bezier{500}(396,108)(250,80)(150,102)
      \put(148,102.5){\vector(-4,1){2}}
      \bezier{100}(396,112)(378,116)(333,114)
      \put(335,114){\vector(-1,0){2}}
      \linethickness{.2pt}
      \put(230,117.6){\Large$\ast$}
      \put(160.00,103.00){\framebox(151.00,18.00)[cc]{\mbox{}}}
      }
    \put(300,125){
      \put(0,-55){
        \put(-160,90){\Verma}
        \put(-151,90){\vector(3,-1){150}}
        }
      \put(0,-110){
        \put(-160,90){\Verma}\put(-3,93){\vector(-1,0){148}}
        \put(0,90){\Verma}
        \put(-151,90){\vector(3,-1){150}}
        }
      \put(0,-165){
        \put(-160,90){\Verma}\put(-3,93){\vector(-1,0){148}}
        \put(0,90){\Verma}
        \put(-151,90){\vector(3,-1){150}}
        }
      \put(0,-220){
        \put(-160,90){\Verma}\put(-3,93){\vector(-1,0){148}}
        \put(0,90){\Verma}
        \put(-151,90){\line(3,-1){50}}
        }
      }
    \put(300,125){
      \put(-100,36){\Relaxed}
      \put(-103,40){\vector(-1,0){47}}\put(-91,40){\vector(1,0){106}}
      }
    \put(170,130){
      \put(0,-55){
        \put(150,86){\TVerma}
        \put(148,88){\vector(-3,-1){144}}
        }
      \put(0,-110){
        \put(154,90){\TVerma}\put(3,93){\vector(1,0){148}}
        \put(-7,90){\TVerma}
        \put(151,90){\vector(-3,-1){150}}
        }
      \put(0,-165){
        \put(153,90){\TVerma}\put(3,93){\vector(1,0){148}}
        \put(-7,90){\TVerma}
        \put(151,90){\vector(-3,-1){150}}
        }
      \put(0,-220){
        \put(153,90){\TVerma}\put(3,93){\vector(1,0){148}}
        \put(-7,90){\TVerma}
        \put(151,90){\line(-3,-1){60}}
        }
      }
  \end{picture}
\end{equation}

\bigskip

This is finite or infinite depending on whether $t$ is negative or
positive respectively. In the negative zone where $t=-\frac{\tp}{q}$,
the diagram terminates differently depending on whether
$(\mu_1-\mu_2)/\tp$ is odd or even:
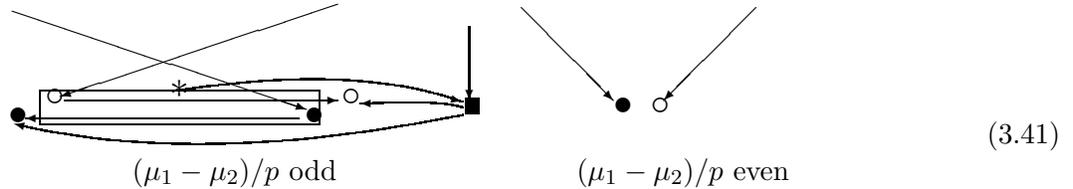
\begin{equation}
  \label{last-bottom}
  \unitlength=0.7pt
  \begin{picture}(500,75)
    \put(70,150){
      \put(0,-165){
        \put(151,90){\vector(-3,-1){150}}
        }
      \put(-20,-230){
        \put(153,90){\Verma}\put(150,93){\vector(-1,0){148}}
        \put(-7,90){\Verma}
        \put(-5,151){\vector(3,-1){159}}
        }
      \put(0,-220){
        \put(153,90){\TVerma}\put(3,93){\vector(1,0){148}}
        \put(-7,90){\TVerma}
        }
      \put(40,-170){$(\mu_1-\mu_2)/p~{\rm odd}$}
      }
    \put(310,-70){
      \put(10,143){\vector(1,-1){50}}
      \put(60,85){\Verma}
      \put(137,143){\vector(-1,-1){50}}
      \put(80,85){\TVerma}
      \put(40,50){$(\mu_1-\mu_2)/p~{\rm even}$}
      }
\put(-100,-93){
      \bezier{300}(233,121)(330,136)(388,115)
      \put(386,116){\vector(4,-1){2}}
      \bezier{500}(388,108)(250,80)(150,102)
      \put(148,102.5){\vector(-4,1){2}}
      \bezier{100}(388,112)(378,116)(333,114)
      \put(335,114){\vector(-1,0){2}}
      \put(392.00,156.00){\vector(0,-1){40}}
      \linethickness{.2pt}
      \put(230,117.6){\Large$\ast$}
      \put(160.00,103.00){\framebox(151.00,18.00)[cc]{\mbox{}}}
      \put(390.00,108.00){\Relaxed}
      }
  \end{picture}
\end{equation}

\bigskip

\noindent
In the second case, {\it there is no \RV{} module in the bottom
  floor}, as described in Part~\ref{no-relaxed} of
Theorem~\ref{thm:Verma}. The distance between the two Verma modules
(as well as between two twisted Verma modules) in the first of these
diagrams is $\tp$, while the distance between the adjacent Verma and
twisted Verma modules is always~1. When $\xi=2$ in~\req{L-condition},
we now have $\mu_2-\mu_1=2\tp$ and the embeddings terminate already at
the second level. The embedding diagram then takes the following
exceptional form:
\begin{equation}
  \label{exceptional}
  \unitlength=0.8pt
  \begin{picture}(500,70)
    \put(40,0){
      \put(200,60){\Relaxed}
      \put(195,64){\vector(-1,0){53}}
      \put(210,64){\vector(1,0){53}}
      \put(128,60){\Verma}
      \put(138,57){\vector(1,-1){50}}
      \put(266,60){\TVerma}
      \put(267,57){\vector(-1,-1){50}}
      \put(206,0){\TVerma}
      \put(190,0){\Verma}
      }
  \end{picture}
\end{equation}

The structure of the module is easily understood in terms of the {\it
extremal\/} diagram
\begin{equation}
  \unitlength=1pt
  \begin{picture}(250,110)
    \put(-20,10){\thicklines
      \put(0,-40){
        \put(130,133){\Large$\star$}
        \put(138,135){\vector(1,0){180}}
        \put(129,135){\vector(-1,0){170}}
        \put(82,135){\vector(3,-2){44}}
        \put(228,135){\vector(-3,-2){44}}
        \put(70, 120){$\mC_-$}
        \put(230, 120){$\mC_+$}
        }
        \put(140,20){\vector(-1,0){160}}
        \put(140,20){\vector(3,-2){44}}
        \put(160,20){\vector(1,0){150}}
        \put(160,20){\vector(-3,-2){44}}
        \put(160,20){\linethickness{.2pt}
          \put(-20,-4){\line(0,1){20}}
          \put(0,-4){\line(0,1){20}}
          \put(-15,12){\vector(1,0){15}}
          \put(-17,12){\vector(-1,0){3}}
          \put(-11,13){${}^1$}
          }
      }
  \end{picture}
  \label{exception-extremal}
\end{equation}
Thus, there are no states that would generate a \RV{} submodule,
instead the lower floor consists of a direct sum of the Verma and the
twisted Verma modules. However, this configuration corresponds to a
zero of the Ka\v c determinant~\cite{[BFK]}; in fact, there are as
many states on that level satisfying the relaxed \hw{} conditions as
in a \RV{} module.

\section{The $\N2$ side: Massive and topological Verma
  modules\label{sec:N2}}\lvm In this section, we introduce Verma-like
modules over the $\N2$ superconformal algebra---the massive and
topological Verma modules. We then explain how the embedding diagrams
constructed in the previous section can be read in the $\N2$ terms, as
the embedding diagrams of massive $\N2$ Verma modules (while the
topological Verma-module embedding diagrams are isomorphic to the
standard embedding diagrams of $\tSL2$ Verma modules). This does not
cover the embedding diagrams of $\N2$ modules with the central charge
$\ctop=3$, which are considered separately in Sec.~\ref{subsec:c3}.

\subsection{The $\N2$ algebra}\lvm
The $\N2$ superconformal algebra contains two fermionic currents,
$\cQ$ and $\cG$, in addition to the Virasoro generators $\cL$ and the
$U(1)$ current $\cH$. The commutation relations can be chosen as
\begin{equation}\new
  \begin{array}{lclclcl}
    \left[\cL_m,\cL_n\right]&=&(m-n)\cL_{m+n}\,,&\qquad&[\cH_m,\cH_n]&=
    &\frac{\Ctop}{3}m\delta_{m+n,0}\,,\\

    \left[\cL_m,\cG_n\right]&=&(m-n)\cG_{m+n}\,,&
    \qquad&[\cH_m,\cG_n]&=&\cG_{m+n}\,,
    \\
    \left[\cL_m,\cQ_n\right]&=&-n\cQ_{m+n}\,,&
    \qquad&[\cH_m,\cQ_n]&=&-\cQ_{m+n}\,,\\

    \left[\cL_m,\cH_n\right]&=&
    \multicolumn{5}{l}{-n\cH_{m+n}+\frac{\Ctop}{6}(m^2+m)
      \delta_{m+n,0}\,,}\\
    \left\{\cG_m,\cQ_n\right\}&=&
    \multicolumn{5}{l}{2\cL_{m+n}-2n\cH_{m+n}+
      \frac{\Ctop}{3}(m^2+m)\delta_{m+n,0}\,,}
  \end{array}\qquad m,~n\in\oZ\,.
  \label{topalgebra}
\end{equation}
The element $\Ctop$ is central; in representations, we will not
distinguish between $\Ctop$ and its eigenvalue $\ctop\!\in\!\oC$,
which it will be convenient to parametrize as $\ctop=3\,\frac{t-2}{t}$
with $t\!\in\!\oC\setminus\{0\}$.  Then, however, the special point
$\ctop=3$ requires an additional investigation. The central charge
appears as the anomaly of the $\cH$ current rather than in the
Virasoro commutation relations, which is simply a matter of choosing
the basis in the algebra.

The spectral flow transform~\cite{[SS],[LVW]}, which acts as
\begin{equation}
  {\cal U}_\theta:\new
  \begin{array}{rclcrcl}
    \cL_n&\mapsto&\cL_n+\theta\cH_n+\frac{\ctop}{6}(\theta^2+\theta)
    \delta_{n,0}\,,&{}&
    \cH_n&\mapsto&\cH_n+\frac{\ctop}{3}\theta\delta_{n,0}\,,\\
    \cQ_n&\mapsto&\cQ_{n-\theta}\,,&{}&\cG_n&\mapsto&\cG_{n+\theta}\,,
  \end{array}
  \label{U}
\end{equation}
produces isomorphic images of the above algebra. The family of
algebras thus obtained includes the Neveu--Schwarz and Ramond $\N2$
algebras, as well as the algebras in which the fermion modes range
over $\pm\theta+\oZ$, $\theta\in\oC$. Any `invariant' assertion
regarding representations of the algebra~\req{topalgebra} is therefore
valid for any of the spectral-flow-transformed algebras as well.
The spectral flow transform is an automorphism for~$\theta\in\oZ$.

\subsection{Massive $\N2$ Verma modules\label{subsec:massive}}\lvm
A massive Verma module $\mU_{h,\ell,t}$ is freely generated by the
generators $\cL_{-m}$, $\cH_{-m}$, $\cG_{-m}$, $m\in\oN$, and
$\cQ_{-m}$, $m\in\oN_0$ from a {\it massive \hw{} vector\/}
$\ket{h,\ell,t}$ satisfying the following set of highest-weight
conditions:
\begin{equation}\new
  \begin{array}{rcl}
    \cQ_{\geq1}\,\ket{h,\ell,t}\kern-4pt&=&
    \kern-4pt\cG_{\geq0}\,
    \ket{h,\ell,t}= \cL_{\geq 1}\,\ket{h,\ell,t}=
    \cH_{\geq1}\,\ket{h,\ell,t}=0\,,\\
    \cH_0\,\ket{h,\ell,t}\kern-4pt&=&
    \kern-4pt
    h\,\ket{h,\ell,t}\,,\qquad
    \cL_0\,\ket{h,\ell,t}
    = \ell\,\ket{h,\ell,t}\,.
  \end{array}
  \label{masshw}
\end{equation}
Using the bigrading implied by (charge,\,level), or more precisely, by
the eigenvalues of $(-\cH_0,\,\cL_0)$, the extremal diagram of the
massive Verma module reads as
\begin{equation}
  \unitlength=1.00mm
  \begin{picture}(140,40)
    \put(50.00,5.00){
      \put(00.00,00.00){$\state$}
      \put(10.00,20.00){$\state$}
      \put(10.00,20.00){$\state$}
      \put(20.00,30.00){$\star$}
      \put(30.00,30.00){$\state$}
      \put(40.00,20.00){$\state$}
      \put(50.00,00.00){$\state$}
      \put(10.00,18.70){\vector(-1,-2){8}}
      \put(19.50,29.50){\vector(-1,-1){7}}
      \put(22.40,31.10){\vector(1,0){7}}
      \put(32.50,29.80){\vector(1,-1){7}}
      \put(42.00,19.00){\vector(1,-2){8}}
      \put(00.00,13.00){${}_{\cG_{-2}}$}
      \put(09.00,26.50){${}_{\cG_{-1}}$}
      \put(23.90,33.50){${}_{\cQ_{0}}$}
      \put(37.00,27.50){${}_{\cQ_{-1}}$}
      \put(47.00,13.00){${}_{\cQ_{-2}}$}
      \put(11.00,32.00){${}_{\ket{h_,\ell,t}}$}
      \put(00.50,-06.00){$\vdots$}
      \put(50.50,-06.00){$\vdots$}
      }
  \end{picture}
  \label{massdiagramdouble}
\end{equation}
It has the shape of a parabola for the simple reason that, once one
has acted on the \hw{} vector with, say, $\cQ_0$, applying the same
operator once again would give identical zero, and `the best one can
do' to construct a state with the extremal bigrading is to act with
the $\cQ_{-1}$ mode, etc.

An important fact is that all of the states on the extremal diagram
satisfy the annihilation conditions
\begin{equation}
  \cQ_{-\theta+m+1}\approx \cG_{\theta+m}\approx
  \cL_{m+1}\approx\cH_{m+1}\approx0\,,\quad m\in\oN_0
  \label{twistedannihil}
\end{equation}
for $\theta$ ranging over the integers, from $-\infty$ in the left end
to $+\infty$ in the right end of the parabola\footnote{Anticipating
  the discussion in Sec.~\ref{subsec:anti-KS}, it is instructive to
  compare~\req{massdiagramdouble} with relaxed-$\tSL2$ extremal
  diagram~\req{floor}. There, all of the extremal states satisfy {\it
    the same\/} annihilation conditions. This difference between the
  behaviour of $\tSL2$ and $\N2$ extremal diagrams is the source of
  several complications arising on the $\N2$ side, even though, as we
  will see in Sec.~\ref{subsec:anti-KS}, the two representation
  theories are essentially isomorphic.}.  Further, there can be {\it
  two\/} different types of Verma submodules in~$\mU_{h,\ell,t}$,
which can conveniently be represented as
\begin{equation}
  \unitlength=1pt
  \begin{picture}(350,80)
    \bezier{600}(0,0)(60,160)(120,0)
    \bezier{600}(20,0)(70,110)(110,0)
    \put(180,40){or}
    \put(230,0){
      \bezier{600}(0,0)(60,160)(120,0)
      \bezier{400}(10,25)(60,100)(100,0)
      \put(10,25){$\bullet$}
      }
  \end{picture}
  \label{pic:compare}
\end{equation}
In the first case, we have a massive Verma {\it sub\/}module, all of
the states on its extremal diagram satisfying annihilation
conditions~\req{twistedannihil}, while in the other case there is a
distinguished state, namely the one that satisfies twisted topological
\hw{} conditions, which we consider in the next subsection.

\subsection{Topological $\N2$ modules\label{subsec:top}}\lvm
The {\it twisted topological \hw{} vector\/} $\ket{h,t;\theta}_{\rm
  top}$ satisfies the
annihilation conditions
\begin{equation}
  \cQ_{-\theta+m}\ket{h,t;\theta}_{\rm top}=
  \cG_{\theta+m}\ket{h,t;\theta}_{\rm top}=
  \cL_{m+1}\ket{h,t;\theta}_{\rm top}=
  \cH_{m+1}\ket{h,t;\theta}_{\rm top}=0\,,\quad m\in\oN_0
  \label{annihiltop}
\end{equation}
for some $\theta\in\oZ$, with the following eigenvalues of the Cartan
generators:
\begin{equation}\label{Cartantheta}\new
  \begin{array}{rcl}
    (\cH_0+\frac{\ctop}{3}\theta)\,\ket{h,t;\theta}_{\rm
      top}&=&
    h\,\ket{h,t;\theta}_{\rm top}\,,\\
    (\cL_0+\theta\cH_0+\frac{\ctop}{6}(\theta^2+\theta))
    \,\ket{h,t;\theta}_{\rm top}&=&0
  \end{array}
\end{equation}
(the second equation in~\req{Cartantheta} follows from the annihilation
conditions).  The state $\ket{h,t;\theta}_{\rm top}$ is defined in
accordance with the action of the automorphism~\req{U}:
$\ket{h,t;\theta}_{\rm top}=\cU_\theta\,\ket{h,t;0}_{\rm top}$.  Then
$\cU_{\theta'}\kettop{h,t;\theta}=\kettop{h,t;\theta+\theta'}$.  We
will also write $\ket{h,t}_{\rm top}\equiv \ket{h,t;0}_{\rm top}$ for
the `untwisted' case of $\theta=0$; then, in particular
\begin{equation}
  \cQ_0\,\kettop{h,t}=0\,,\qquad
  \cG_0\,\kettop{h,t}=0\,,\qquad
  \cL_0\,\kettop{h,t}=0\,.
\end{equation}

The {\it twisted topological Verma module\/} $\smV_{h,t;\theta}$ is
freely generated from $\kettop{h,t;\theta}$ by $\cQ_{\leq-1-\theta}$,
$\cG_{\leq-1+\theta}$, $\cL_{\leq-1}$, and $\cH_{\leq-1}$.  We also
denote by $\mV_{h,t}\equiv\smV_{h,t;0}$ the untiwsted module.  The
extremal diagram of a topological Verma module reads (in the
`untwisted' case of $\theta=0$ for simplicity)
\begin{equation} \unitlength=1.00mm
  \begin{picture}(140,40)
    \put(50.00,5.0){
      \put(00.00,00.00){$\state$}
      \put(10.00,20.00){$\state$}
      \put(10.00,20.00){$\state$}
      \put(20.00,30.00){$\bullet$}
      \put(29.70,20.00){$\state$}
      \put(40.00,00.00){$\state$}
      \put(9.70,19.00){\vector(-1,-2){8}}
      \put(19.70,29.70){\vector(-1,-1){7}}
      \put(22.00,29.70){\vector(1,-1){7}}
      \put(32.00,19.00){\vector(1,-2){8}}
      \put(00.00,13.00){${}_{\cG_{-2}}$}
      \put(11.00,28.00){${}_{\cG_{-1}}$}
      \put(27.00,28.00){${}_{\cQ_{-1}}$}
      \put(37.00,13.00){${}_{\cQ_{-2}}$}
      \put(19.00,34.00){${}_{\ket{h,t}_{\rm top}}$}
      \put(00.50,-06.00){$\vdots$}
      \put(40.50,-06.00){$\vdots$}
      }
  \end{picture}
  \label{topdiag}
\end{equation}
As before, the extremal states satisfy annihilation
conditions~\req{twistedannihil}.  A characteristic feature of the
topological extremal diagram, however, is the existence of a `cusp',
i.e.\ the (twisted) topological \hw{} state, which satisfies stronger
annihilation conditions than the other extremal states. As a result,
the extremal diagram is narrower than that of a massive Verma module.
This can be formalized as the following {\it criterion of terminating
fermionic chains\/}~\cite{[FST]}, which singles out all the modules of
the topological \hw-{\it type\/}: \ Given a vector $\ket{X}$ in the
module and any $n\in\oZ$, by the `massive' parabola $\cP(n,X)$ running
through $\ket{X}$ we understand the set of states
\begin{equation}
  \cQ_{n-N}\,\ldots\,\cQ_{n-1}\,\cQ_{n}\,\ket X\,,\quad
  \cG_{-n-M}\,\ldots\,\cG_{-n-2}\,\cG_{-n-1}\,\ket X
  \qquad N, M\in\oN\,.
  \label{terminate}
\end{equation}
Then, a module is of the topological \hw-type if and only if {\it any
  `massive' parabola intersects the boundary (the extremal diagram of
  the module) on at least one end,} which means that the
states~\req{terminate} become zero in at least one branch, either
for $N\gg1$ or for $M\gg1$.
That the massive $\N2$ Verma modules do not satisfy this criterion is a
formal way to express the fact that the extremal diagrams of massive
Verma modules are wider than those of the topological Verma modules.

Another way to understand the difference between the massive and the
topological $\N2$ Verma modules is to recall that, in general,
Verma-like modules can be defined in terms of induced representations.
In the case of the $\N2$ algebra, one uses representations of Lie
superalgebra $gl(1|1)$, all of whose irreducible representations are
$(1,1)$-dimensional. A proper Verma module is then the one induced from
the trivial representation, which leaves us with precisely one extremal
state at the top of the extremal diagram and, thus, singles out the
{\it topological\/} Verma modules.  On the other hand,
$(1,1)$-dimensional representations of $gl(1|1)$ correspond, in an
obvious way, to the massive `Verma' modules\footnote{We thank
  I.~Shchepochkina for this observation.}.  Thus, although only the
topological Verma modules are the `true' {\it Verma\/} modules from
this point of view, we still use the name of Verma modules also for the
massive ones, since this has become traditional in the literature.

\subsection{From $\N2$ to $\tSL2$\label{subsec:anti-KS}}\lvm
We now recall an operator construction~\cite{[FST]} that allows one to
build the $\tSL2$ currents out of the $\N2$ generators and a free
`Liouville' scalar with the operator product
$\phi(z)\phi(w)=-\ln(z-w)$.  As a necessary preparation, we `pack' the
modes of the $\N2$ generators into the corresponding fields,
$\cT(z)=\sum_{n\in\oZ}\cL_n z^{-n-2}$, \ $\cG(z)=\sum_{n\in\oZ}\cG_n
z^{-n-2}$, $\cQ(z)=\sum_{n\in\oZ}\cQ_n z^{-n-1}$, and
$\cH(z)=\sum_{n\in\oZ}\cH_n z^{-n-1}$, and similarly with the $\tSL2$
currents. We also define vertex operators $\psi=e^\phi$ and
$\spsi=e^{-\phi}$.  Then, given the generators of the $\N2$ algebra
with central charge $\ctop\neq3$, the currents
\begin{equation}
  J^+= \cQ\psi\,,\qquad J^-=\frac{3}{3-\ctop}\, \cG\spsi\,,\qquad
  J^0=-\frac{3}{3-\ctop}\,\cH+\frac{\ctop}{3-\ctop}\,\d\phi
  \label{invKS}
\end{equation}
satisfy the $\tSL2$ algebra of level $k={2\ctop\over3-\ctop}$ (or, in
terms of $t=k+2$, we have the familiar relation $\ctop=3(t-2)/t$).  We
also obtain a free scalar, with signature $-1$, whose modes commute
with the $\tSL2$ generators~\req{invKS}:
\begin{equation}
  I^-=\sqrt{\frac{t}{2}}(\cH-\d\phi)\,.
  \label{dF}
\end{equation}
The modes $I^-_n$ introduced as $I^-(z)=\sum^{\infty}_{n=-\infty}I^-_n
z^{-n-1}$ generate a Heisenberg algebra
$[I^-_n,\,I^-_m]=-n\delta_{m+n,0}$. Let $\mF^-_q$ be the Fock module
over this Heisenberg algebra with the \hw{} vector defined by
\begin{equation}
  I^-_n\ket{q}^-=0\,,\quad n\geq1\,,\qquad I^-_0\ket{q}^-=q\ket{q}^-.
\end{equation}

The behaviour of {\it representations\/} under operator mappings used
in conformal field theory can be quite complicated\footnote{Recall,
  for instance, how the $\tSL2$ Verma modules are rearranged under the
  Wakimoto bosonization~\cite{[FFr]} --- the Wakimoto modules more or
  less `interpolate' between the Verma and contragredient Verma
  modules.}. In our case, we take a topological Verma module
$\mV_{h,t}$ and tensor it with the module~$\Xi$ of the `Liouville'
scalar.  This is defined as $\Xi=\oplus_{n\in\oZ}\mF_n$, where $\mF_n$
is a Verma module with the highest-weight vector $\ket{n}_\phi$ such
that
\begin{equation}
  \phi_m\ket{n}_\phi=0\,,~ m\geq1\,,\quad
  \psi_m\ket{n}_\phi=0\,,~ m\geq n+1\,,\quad
  \spsi_m\ket{n}_\phi=0\,,~ m\geq -n+2\,,
\end{equation}
and $\phi_0\ket{n}_\phi=-n\ket{n}_\phi$. We then have the following
Theorem:
\begin{thm}[\cite{[FST]}]\label{thm:topequiv}\mbox{}

  \begin{enumerate}
    \addtolength{\parskip}{-4pt}
  \item There is an isomorphism of $\tSL2$ representations
    \begin{equation}
      \mV_{h,t}\tensor\Xi\approx\bigoplus_{\theta\in\oZ}\,
      \smM_{-\frac{t}{2}h,t;\theta}\tensor
      \mF^-_{\sqrt{\frac{t}{2}}(h+\theta)}
      \label{idspaces2}
    \end{equation}
    where on the left-hand side, the $\tSL2$ algebra acts by
    generators~\req{invKS}, while on the right-hand side it acts
    naturally on the twisted Verma
    module~$\smM_{-\frac{t}{2}h,t;\theta}$.

  \item A singular vector exists in the topological Verma module
    $\mV_{h,t}$ if and only if a singular vector exists in one (hence,
    in all) of the twisted $\tSL2$ Verma modules
    $\smM_{-\frac{t}{2}h,t;\theta}$, $\theta\in\oZ$. Whenever this is
    the case, moreover, the respective submodules associated with the
    singular vectors, in their own turn, satisfy an equation of the
    same type as~\req{idspaces2}.

  \end{enumerate}
\end{thm}
The Theorem means that, as regards the existence and the embedding
structure of submodules, the topological $\N2$ modules are {\it
  equivalent\/} to $\tSL2$ Verma modules\footnote{In particular, the
  $\Xi$ and $\mF^-_{\ldots}$ modules in~\req{idspaces2} are really
  `auxiliary', since nothing interesting can happen with these modules
  that would violate the balance between the topological $\N2$ and the
  Verma $\tSL2$ modules.}. The statement about the singular vectors
appeared, in a rudimentary form, in~\cite{[S-sing]}.

While the submodules appear in a twisted topological Verma modules
simultaneously with submodules in the corresponding $\tSL2$ Verma
module, yet the submodules in a topological Verma module are
necessarily {\it twisted\/} by some $\theta\neq0$\,\footnote{A common
  feature of the $\tSL2$ Verma modules and the topological $\N2$ Verma
  modules is that all of them are {\it freely\/} generated from a
  state that satisfies stronger annihilation conditions than the other
  states in the extremal diagram; {\it when there is no additional
    degeneration}, one {\it can\/} generate the same submodule from
  the state marked with a $\times$, but there are hardly any reasons
  to do so in the $\tSL2$ case. The point of~\cite{[ST4]} is that
  doing so in the $\N2$ case is equally inconvenient.  }:
$$
\unitlength=.5pt
\begin{picture}(600,170)
  \put(0,10){
    \bezier{400}(0,0)(50,120)(118,166)
    \bezier{400}(118,166)(190,120)(240,0)
    \bezier{200}(20,0)(40,50)(70,85)
    \bezier{300}(70,85)(170,140)(230,0)
    \put(115,95){$\times$}
    \put(240,100){$\Longleftrightarrow$}
    \put(300,160){\line(1,0){180}}
    \put(480,160){\line(1,-1){120}}
    \put(310,100){\line(1,0){100}}
    \put(410,100){\line(1,-1){100}}
    \put(480,17){$\times$}
    }
\end{picture}
$$
More precisely, to the MFF singular vector $\ket{{\rm
    MFF}(r,s,t)}^\pm$, $r,s\in\oN$ (see~\req{mff}), there corresponds
the {\it topological singular vector\/}~\cite{[ST2],[ST3]}
$\ket{E(r,s,t)}^\pm$ that satisfies the $\theta=\mp r$-twisted
topological \hw{} conditions
\begin{equation}
  \cQ_{\geq\pm r}\ket{E(r,s,t)}^\pm=
  \cG_{\geq\mp r}\ket{E(r,s,t)}^\pm=
  \cL_{\geq1}\ket{E(r,s,t)}^\pm=
  \cH_{\geq1}\ket{E(r,s,t)}^\pm=0\,.
  \label{twistedtophw}
\end{equation}
As we see from the twist, the submodule generated from
$\ket{E^\pm(r,s,t)}^\theta\in\smV_{h,t;\theta}$ is the twisted
topological Verma module $\smV_{h\pm r\frac{2}{t},t;\theta\mp r}$.
Equivalently, one may choose to describe the positions of
$\ket{E^\pm(r,s,t)}^\theta\in\smV_{h,t;\theta}$ in the (charge, level)
lattice by using the {\it eigenvalues\/} of $\cH_0$ and $\cL_0$:
\begin{equation}
  \cH_0\,\ket{E^\pm(r,s,t)}^\theta
  = h_0^\pm\,\ket{E^\pm(r,s,t)}^\theta\,,\qquad
  \cL_0\,\ket{E^\pm(r,s,t)}^\theta
  =\ell_0^\pm\,\ket{E^\pm(r,s,t)}^\theta\,,
\end{equation}
then
\begin{equation}
  h_0^\pm=h_0 \pm r\,,\qquad
  \ell_0^\pm = \ell_0 + \half r(r\mp2\theta+2s-1)
\end{equation}
where $h_0$ and $\ell_0$ are the {\it eigenvalues\/} of $\cH_0$ and
$\cL_0$, respectively, on the \hw{} vector of the twisted topological
Verma module~\,$\smV_{h,t;\theta}$: according to~\req{Cartantheta}, we
have $h_0=h-\frac{\ctop}{3}\theta$ and $\ell_0=-\theta h +
\frac{\ctop}{6}(\theta^2-\theta)$.  An important consequence of these
observations is
\begin{lemma}\label{lemma:same-top}
  All singular vectors in the twisted topological Verma module
  $\smV_{h,t;\theta}$ have the same value of $h_0^\pm+\theta^\pm$,
  which equals $h_0+\theta$ for the \hw{} state of
  $\smV_{h,t;\theta}$.
\end{lemma}

The topological singular vectors occur in the (twisted) topological
Verma module $\smV_{h,t;\theta}$ whenever there exist $r,s\in\oN$ such
that the $h$ parameter can be represented as $h=\hminus(r,s,t)$ or
$h=\hplus(r,s,t)$, where
\begin{equation}
  \htop^-(r,s,t)=\frac{r+1}{t}-s\,,\qquad
  \htop^+(r,s,t)=-\frac{r-1}{t}+s-1\,.
  \label{htop}
\end{equation}

The explicit construction for these singular vectors can be outlined
as follows~\cite{[ST2],[ST3]}.  As an analogue of the complex powers
of generators used in the $\tSL2$ case, one now performs the
continuation in terms of `dense' products of modes of the fermionic
generators $\cG$ and $\cQ$: one introduces operators $g(a,b)$ and
$q(a,b)$ that can be thought of as a continuation of
$\cG_{b-N}\,\cG_{b-N+1}\ldots\cG_{b}$ and
$\cQ_{b-N}\,\cQ_{b-N+1}\ldots\cQ_{b}$, respectively, to a complex
number of factors. In particular, whenever the {\it length\/} $b-a+1$
of \ $g(a,b)$ or $q(a,b)$ is a non-negative integer, the corresponding
operator becomes, by definition, the product of the corresponding
modes:
\begin{equation}
  g(a,b)=\prod_{i=0}^{L-1}\cG_{a+i}\,,\quad q(a,b)=
  \prod_{i=0}^{L-1}\cQ_{a+i}\,,
  \quad{\rm iff}\quad L\equiv b-a+1=0,1,2,\ldots
  \label{integrallength}
\end{equation}
It is possible to postulate a number of algebraic properties of the
new operators in such a way that these properties become identities
whenever the operators reduce to elements of the universal enveloping
algebra. This is directly analogous to the rules~\req{properties} used
to operate with complex powers, however some of the algebraic rules
that are not even formulated explicitly for the complex powers (e.g.,
$(J^+_n)^\alpha\,(J^+_n)^\beta=(J^+_n)^{\alpha+\beta}$) become less
trivial in the $\N2$ case, see~\cite{[ST3],[ST4]}.  Then the
topological singular vectors in the twisted topological Verma modules
read as
\begin{equation}
  \label{Epm}\kern-6pt\new
  \begin{array}{rcl}
    \ket{E^+(r,s,t)}^\theta\kern-6pt&=&\kern-6pt
    g(-\beta^+_s,\alpha^+_s-1)\,\ldots\,
    g(-\beta^+_2,\alpha^+_2-1)\,q(-\alpha^+_2,\beta^+_1-1)\,
    g(-\beta^+_1,\alpha^+_1-1)\,\kettop{\hplus(r,s,t),t;\theta}\,,\\
    \ket{E^-(r,s,t)}^\theta\kern-6pt&=&\kern-6pt
    q(-\beta^-_s,\alpha^-_s-1)\,\ldots\,
    q(-\beta^-_2,\alpha^-_2-1)\,g(-\alpha^-_2,\beta^-_1-1)\,
    q(-\beta^-_1,\alpha^-_1-1)\,\kettop{\hminus(r,s,t),t;\theta}
  \end{array}
\end{equation}
where
\begin{equation}
  \alpha^\pm_i=(i-1)t \pm \theta\,,\qquad
  \beta^\pm_i=r - (s-i)t \mp \theta\,.
\end{equation}
Using the algebraic rules satisfied by the $g$ and $q$ operators, one
checks that these vectors do indeed satisfy the twisted topological
\hw{} conditions with the twist parameter $\theta^\pm=\theta\mp r$
(which become~\req{twistedtophw} in the $\theta=0$ case).  Singular
vectors~\req{Epm} evaluate as the elements of the topological Verma
module~\,$\smV_{\htop^\pm(r,s,t),t;\theta}$, with no continued
operators left after the algebraic rearrangements.

\medskip

The idea regarding the correspondence between submodules in $\tSL2$ and
$\N2$ modules can be developed in the direction of category
theory~\cite{[FST]}.  Taking the objects to be all the twisted
topological $\N2$ Verma modules, the morphisms would have to be the
{\it embeddings\/}.  We have just seen that the embeddings --- i.e.,
the occurrence of {\it sub\/}modules --- are `synchronized' between the
topological Verma modules over $\N2$ and the $\tSL2$ Verma modules.
The corresponding representation categories can be compared after one
effectively takes the factor over the spectral flows on the $\N2$ as
well as the $\tSL2$ sides, see~\cite{[FST]} for a rigorous statement.
One eventually concludes that the category of {\it chains\/} of twisted
topological $\N2$ Verma modules is equivalent to the category of chains
of twisted $\tSL2$ Verma modules. An immediate consequence of this
equivalence is that {\it embedding diagrams of topological $\N2$ Verma
modules are isomorphic to the embedding diagrams of $\tSL2$ Verma
modules}.  The only additional information that may be interesting in
the $\N2$ case is that regarding the twists of the modules making up an
embedding diagram; however, we saw in~\req{twistedtophw} that the
relative twist of a submodule generated from the topological singular
vector $\ket{E(r,s,t)}^\pm$ is~$\mp r$, therefore the twists are easy
to reconstruct \hbox{from the standard $\tSL2$ embedding diagrams.}

Now, what is more important for our present purposes of describing
embedding diagrams of the {\it massive\/} $\N2$ Verma modules, is that
Theorem~\ref{thm:topequiv} can be extended to a similar statement
involving massive $\N2$ Verma modules and (twisted) \RV{} $\tSL2$
modules.
\begin{thm}[\cite{[FST]}]\mbox{}\label{thm:massequiv}\nopagebreak
  \begin{enumerate}
    \addtolength{\parskip}{-6pt}
  \item There is an isomorphism of $\tSL2$ representations
    \begin{equation}
      \mU_{h,\ell,t}\tensor\Xi\approx\bigoplus_{\theta\in\oZ}\,
      \smR_{\mu_1,\mu_2,t;\theta}\tensor
      \mF^-_{\sqrt{\frac{t}{2}}(h+\theta)}\,,\qquad
      \left\{\new
        \begin{array}{rcl}
          \mu_1\cdot\mu_2&=&-t\ell\,,\\
          \mu_1 + \mu_2 &=&ht - 1\,.
        \end{array}
      \right.
      \label{relaxedidspaces2}
    \end{equation}
    where on the left-hand side the $\tSL2$ algebra acts by
    generators~\req{invKS}, while on the right-hand side it acts
    naturally on twisted relaxed Verma module
    $\smR_{\mu_1,\mu_2,t;\theta}$.

  \item A singular vector exists in the massive Verma module
    $\mU_{h,\ell,t}$ if and only if a singular vector exists in one
    (hence, in all) of the \RV{} modules
    $\smR_{\mu_1,\mu_2,t;\theta}$, $\theta\in\oZ$. Whenever this is
    the case, moreover, the respective submodules associated with the
    singular vectors, in their own turn, satisfy an equation of the
    same type as~\req{relaxedidspaces2} if these are massive/relaxed
    submodules, and (the twist of) Eq.~\req{idspaces2} if these are
    twisted topological/usual-Verma submodules.
  \end{enumerate}
\end{thm}
As a consequence, {\it the embedding diagram of any massive $\N2$
  Verma module $\mU_{h,\ell,t}$ is isomorphic to the embedding diagram
  of the \RV{} module $\mR_{\mu_1,\mu_2,t}$}, where the parameters are
related as in~\req{relaxedidspaces2}.

The appearance of the {\sf massive/relaxed} and
topological/usual-Verma submodules can be illustrated in the following
extremal diagrams:
$$
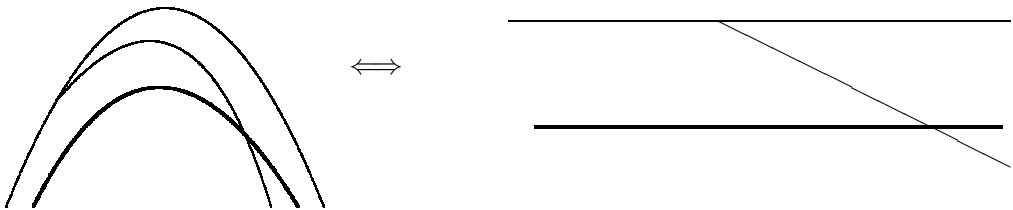
\unitlength=1pt
\begin{picture}(440,80)
  \put(20,0){
    \bezier{800}(0,0)(60,150)(120,0)
    {\linethickness{.9pt}
      \bezier{400}(10,0)(55,90)(110,0)
      }
    \bezier{500}(19,40)(70,100)(100,0)
    \put(130,50){$\Longleftrightarrow$}
    \put(190,70){\line(1,0){190}}
    \put(270,70){\line(2,-1){110}}
    {\linethickness{1.1pt}
      \put(200,30){\line(1,0){177}}
      }
    }
\end{picture}
$$
Accordingly, the $\N2$ counterparts of singular vectors in the relaxed
$\tSL2$ Verma modules are as follows.

First, charged singular vectors~\req{chargedsl2} translate into the
$\N2$ singular vectors that exist in $\mU_{h,\ell,t}$ if and only if \
$\ell=\ellch(n,h,t)$, where
\begin{equation}
  \ellch(n,h,t)=-n(h-\frac{n+1}{t})\,,\quad n\in\oZ\,,          
  \label{Lambdach}
\end{equation}
which reproduces a series of zeros of the Ka\v c
determinant~\cite{[BFK]}.  Just as the charged $\tSL2$ singular
vectors, {\it the charged $\N2$ singular vectors\/} are given by the
simple construction~\cite{[ST3]}
\begin{equation}
  \ket{E(n,h,t)}_{\rm ch}=\left\{\kern-4pt\new
    \begin{array}{ll}
      \cQ_{-n}\,\ldots\,\cQ_0\,\ket{h,\ellch(n,h,t),t}&n\geq0\,,\\ 
      \cG_{n}\,\ldots\,\cG_{-1}\,\ket{h,\ellch(n,h,t),t}\,,&      
      n\leq-1\,.                                                  
    \end{array}\right.
  \label{ECh}
\end{equation}
It is elementary to verify in intrinsic $\N2$ terms that every such
vector satisfies twisted topological \hw{} conditions
\req{twistedtophw} with
$\theta=n$.  Thus, in accordance with the      
Theorem, the submodule generated from this vector is a twisted
topological Verma module.

Describing the positions of singular vectors in the (charge, level)
lattice, let $h'_0$ be the {\it eigenvalue\/} of $\cH_0$ on the
charged singular vector satisfying the twisted topological \hw{}
conditions with the twist parameter $\theta'$. It is easy to see that
for the charged singular vectors
labelled by $n\leq-1$, the value of   
$h'_0+\theta'$ (where $\theta'=n$) is equal to the
eigenvalue $h_0$    
of $\cH_0$ on the (untwisted) massive \hw{} vector of the massive
Verma module. On the other hand, for the charged singular vectors
labelled
by $n\geq0$, the value of $h'_0+\theta'$ equals~$h_0-1$. In  
combination with Lemma~\ref{lemma:same-top} and
Theorem~\ref{thm:3types} (translated into the $\N2$ language in
accordance with Theorems~\ref{thm:topequiv} and~\ref{thm:massequiv}),
this shows that all of the twisted topological Verma modules that can
appear in a massive Verma module are divided into those for which
\begin{equation}
  \label{filled}
  \mbox{\parbox{.9\textwidth}{the \hw{} vector $\ket{e'}$ of the
      submodule satisfies twisted topological \hw{} conditions with the
      twist parameter $\theta'$ and $\cH_0\ket{e'}=(h_0-\theta')
      \ket{e'}$, with $h_0$ being the
      eigenvalue of $\cH_0$ on the \hw{} vector of the massive Verma
      module,}}
\end{equation}
and those for which
\begin{equation}
  \label{open}
  \mbox{\parbox{.9\textwidth}{the \hw{} vector $\ket{e'}$ of the
      submodule satisfies twisted topological \hw{} conditions with the
      twist parameter $\theta'$ and $\cH_0\ket{e'}=(h_0-\theta'-1)
      \ket{e'}$, with $h_0$ being the eigenvalue
      of $\cH_0$ on the \hw{} vector of the massive Verma module.}}
\end{equation}

\medskip

As regards those singular vectors that generate \RV{} submodules in
\RV{} $\tSL2$ modules, they translate into the {\it massive\/}
singular vectors in $\N2$ Verma modules. The massive singular vectors
are in the same relation to topological singular vectors in certain
{\it auxiliary topological Verma\/} modules as the relaxed-$\tSL2$
singular vectors to the auxiliary Verma-module singular vectors.  The
method applied in Sec.~\ref{subsec:r-singular},
Eqs~\req{auxVerma}--\req{sigmaplus}, now works for the $\N2$ algebra
as follows~\cite{[ST3],[ST4]}.  One constructs singular vectors in the
`auxiliary' {\it topological\/} Verma modules, the mappings between
the given massive $\N2$ Verma module and the auxiliary topological
ones being performed with the help of continued operators $g(a,b)$ and
$q(a,b)$.  When applied to a massive \hw{} state $\ket{h,\ell,t}$, the
continued operators $g$ and $q$ map it into a state that, generically,
is a twisted massive \hw{} state. However, whenever
\begin{equation}
  \ell=-\theta h+\frac{1}{t}(\theta^2+\theta)\,,
  \label{ell}
\end{equation}
then
\begin{eqnarray}
  g(\theta,-1)\,\ket{h,\ell,t}&\sim&
  \kettop{h-\frac{2}{t}\theta,t;\theta}
  \label{hnew}\,,\\
\noalign{\noindent\mbox{and}}
  q(-\theta,0)\ket{h,\ell,t}&\sim&
  \kettop{h - \frac{2}{t}\theta-1,t;\theta}\,.
  \label{htopnew}
\end{eqnarray}

Further, once the twisted topological \hw{} conditions are thus
insured, one demands that the corresponding twisted topological \hw{}
state admit a topological singular vector $\ket{E^\pm(r,s,t)}$. Thus,
requiring that $h-\frac{2}{t}\theta=\hminus(r, s, t)$, we see
that $\ell=\theell(r, s, h, t)$.  where
\begin{equation}
    \theell(r, s, h, t) =
    -\frac{t}{4}(h-\hminus(r,s,t))(h-\hplus(r,s+1,t))\,.
  \label{theell}
\end{equation}
The same expression for $\ell$ is arrived at in the case where a
$\ket{E(r,s,t)}^+$ singular vector exists in the auxiliary topological
Verma module.  Finally, one maps the singular vectors in the auxiliary
Verma module back to the original massive Verma module. In this way,
massive singular vectors are of the following structure:
\begin{eqnarray}
  &&
  g(-rs,r+\theta'-1)\cdot
  \cE(r,s,t)^{\pm,\theta'}\cdot g(\theta',-1)\,
  \ket{h,\ell,t}\,,\\
  \noalign{\noindent{\mbox{or}}}
  &&
  q(1-rs,r-\theta''-1)\,
  \cE(r,s,t)^{\pm,\theta''}\,q(-\theta'',0)\,
  \ket{h,\ell,t}\,,
\end{eqnarray}
where $\cE(r,s,t)^{\pm,\theta}$ is the spectral flow transform of the
topological singular vector operator $\cE(r,s,t)^\pm$ whenever $\ell$
equals $\theell(r, s, h, t)$ from~\req{theell}. Here, $\theta'$ and
$\theta''$ are the roots of~\req{ell}, where we substitute
$\ell=\theell(r, s, h, t)$.

We, thus, conclude that a massive singular vector occurs in the module
$\mU_{h,\ell,t}$ whenever $\ell=\theell(r, s, h, t)$. Together with
the condition $\ell=\theell_{\rm ch}(n, h, t)$, this reproduces the
zeros of the Ka\v c determinant~\cite{[BFK]}. The representatives of
the massive singular vector read as
\begin{eqnarray}
  \mbox{}\kern-25pt
  \ket{S(r,s,h,t)}^-\kern-8pt&=&\kern-7pt
  g(-rs,r+\theta^-(r,s,h,t)-1)\,
  \cE^{-,\theta^-(r,s,h,t)}(r,s,t)\,g(\theta^-(r,s,h,t),-1)\,
  \ket{h,\theell(r,s,h,t),t}\,,
  \label{Sminus}\\
  \mbox{}\kern-25pt\ket{S(r,s,h,t)}^+\kern-8pt&=&\kern-7pt
  q(1-rs,r-\theta^+(r,s,h,t)-1)\,
  \cE^{+,\theta^+(r,s,h,t)}(r,s,t)\,q(-\theta^+(r,s,h,t),0)\,
  \ket{h,\theell(r,s,h,t),t},
  \label{Splus}
\end{eqnarray}
It is now straightforward to check that
\begin{equation}\new
  \begin{array}{l}
    \cQ_{\geq1\mp rs}\,\ket{S(r,s,h,t)}^\pm=
    \cH_{\geq1}\,\ket{S(r,s,h,t)}^\pm=\cL_{\geq1}\,\ket{S(r,s,h,t)}^\pm=
    \cG_{\geq\pm rs}\,\ket{S(r,s,h,t)}^\pm=0\,,\\
    \cL_0\,\ket{S(r,s,h,t)}^\pm=
    \theell^\pm(r,s,h,t)\,\ket{S(r,s,h,t)}^\pm\,,\\
    \cH_{0}\,\ket{S(r,s,h,t)}^\pm=(h \mp r s)\,\ket{S(r,s,h,t)}^\pm
  \end{array}
  \label{hwsingN2}
\end{equation}
with
\begin{equation}
  \theell^\pm(r,s,h,t)=
  \theell(r,s,h,t) + \half r s (r s + 2 \mp 1)\,.
  \label{theellpm}
\end{equation}
In the general position (for generic $(r,s,h,t)$, see~\cite{[ST4]}),
the vectors $\ket{S(r,s,h,t)}^-$ and $\ket{S(r,s,h,t)}^+$ generate the
same submodule. It may be useful to note that the top of the extremal
diagram containing $\ket{S(r,s,h,t)}^-$ and $\ket{S(r,s,h,t)}^+$ is
the state $\ket{S(r,s,h,t)}^0$ that satisfies massive \hw{}
conditions~\req{masshw} and such that
$\cH_0\ket{S(r,s,h,t)}^0=h\,\ket{S(r,s,h,t)}^0$ and
$\cL_0\ket{S(r,s,h,t)}^0=(\theell(r,s,h,t)+rs)\cdot\ket{S(r,s,h,t)}^0$.

\subsection{Embedding diagrams of the massive $\N2$ Verma
  modules\label{subsec:comments}}\lvm For $\ctop\neq3$, the problem of
classifying and constructing the $\N2$ embedding diagrams is solved by
virtue of the above results on the relaxed-$\tSL2$ embedding diagrams.
One can parametrize massive $\N2$ Verma modules by $\mu_1$, $\mu_2$,
and $t$ from \req{relaxedidspaces2}, which is always possible for
$\ctop\neq3$.  Then, depending on the values of $(\mu_1,\mu_2,t)$, we
have for each massive Verma module exactly the same embedding diagram
as for the respective \RV{} module $\mR_{\mu_1,\mu_2,t}$.

In the embedding diagrams of Sec~\ref{subsec:diagrams}, {\Relaxed} now
denote the massive $\N2$ Verma modules, while {\Verma} are the twisted
topological Verma modules satisfying condition~\req{filled} and
{\TVerma} are the twisted topological Verma modules satisfying
condition~\req{open}.  The distances in the units of $J^0_0$-charge
map into distances in the units of the $\cH_0$-charge. However, as we
have seen, the level of states along $\N2$ extremal diagrams is never
constant --- which, as a matter of fact, is the main reason making the
$\N2$ description more complicated than the relaxed-$\tSL2$
description of the same structures. Therefore, a horizontal arrow
connecting two modules does {\it not\/} mean in the $\N2$ case that
the \hw{} vectors of the modules are on the same level. However, the
horizontal arrows do have a meaning in the intrinsic $\N2$ terms: a
horizontal arrow leads from an $\N2$ module \,$\mW$ (either twisted
topological or massive) to a {\Verma}-module
\,$\mV^{\phantom{y}}_\bullet$ (see~\req{filled}) if the embedding of
the \hw{} vector $\ket{X}\in\mV^{\phantom{y}}_\bullet$ reads as
\begin{equation}
  \ket{X}=\cG_{\theta-N}\,\ldots\,\cG_{\theta-1}\,\ket{Y}\,,\quad
  N\in\oN\,,
  \label{eq:G}
\end{equation}
where $\ket{Y}$ is the \hw{} state of \,$\mW$ that satisfies the
$\theta$-twisted massive \hw{} conditions. Similarly, a horizontal
arrow leads from an $\N2$ module \,$\mW$ to a {\TVerma}-module
\,$\mV^{\phantom{y}}_\circ$ (see~\req{open}) if the embedding of the
\hw{} vector $\ket{X}\in\mV^{\phantom{y}}_\circ$ reads as
\begin{equation}
  \ket{X}=\cQ_{-\theta-M}\,\ldots\,\cQ_{-\theta}\,\ket{Y}\,,\quad
  M\in\oN_0\,,
  \label{eq:Q}
\end{equation}
where $\ket{Y}$ is the \hw{} state of \,$\mW$ that satisfies the
$\theta$-twisted massive \hw{} conditions, or
\begin{equation}
  \ket{X}=\cQ_{-\theta-M}\,\ldots\,\cQ_{-\theta-1}\,\ket{Y}\,,\quad
  M\in\oN_0\,,
  \label{eq:Q0}
\end{equation}
if $\ket{Y}$ satisfies the $\theta$-twisted topological \hw{}
conditions.

\medskip

We now reproduce the list from Sec.~\ref{subsec:list} in the form
directly applicable to the classification of the $\N2$ embedding
diagrams in terms of the parameters $h$ and $\ell$ of the massive
Verma module $\mU_{h,\ell,t}$.
Equations~\req{theell} and~\req{Lambdach} exhaust the zeros of the
Ka\v c determinant~\cite{[BFK]}. The condition that
$\ell=\theell^\pm(r,s,h,t)$ for some $r,s\in\oN$ is, obviously, the
analogue of the $\tSL2$ condition that $\mu_1-\mu_2\in\oK(t)$.
Accordingly, whenever we write $\ell\neq\theell(r,s,h,t)$, we mean
that there do not exist $r,s\in\oN$ such that this would become an
equality for a given triple $(h, \ell, t)$, and similarly with the
condition $\ell\neq\ellch(n,h,t)$,~$n\in\oZ$.
The comments to diagrams~\req{d:I}--\req{last-bottom} on
pp.~\pageref{d:I}--\pageref{last-bottom} apply in the $\N2$ case with
the following replacements:

\smallskip

\begin{tabular}{rcl}
  Verma module &$\longrightarrow$&
  twisted topological Verma module satisfying Eq.~\req{filled}\\
  twisted Verma module with $\theta=1$ &$\longrightarrow$&
  twisted topological Verma module satisfying Eq.~\req{open}\\
  charged singular vector~\req{chargedsl2} &$\longrightarrow$&
  charged singular vector~\req{ECh}\\
  \RV{} (sub)module &$\longrightarrow$& massive Verma (sub)module\\
  embedding onto the same level &$\longrightarrow$& embedding via
  \req{eq:G} or \req{eq:Q}, \req{eq:Q0}\\
  $\ket{{\rm MFF}^\pm(r,s,t)}$ &$\longrightarrow$&
  $\ket{E(r,s,t)}^\pm$\\
  relative $J^0_0$-charge &$\longrightarrow$& minus the relative
  $\cH_0$-charge
\end{tabular}

\noindent
Note also that the positive (negative) zone can now be described as
$\ctop<3$ (respectively, $\ctop>3$).

\medskip

We, thus, have the following patterns of embedding diagrams, where we
refer to the diagrams from Sec.~\ref{subsec:diagrams} that are to be
read using the $\N2$ conventions discussed above.
\begin{enumerate}

  \def\theenumi{\Roman{enumi}} \renewcommand\labelenumi{\theenumi:}
  \renewcommand\labelenumii{\theenumi\theenumii:}
  \renewcommand\labelenumiii{\theenumi\theenumii,\theenumiii):}
\item{}\label{l2:I} $\ell\neq\theell(r, s, h, t)$ with $r\in\oN$,
  $s\in\oN$. The embedding diagrams shown in~\req{d:I} correspond to
  the following subcases:
  \vspace{-4pt}

  \setlength{\leftmarginii}{27pt}
  \begin{enumerate}

    \addtocounter{enumii}{-1} \def\theenumii{(\arabic{enumii})}
  \item \label{l2:I(0)}
    $\ell\neq\ellch(n,h,t)$ with $n\in\oZ$, i.e., 
    $\frac{1}{2}\left(ht-1 \pm \sqrt{4\ell t +
        (ht-1)^2}\right)\notin\oZ$. This is the trivial case.

  \item \label{l2:I(1)}$\ell=\ellch(n,h,t)$ for some $n\in\oZ$,   
    $ht\notin\oZ$.  We then have

    \def\theenumii{(\arabic{enumii}}

    \begin{enumerate}
      \addtolength{\itemindent}{18pt} \def\theenumiii{${-}$}
    \item a twisted topological Verma module satisfying
      Eq.~\req{filled} if $n\in-\oN$, or

      \def\theenumiii{${+}$}
    \item a twisted topological Verma module satisfying
      Eq.~\req{open} if $n\in\oN_0$,
    \end{enumerate}
     embedded via a charged singular vector, the corresponding
     diagrams being given Eq.~\req{d:I}.

     \def\theenumii{(\arabic{enumii})}
   \item \label{l2:I(2)} $h=\frac{1+n+m}{t}$ and $\ell=-\frac{nm}{t}$
     for some $m,n\in\oZ$.

    \def\theenumii{(\arabic{enumii}} 

        \setlength{\leftmarginiii}{52pt}
    \begin{enumerate}

      \def\theenumiii{${-}{-}$}
    \item $n,m\in-\oN$, the first diagram in~\req{d:I(2)}.

      \def\theenumiii{${+}{+}$}
    \item $n,m\in\oN_0$, the second diagram in~\req{d:I(2)}.

      \def\theenumiii{${-}{+}$}
    \item \label{l2:II(2,mp)} $n\in-\oN$, $m\in\oN_0$, the third
      diagram in~\req{d:I(2)}.

    \end{enumerate}

  \end{enumerate}
  \vspace{-6pt}

\item \label{l2:II} $\ell=\theell(r, s, h, t)$ for some $r,s\in\oN$
  and $t\notin\oQ$. All of the following items in case~II refer to the
  chosen $r$ and $s$.
  \vspace{-4pt}

    \setlength{\leftmarginii}{32pt}
  \begin{enumerate}

    \addtocounter{enumii}{-1}
    \def\theenumii{(\arabic{enumii})}
  \item \label{l2:II(0)} $ht - 1 + st-r\notin 2\oZ$ and
    $ht - 1 - (st-r)\notin 2\oZ$.  We have the first of the
    diagrams~\req{d:II}.

  \item \label{l2:II(1)} $h=\frac{1+2n + r-st}{t}$ or $h=\frac{1+2n -
      (r-st)}{t}$ for some $n\in\oZ$ (then $\ell=-\frac{n(n \pm
      (r-st))}{t}$, respectively).  We have the respective
    diagrams~\req{d:II}, depending on the sign of~$n$:

    \def\theenumii{(\arabic{enumii}} 

    \setlength{\leftmarginiii}{43pt}
    \begin{enumerate}

      \def\theenumiii{${-}$}
    \item $n\in-\oN$,

      \def\theenumiii{${+}$}
    \item $n\in\oN_0$.
    \end{enumerate}
    As before, these cases differ by which of the
    relations~\req{filled} or~\req{open} is satisfied for the twisted
    topological Verma submodule.

  \end{enumerate}
  \vspace{-6pt}

\item{} $\ell=\theell(r, s, h ,t)$ for some $r,s\in\oN$,
  $t=\frac{p}{q}\in\oQ$. In the following items in case~III, we use
  the chosen $r$ and~$s$.
  \vspace{-4pt}

      \setlength{\leftmarginii}{28pt}
  \begin{enumerate}

    \def\theenumii{$_\pm$}
  \item{} $\frac{r}{t}\notin\oZ$, $st\notin\oZ$. This means that
    $r\neq\alpha p$, $s\neq\beta q$ for $\alpha,\beta\in\oZ$.

    \setlength{\leftmarginiii}{42pt}
    \begin{enumerate}
      \addtocounter{enumiii}{-1}
      \def\theenumiii{(\arabic{enumiii})}
      \renewcommand\labelenumiii{\theenumi\theenumii\theenumiii:}

    \item{} \label{l2:IIIpm(0)} $ht - 1 + st-r\notin2\oZ$ and
      $ht - 1 - (st-r)\notin2\oZ$.  We
      have the double-chains of massive Verma modules shown in
      diagram~\req{d:IIIpm(0)}. These look identical to the familiar
      embedding diagrams of the topological Verma modules. Indeed, the
      structure of the massive Verma module repeats in this case the
      structure of the ``auxiliary'' topological Verma
      module~\cite{[ST4]}.
    \end{enumerate}

    Further degenerations are given by (various combinations of) the
    following effects. First, the auxiliary topological Verma module
    may become an actual submodule in the massive Verma module, in
    which case the entire {\it topological Verma\/}-module embedding
      diagram joins the diagram~\req{d:IIIpm(0)} via embeddings
    performed by charged singular vectors.  Second, the embedding
    diagram growing from the auxiliary topological Verma module may
    acquire a special form, which would also affect the `massive'
    embedding diagram.

    \begin{enumerate}
      \def\theenumiii{(\arabic{enumiii})}
      \renewcommand\labelenumiii{\theenumi\theenumii\theenumiii:}

    \item{} \label{l2:IIIpm(1)} either $h=\frac{1+2n+(r-st)}{t}$ or
      $h=\frac{1+2n-(r-st)}{t}$, $n\in\oZ$. \ Here, the auxiliary
      topological Verma module becomes a submodule in the massive Verma
      module; we then have the usual subcases depending on the sign
      of~$n$:

      \setlength{\leftmarginiv}{54pt}
      \begin{enumerate}

        \def\theenumiii{(\arabic{enumiii}}
        \renewcommand\labelenumiv{\theenumi\theenumii\theenumiii,\theenumiv):}

        \def\theenumiv{${-}$}
      \item \label{l2:IIIpm(1,m)} $n\in-\oN$, diagram~\req{d:IIIpm(1)}.
      \pagebreak[3]

        \def\theenumiv{${+}$}
      \item \label{l2:IIIpm(1,p)} $n\in\oN_0$, diagram~\req{d:RTV}.
      \end{enumerate}
      These diagrams are finite or infinite depending on whether $t<0$
      (III$_-$) or $t>0$~(III{}$_+$), respectively. In~\req{d:IIIpm(1)}
      and~\req{d:RTV}, the top topological Verma module is the
      submodule in the massive Verma module associated with a charged
      singular vector. Each of the subsequent topological Verma
      modules is embedded via a charged singular vector into the
      corresponding massive Verma module.

    \end{enumerate}

    \def\theenumii{$_\pm^0$}
  \item{} \label{l2:IIIpm0} Either ($\frac{r}{t}\in\oZ$,
    $st\notin\oZ$) or ($\frac{r}{t}\notin\oZ$, $st\in\oZ$). Thus, we
    have either, {\it but not both}, of the conditions \ $r=\alpha p$
    \ and \ $s=\beta q$, \ $\alpha,\beta\in\oZ$. \ The interplay of
    these cases is rather interesting.  Recall that, for the Virasoro
    algebra, the condition that singles out the III$^0$ case is that
    (in the current notations) the line $y=t\,x - \mu_1 + \mu_2$
    intersect one of the $x$, $y$ axes at an integral
    point~\cite{[FF]}.  For the usual $\tSL2$ Verma modules, on the
    other hand, this means two {\it different\/} cases: either
    $2j+1\in\oZ$ or $(2j+1)/t\in\oZ$, where $j$ is the spin of the
    \hw{} vector; it is only the Hamiltonian reduction that erases the
    difference between the two cases. However, since the Virasoro
    algebra is a subalgebra of the $\N2$ algebra, one may expect these
    to become again a single $\N2$/relaxed-$\tSL2$ pattern,
    which is indeed the case.

    \begin{enumerate}{}
      \addtocounter{enumiii}{-1}
      \def\theenumiii{(\arabic{enumiii})}
      \renewcommand\labelenumiii{\theenumi\theenumii\theenumiii:}

    \item{} \label{l2:IIIpm0(0)} $ht - 1 + st-r\notin2\oZ$ and $ht - 1
      - (st-r)\notin2\oZ$.  We then have a single-chain of massive
      Verma modules, diagram~\req{d:IIIpm0(0)}.  The single chain
      takes the place of the double chain because each submodule in
      $\mU$ corresponds to a {\it pair\/} of submodules in the
      auxiliary topological Verma module~$\aV$~\cite{[ST4]}.

    \item{} \label{l2:IIIpm0(1)} either $h=\frac{1+2n+(r-st)}{t}$ or
      $h=\frac{1+2n-(r-st)}{t}$, with $n\in\oZ$ and $st\notin\oZ$
      (hence we should have $\frac{r}{t}\in\oZ$; then, respectively,
      $h=\frac{q}{p}(1+2n) \pm (\alpha q - s)$ and $\ell=-\frac{n^2
        q}{p} \mp n (\alpha q - s)$).

      \def\theenumiii{(\arabic{enumiii}}
      \renewcommand\labelenumiv{\theenumi\theenumii\theenumiii,\theenumiv):}
      \setlength{\leftmarginiv}{54pt}
      \begin{enumerate}

        \def\theenumiv{${-}$}
      \item $n\in-\oN$, diagram~\req{d:IIIpm0(1)};

        \def\theenumiv{${+}$}
      \item $n\in\oN_0$, the vertical mirror of~\req{d:IIIpm0(1)}
        (where, as we have discussed, the modules satisfying
        Eq.~\req{filled} are replaced by those satisfying~\req{open}).
      \end{enumerate}

      \def\theenumiii{(\arabic{enumiii})}
    \item{} \label{l2:IIIpm0(2)} $h=\frac{q}{p}(1 + 2n \pm (r - \beta
      p))$ with $\frac{r}{p}\notin\oZ$. Then $\ell=-\frac{q}{p}n(n \pm
      (r - \beta p))$. This can also be expressed as
      $h=\frac{1+n+m}{t}$, $\ell=-\frac{nm}{t}$ for some $m,n\in\oZ$
      such that $\frac{m-n}{t}\notin\oZ$. Accordingly, there are two
      charged singular vectors in the massive Verma module on one side
      of the \hw{} vector.  The following subcases depend on the sign
      of these two integers:

      \setlength{\leftmarginiv}{62pt}
      \begin{enumerate}
        \def\theenumiii{(\arabic{enumiii}}

        \def\theenumiv{${-}{-}$}
      \item \label{l2:IIIpm0(2,mm)} $n\in-\oN$, $m\in-\oN$,
        diagram~\req{d:IIIpm0(2)}.

        \def\theenumiv{${+}{+}$}
      \item \label{l2:IIIpm0(2,pp)} $n\in\oN_0$, $m\in\oN_0$, a mirror
        diagram of~\req{d:IIIpm0(2)}.

        \def\theenumiv{${-}{+}$}
      \item{} \label{l2:IIIpm0(2,mp)} $n\in-\oN$, $m\in\oN_0$,
        diagram~\req{d:IIIpm0(2,mp)}.

      \end{enumerate}
      The diagrams are finite or infinite depending on whether $t<0$
      or $t>0$ respectively. In the \IIIpmziimp{} case, there
      are charged singular vectors on different sides of the \hw{}
      vector of the massive Verma module. One of the charged singular
      vectors comes with the embedding diagram of topological Verma
      modules, while the other contributes a similar (in fact,
      mirror-symmetric) diagram of twisted topological Verma modules

    \end{enumerate}

    \def\theenumii{$_\pm^{00}$}
  \item $\frac{r}{t}\in\oZ$, $st\in\oZ$. In the following, we use
    $\alpha,\beta\in\oZ$ such that $r=\alpha p$ and $s=\beta q$ (thus,
    $\ell= \half h - \frac{q}{4 p} - \frac{h^2 p}{4 q} +
    \frac{1}{4}(\alpha - \beta)^2 q p$).

    \setlength{\leftmarginiv}{24pt}
    \begin{enumerate}
      \addtocounter{enumiii}{-1} \def\theenumiii{(\arabic{enumiii})}
      \def\labelenumiii{\theenumi\theenumii\theenumiii:}

    \item \label{l2:IIIpm00(0)} $ht - 1 + st-r\notin2\oZ$ and $ht - 1
      - (st-r)\notin2\oZ$. The embedding diagram looks identical
      to~\req{d:IIIpm0(0)}, however it is half that long in the
      negative zone.

      \addtocounter{enumiii}{1}
      \def\labelenumiv{\theenumi\theenumii\theenumiii,\theenumiv):}
    \item \label{l2:IIIpm00(2)} $h=\frac{q}{p}(1+2n) \pm
      q(\alpha-\beta)$, then $\ell=\mp(\alpha-\beta)nq -
      \frac{n^2q}{p}$. This can also be written as $h=\frac{1+n+m}{t}$
      and $\ell=-\frac{nm}{t}$ for some $n,m\in\oZ$ such that
      $\frac{m-n}{p}\in\oZ$.  The following subcases depend on the
      signs of these two integers:

      \setlength{\leftmarginiv}{64pt}
      \begin{enumerate}
        \def\theenumiii{(\arabic{enumiii}}
        \def\theenumiv{${-}{-}$}

      \item \label{l2:IIIpm00(2,mm)} $n\in-\oN$, $m\in-\oN$,
        diagram~\req{d:IIIpm00(2,mm)}, which can be viewed as a
        degeneration of diagrams~\req{d:IIIpm0(1)}.  It is finite or
        infinite depending on whether $t$ is negative or positive
        respectively.  {\it In the negative zone}, we can further
        distinguish the following two cases depending on how the
        modules arrange near the bottom of the embedding diagram:

        \begin{itemize}
          \addtolength{\itemindent}{4pt}

        \item[i)] $\beta - \alpha$ is odd ($\frac{m-n}{p}$ is odd),
          the first diagram in~\req{d:terminate1}.

        \item[ii)] $\beta - \alpha$ is even ($\frac{m-n}{p}$ is even),
          the second diagram in~\req{d:terminate1}.
        \end{itemize}

        \def\theenumiv{${+}{+}$}
      \item \label{l2:IIIpm00(2,pp)} $n\in\oN_0$, $m\in\oN_0$, the
        diagram is the mirror of~\req{d:IIIpm00(2,mm)}. {\it In the
          negative zone}, in complete similarity with~\IIIpmzziimm, we
        can distinguish two cases,
        \begin{itemize}
          \addtolength{\itemindent}{4pt}

        \item[i)] $\beta - \alpha$ odd ($\frac{m-n}{p}$ odd),

        \item[ii)] $\beta - \alpha$ even ($\frac{m-n}{p}$ even),
        \end{itemize}
        which, again, are the mirror of~\req{d:terminate1}.

        \setlength{\leftmarginiv}{62pt}

        \def\theenumiii{(\arabic{enumiii}}
        \def\labelenumiv{\theenumi\theenumii\theenumiii,\theenumiv):}
        \def\theenumiv{${-}{+}$}

      \item \label{l2:IIIpm00(2,mp)} $n\in-\oN$, $m\in\oN_0$,
        diagram~\req{d:IIIpm00(2,mp)}. {\it In the negative zone}, we
        have to distinguish two possibilities of its structure near
        the bottom:
        \begin{itemize}
          \addtolength{\itemindent}{4pt}

        \item[i)] $\beta - \alpha$ is odd ($\frac{m-n}{p}$ is odd),

        \item[ii)] $\beta - \alpha$ is even ($\frac{m-n}{p}$ is even),
          shown in \req{last-bottom}.

          Setting $t=-\frac{\tp}{q}$, we have an exceptional case
          whenever $|\alpha-\beta|=2$, therefore
          \begin{equation}
            h=-\frac{2m-2\tp+1}{\tp}\,q\,,\qquad
            \ell=\frac{m(m-2\tp)}{\tp}\,q
          \end{equation}
          This is described by diagram~\req{exceptional} {\it with no
            massive Verma submodules}.

        \end{itemize}

      \end{enumerate}

    \end{enumerate}

  \end{enumerate}

\end{enumerate}
Note that the Virasoro embedding diagrams are obtained simply  by
contracting all of the horizontal arrows in the diagrams (in
particular, by dropping all of the {\Verma} and {\TVerma} dots).

\subsection{$\ctop=3$ singular vectors and embedding diagrams
  \label{subsec:c3}}\lvm
The case of $\ctop=3$ (i.e., $t\to\infty$) has to be considered
separately, since the correspondence with $\tSL2$ modules then breaks
down and the $\N2$ Verma modules have to be analysed directly.  We,
thus, do not prove that the set of singular vectors we consider is
complete; we simply apply the approach outlined at the end of
Sec.~\ref{subsec:anti-KS} in the $t\to\infty$ limit.

First, let us consider topological Verma modules $\mV_{h,\infty}$.
Observe first of all that any submodule $\mV_{h,\infty}$ is isomorphic
to a twist of $\mV_{h,\infty}$, i.e., the value of the $h$ parameter
does not change. A singular vector exists in $\mV_{h,\infty}$ whenever
$h\in\oZ$. For $h\in\oN_0$, we denote $h=s-1$ with $s\in\oN$, then we
have the singular vectors
\begin{equation}\new
  \label{c3topplus}
  \begin{array}{rcl}
    \ket{e(r,s)}^+ &=& \lim_{t\to\infty}\bigl(\frac{1}{t^{r(s-1)}}\,
    \ket{E(r,s, t}^+\bigr)\\
    {}&\equiv& e^+(r,s)\,\kettop{s-1, \infty}
  \end{array}
\end{equation}
for all $r\in\oN$. The vector $\ket{e(r,s)}^+$ satisfies the twisted
topological \hw{} conditions with the twist parameter $\theta=-r$; its
$\cH_0$-charge and level are given by
\begin{equation}\new
  \begin{array}{rcl}
    \cH_0\,\ket{e(r,s)}^+&=&(s-1+r)\ket{e(r,s)}^+,\\
    \cL_0\,\ket{e(r,s)}^+&=&\half r(r+2s-1)\,\ket{e(r,s)}^+.
  \end{array}
\end{equation}
Moreover, the singular vector {\it operators\/} compose as follows:
\begin{equation}
  e(1, s)^{+,-1}\,e^+(1, s) = e^+(2, s)
\end{equation}
(where $e(r,s)^{+,\theta}$ is the spectral flow transform of the
singular vector operator) and, thus, all of the singular
vectors~\req{c3topplus} belong to the submodule generated from
$\ket{e(1, s)}^+$.

Similarly, for $h=-s\in-\oN$, we have the singular vectors
\begin{equation}\new
  \label{c3topminus}
  \begin{array}{rcl}
    \ket{e(r,s)}^- &=& \lim_{t\to\infty}\bigl(\frac{1}{t^{r(s-1)}}\,
    \ket{E(r,s, t}^-\bigr)\\
    {}&\equiv& e^-(r,s)\,\kettop{-s, \infty}
  \end{array}
\end{equation}
for all $r\in\oN$. These satisfy the twisted topological \hw{}
conditions with the twist parameter $\theta=r$, while
\begin{equation}\new
  \begin{array}{rcl}
    \cH_0\,\ket{e(r,s)}^-&=&-(s+r)\ket{e(r,s)}^-,\\
    \cL_0\,\ket{e(r,s)}^-&=&\half r(r+2s-1)\,\ket{e(r,s)}^-.
  \end{array}
\end{equation}
The singular vector operators compose as follows:
\begin{equation}
  e(1, s)^{-,1}\,e^-(1, s) = e^-(2, s)
\end{equation}
hence all of the singular vectors~\req{c3topminus} belong to the
submodule generated from $\ket{e(1, s)}^-$.

Thus, the embedding diagram of $\mV_{h,\infty}$ is a single-chain of
modules $(\smV_{h,\infty;-r})_{r\in\oN}$ if $h\in\oN_0$ and
$(\smV_{h,\infty;r})_{r\in\oN}$ if $h\in-\oN$.

\medskip

As regards singular vectors in massive Verma modules
\,$\mU_{h,\ell,\infty}$, we begin with the simplest case of the
charged singular vectors. These exist in \,$\mU_{h,\ell,\infty}$
whenever
$\ell=-nh$, $n\in\oZ$, and read                                   
\begin{equation}
  \ket{E(n,h,\infty)}_{\rm ch}=\left\{\kern-4pt\new
    \begin{array}{ll}
      \cQ_{-n}\,\ldots\,\cQ_0\,\ket{h,-nh,\infty}&n\geq0\,,\\     
      \cG_{n}\,\ldots\,\cG_{-1}\,\ket{h,-nh,\infty}\,,&           
      n\leq-1\,.                                                  
    \end{array}\right.
  \label{EChN2}
\end{equation}
If $h=0$, the module \,$\mU_{0,0,\infty}$ contains all singular
vectors $\ket{E(n,0,\infty)}_{\rm ch}$, $n\in\oZ$. The embedding
diagram has the following, rather curious, form:
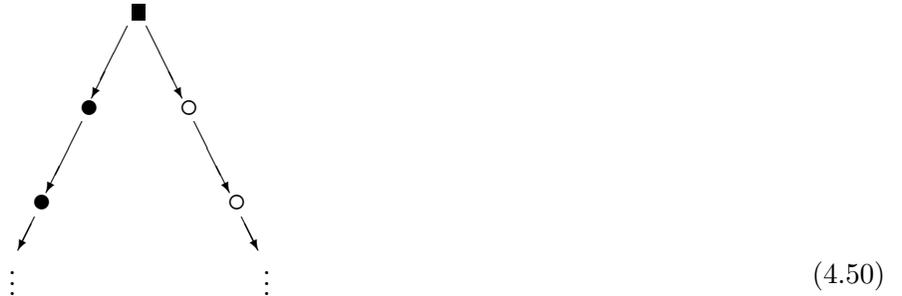
\begin{equation}
  \label{c3-00}
  \unitlength=0.9pt
  \begin{picture}(440,115)
    \put(0,30){
      \put(200,80){\Relaxed}
      \put(198,78){\vector(-1,-2){15}}
      \put(178,40){\Verma}
      \put(206,78){\vector(1,-2){15}}
      \put(220,40){\TVerma}
      \put(-30,-40){
        \put(188,40){\Verma}
        \put(209,78){\vector(-1,-2){15}}
        }
      \put(-50,-80){
        \put(209,78){\vector(-1,-2){7}}
        }
      \put(-40,-75){
        \put(188,40){$\vdots$}
        }
      \put(20,-40){
        \put(206,78){\vector(1,-2){15}}
        \put(220,40){\TVerma}
        }
      \put(40,-80){
        \put(206,78){\vector(1,-2){7}}
        }
      \put(53,-75){
        \put(202,40){$\vdots$}
        }
      }
  \end{picture}
\end{equation}

\noindent
where we do {\it not\/} follow the convention that all of the
embeddings via charged singular vectors are shown with horizontal
arrows; that convention was particularly natural on the $\tSL2$ side,
however in the present diagram---which has no $\tSL2$ analogues---we
place the submodules on different levels, in accordance with the
actual expression for the charged singular vectors. In fact, the {\it
  embedding\/} diagram looks in this case very much like the {\it
  extremal\/} diagram of the module, because a submodule is generated
from every extremal state of the module (except the \hw{} one); thus,
it may be even more natural to place the dots in~\req{c3-00} on the
two half-parabolas, as in~\req{topdiag}.

\medskip

As to the massive singular vectors, further, we see that (one of) the
states~\req{hnew} and~\req{htopnew} admits a singular vector and at
the same time (one of) the roots $\theta'$ and $\theta''$ of
Eq.~\req{ell} has a finite limit as $t\to\infty$ if and only if $h=\pm
s$.  Thus, massive singular vectors in the massive Verma module
\,$\mU_{h,\ell,\infty}$ exist whenever $h\in\oZ\setminus\{0\}$. For
$h\equiv-s\in-\oN$, the massive singular vectors in $\mU_{-s,\ell,
  \infty}$ are constructed as follows:
\begin{equation}
  \ket{\sigma(r, s, \ell)}^- =
  g(-rs, \frac{\ell}{s} + r - 1)
  e(r, s)^{-,\,\ell/s} g(\frac{\ell}{s},-1)\,
  \ket{-s, \ell, \infty}\,,
\end{equation}
where, as before, $e(r, s)^{-,\theta}$ is the spectral flow transform
of the singular vector operator $e(r, s)^-$.

Whenever the module $\mU_{-s,\ell,\infty}$ contains, in addition, a
charged singular vector, a somewhat special case occurs for
$\frac{\ell}{s}\in-\oN$.  Then the vector $\ket{\sigma(r,s,\ell)}^-$
belongs to the submodule generated from the charged singular vector
\begin{equation}
  \ket{E(\frac{\ell}{s},-s,\infty)}_{\rm ch}=
  \cG_{\frac{\ell}{s}}\,\ldots\,\cG_{-1}\,\ket{-s,\ell,\infty}
\end{equation}
since, obviously, $\ket{\sigma(r, s, sn)}^-$ with $n\in-\oN$ is a  
descendant of the state
\begin{equation}
  \ket{\bar\sigma(r, s, sn)}^-=                                   
  e(r, s)^{-,n}\,                                                 
  \cG_{n}\,\ldots\,\cG_{-1}\,\ket{-s,ns,\infty}                   
\end{equation}
that satisfies twisted topological \hw{} conditions with the twist
parameter $\theta = r+n$.  In that case, there always exists
the state                                   
\begin{equation}
  \ket{x(r, s, sn)}^- = g(r + n + 1,r + n - 1)\,               
  \ket{\bar\sigma(r, s, sn)}^-                                 
\end{equation}
which satisfies twisted massive \hw{} conditions with the twist
parameter $\theta=r + n + 1$ and generates a
{\it massive\/} Verma                                            
submodule.

\smallskip

Next, for $h\equiv s\in\oN$, singular vectors in $\mU_{s,\ell,
  \infty}$ are of the form
\begin{equation}
  \ket{\sigma(r, s, \ell)}^+ =
  q(-rs, r + \frac{\ell}{s} - 1)\,
  e(r, s)^{+,\,-\ell/s} q(\frac{\ell}{s},0)\,
  \ket{s, \ell, \infty}\,.
  \label{c3sigmaplus}
\end{equation}
These satisfy the twisted massive \hw{} conditions with the twist
parameter $rs+1$.

Again, a simultaneous occurrence of a charged singular vector requires
some attention when $\frac{\ell}{h}\in-\oN_0$. Then
vectors~\req{c3sigmaplus} are descendants of the charged singular
vector
\begin{equation}
  \cQ_{\ell/s}\ldots\,\cQ_0\,
  \ket{s, \ell, \infty}\,.
\end{equation}
Moreover, $\ket{\sigma(r, s, -sn)}^+$ with $n\in\oN_0$ is
inside the submodule built on the vector
\begin{equation}
  \ket{\bar\sigma(r, s, -sn)}^+ =
  e(r, s)^{+,\,n} \cQ_{-n}\ldots\,\cQ_0\,
  \ket{s, -sn, \infty}\,,
\end{equation}
which satisfies the twisted topological \hw{} conditions with the
twist parameter $n-r$. In that case, there exist the vector
\begin{equation}
  \ket{x(r, s, -sn)}^+ =
  q(n - r + 1, n - r - 1)
  \ket{\sigma(r, s, -sn)}^+
\end{equation}
from which the respective massive Verma submodule is generated.

\medskip

Therefore, when $h\in\oZ\setminus\{0\}$ but $\frac{\ell}{h}\notin\oZ$,
the embedding diagrams are simply a single chain of massive Verma
modules. When, in addition, $\frac{\ell}{h}\in\oZ$, the embedding
diagram takes the form (again in the convention that the embedding via
charged singular vectors are shown with horizontal arrows)
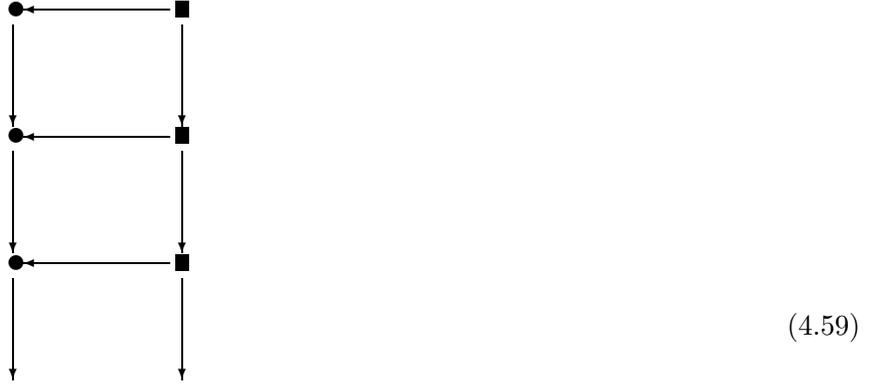
\begin{equation}
  \label{d:c3}
  \unitlength=0.8pt
  \begin{picture}(500,160)
    \put(0,-30){
      \put(260,180){\Relaxed}
      \put(263,177){\vector(0,-1){48}}
      \put(180,180){\Verma}
      \put(257,184){\vector(-1,0){69}}
      \put(183,177){\vector(0,-1){48}}
      \put(0,-60){
        \put(260,180){\Relaxed}
        \put(263,177){\vector(0,-1){48}}
        \put(180,180){\Verma}
        \put(257,184){\vector(-1,0){69}}
        \put(183,177){\vector(0,-1){48}}
        }
      \put(0,-120){
        \put(260,180){\Relaxed}
        \put(263,177){\vector(0,-1){48}}
        \put(180,180){\Verma}
        \put(257,184){\vector(-1,0){69}}
        \put(183,177){\vector(0,-1){48}}
        }
      }
  \end{picture}
\end{equation}

\bigskip

\noindent
or the vertical mirror of this with the {\Verma} modules replaced by
{\TVerma} modules.

We, thus, arrive at the following cases, which can be labelled as
I${}_\infty$ and III${}_\infty$ (no II cases exist at
$\ctop=3$):\setlength{\leftmargini}{24pt}
\begin{enumerate}
\item[I${}_\infty$:] $h\notin\oZ$  or $h=0$.
  \setlength{\leftmarginii}{37pt}
  \begin{enumerate}
  \item[I${}_\infty(0)$:] $h\neq0$, $\ell/h\notin\oZ$.  There are no
    submodules in $\mU_{h,\ell,\infty}$ whatsoever, the embedding
    diagram is the lonely massive Verma module.

  \item[I${}_\infty(1)$:] $h\neq0$, $\ell/h\in\oZ$. Then, we have the
    respective diagrams from~\req{d:I} in the cases where
    \setlength{\leftmarginiii}{52pt}
    \begin{enumerate}

    \item[I${}_\infty(1,{-})$:] $\ell/h\in\oN$.

    \item[I${}_\infty(1,{+})$:] $\ell/h\in-\oN_0$.

    \end{enumerate}

  \item[I${}_\infty(\infty)$:] $\ell=0$, $h=0$. Here, the embedding
    diagram is of the form~\req{d:c3}.
  \end{enumerate}

\item[III${}_\infty$:] $h\in\oZ\setminus\{0\}$.
  \setlength{\leftmarginii}{35pt}
  \begin{enumerate}
  \item[III${}_\infty(0)$:] $\ell/h\notin\oZ$, the embedding diagram
    being the same as in~\req{d:IIIpm0(0)}.
  \item[III${}_\infty(1)$:] $\ell/h\in\oZ$,
    \setlength{\leftmarginiii}{57pt}
    \begin{enumerate}
    \item[III${}_\infty(1,{-})$:] $\ell/h\in\oN$, the embedding
      diagram~\req{d:c3}, which is `topologically' identical to the
      diagrams in~\req{d:IIIpm0(1)}.
    \item[III${}_\infty(1,{+})$:] $\ell/h\in-\oN_0$, with the
      embedding diagram being the vertical mirror of the previous one.
    \end{enumerate}
  \end{enumerate}
\end{enumerate}

\section{Conclusions\label{sec:conclusion}}\lvm
In this paper, we have constructed the $\N2$ embedding diagrams, which
at the same time are the embedding diagrams of the relaxed $\tSL2$
Verma modules.
Let us point out once again that the \hbox{$\N2$/relaxed-$\tSL2$}
embedding diagrams are made up of {\it embeddings}, i.e., of mappings
with trivial kernels.  On the $\N2$ side, this matter appears to have
caused some confusion in the literature, because the existence of
fermions was believed to lead to the vanishing of certain would-be
embeddings. In the $\tSL2$ terms, however, this problem is obviously
absent, hence it is but an artefact on the $\N2$ side as well. The
vanishing of some compositions of ``embeddings'' observed previously
is nothing but the manifestation of two facts: (i)~states
\req{terminate} vanish for $N\gg1$ or for $M\gg1$ once $\ket{X}$ is
inside a (twisted) topological Verma module, and (ii)~every submodule
generated from a charged singular vector is necessarily a twisted
topological Verma module. This is even more transparent on the $\tSL2$
side, where the charged singular vectors generate the usual (i.e.,
{\it not\/} relaxed) Verma modules, which obviously `defermionizes'
the whole picture.

As we have already remarked in the Introduction, an immediate
consequence of the embedding diagrams constructed here would be the
derivation of the `relaxed' BGG-like resolution and, thus, of the
\hbox{$\N2$}/relaxed-$\tSL2$ characters.  It should only be noted that
constructing the BGG-resolution for the $\N2$ representations requires
slightly more work than in the `classical' (affine) cases because of
the two types of submodules existing in the massive $\N2$ Verma
modules.  Unlike the embedding diagrams, the resolutions are
constructed in terms of modules of only one type, therefore, on the way
from the $\N2$ {\it embedding\/} diagrams to the BGG {\it
resolutions\/}, one has to additionally resolve all the topological
Verma modules in terms of the massive ones.

\medskip

In view of the results of~\cite{[S-sl21],[S-sl21sing]} on the
construction of $\tSSL21$ representations out of the $\N2$ Verma
modules and on the evaluation of $\tSSL21$ singular vectors in the
$\N2$ terms, certain elements of the above embedding diagrams must
also be present in the $\tSSL21$ embedding diagrams (cf.~\cite{[BT]}),
where---as in the $\N2$ case---different types of Verma-like modules
have been observed~\cite{[S-sl21sing]}.

\paragraph{Acknowledgements.} We thank I.~Tipunin for collaboration at
an early stage of this work and many useful discussions, and also
B.~Feigin, F.~Malikov, and I.~Shchepochkina for useful remarks. The
work of AMS was supported in part by the RFBR Grant 96-02-16117 and
that of VAS, by the ISSEP grant~a97-442.


\small
\begin{thebibliography}{99999}
  \parindent=0pt \parskip=-2pt

\bibitem[A]{[Ade]}M.~Ademollo, 
  L.~Brink, A.~D'Adda, R.~D'Aura, E.~Napolitano, S.~Sciuto,
  E.~Del~Giudice, P.~Di~Vecchia, S.~Ferrara, F.~Gliozzi, R.~Musto,
  R.~Pettorino and J.~Schwarz, {\it Dual String With U(1) Color
    Symmetry,\/}
  \NPB111 (1976) 77.\\
  M.~Ademollo, 
  L.~Brink, A.~D'Adda, R.~D'Aura, E.~Napolitano, S.~Sciuto,
  E.~Del~Giudice, P.~Di~Vecchia, S.~Ferrara, F.~Gliozzi, R.~Musto and
  R.~Pettorino, {\it Dual String Models With Nonabelian Color and
    Flavor Symmetries,\/} \NPB114 (1976) 297.

\bibitem[BH]{[BH]}K.~Bardak\c ci and M.B.~Halpern, Phys.\ Rev.\ D3
  (1971) 2493.

\bibitem[BGG]{[BGG]}I.~Bernshtein, I.~Gelfand, and S.~Gelfand, Funk.
  An. Prilozh. 10 (1976) 1.


\bibitem[BLNW]{[BLNW]}M.~Bershadsky, W.~Lerche, D.~Nemeschansky, and
  N.P.~Warner, Nucl. Phys. B401 (1993) 304.

\bibitem[BO]{[BO]}M.~Bershadsky and H.~Ooguri,
{\it Hidden $Osp(N,2)$ Symmetries in Superconformal Field Theories},
\PLB229 (1989) 374.

\bibitem[BLLS]{[BLLS]}A. Boresch, K. Landsteiner, W. Lerche, and A.
  Sevrin, {\it Superstrings from Hamiltonian Reduction}, \NPB436
  (1995) 609--637;

\bibitem[BFK]{[BFK]}W.~Boucher, D.~Friedan, and A.~Kent, {\it
    Determinant Formulae and Unitarity For the $N=2$ Superconformal
    Algebras in Two-Dimensions or Exact Results on String
    Compactification\/}, \PLB172 (1986) 316.

\bibitem[BT]{[BT]}P.~Bowcock and A.~Taormina, {\it Representation
    theory of the affine Lie superalgebra $sl(2|1)$ at fractional
    level\/}, Commun. Math. Phys. 185  (1997) 467-493.

\bibitem[CGP]{[C]}S.~Cecotti, L.~Girardello, and A.~Pasquinucci,
  \IJMPA 6 (1991) 2427;\\
  S.~Cecotti, \IJMPA6 (1991) 1749; \NPB355 (1991) 755.

\bibitem[CV]{[CV]}S.~Cecotti  and C.~Vafa, \NPB367 (1991) 359;

\bibitem[Di]{[Dix]}J.~Dixmier, {\sl Alg\'ebres Enveloppantes\/},
  Gauthier-Villars, 1974.

\bibitem[Do]{[Dob]}V.K.~Dobrev, {\it Characters of Unitarizable
    Highest Weight Modules over the $\N2$ Superconformal Algebra},
  \PLB186 (1987) 43.

\bibitem[D]{[Doerr2]}M.~D\"orrzapf, {\it Analytic Expressions for
    Singular Vectors of the $\N2$ Superconformal Algebra}, Commun.\
  Math.\ Phys.\ 180, 195 (1996).

\bibitem[EY]{[EY]}T.~Eguchi and S.-K.~Yang, {\it $N=2$ superconformal
    models as topological field theories}, \MPLA5 (1990) 1653.

\bibitem[EG]{[EG]}W.~Eholzer and M.R.~Gaberdiel, {\it Unitarity of
    rational $N=2$ superconformal theories\/},
  Commun.\ Math. Phys. 186 (1997) 61--85.

\bibitem[FST]{[FST]}B.L.~Feigin, A.M.~Semikhatov, and I.Yu.~Tipunin,
  {\it Equivalence between Chain Categories of Representations of
    Affine $\SL2$ and $N=2$ Superconformal Algebras}, hep-th/9701043.

\bibitem[FF]{[FF]}B.L.~Feigin and D.B.~Fuchs, {\it Representations of
    the Virasoro Algebra}, in: {\sl Representations of Lie Groups and
    Related Topics}, eds. A.M.~Vershik and A.D.~Zhelobenko, Gordon \&
  Breach, 1990.

\bibitem[FFr]{[FFr]}B.L.~Feigin and E.V.~Frenkel, {\it Representations
    of Affine Kac--Moody Algebras and Bosonization}, in: {\sl Physics
    and Mathematics of Strings}, eds. L.~Brink, D.~Friedan, and
  A.M.~Polyakov, World Sci.


\bibitem[FoF]{[Jose]}J.M.~Figueroa-O'Farrill, {\it Untwisting
    Topological Field Theories} (1996);\\
  {\it Affine algebras, $\N2$ superconformal algebras, and gauged WZNW
    models}, \PLB316 (1993) 496;\\
  {\it $\N2$ Structures in String Theories}, hep-th/9507145.

\bibitem[FMS]{[FMS]}D.H.~Friedan, E.J.~Martinec, and S.H.~Shenker,
  {\it Conformal Invariance, Supersymmetry and String Theory\/},
  \NPB271 (1986) 93.

\bibitem[GRS]{[GS2]}B.~Gato-Rivera and A.M.~Semikhatov, {\it
    $d\leq1\cup d \geq 25$ and W Constraints From BRST-Invariance in
    the $c\neq3$ Topological Algebra\/}, \PLB293 (1992) 72.

\bibitem[G]{[G]}D.~Gepner, Phys.\ Lett. B199 (1987) 380; Nucl.\
  Phys.\ B 296 (1988) 757.

\bibitem[Get]{[Get]}E.~Getzler, {\it Manin Triples and Topological
    Field Theory}, hep-th/9509057; {\it Manin Triples and $\N2$
    Superconformal Field Theory}, hep-th/9307041.

\bibitem[Get2]{[Get2]}E. Getzler, {\it Batalin-Vilkovisky algebras and
    two-dimensional topological field theories}, Commun. Math.  Phys.
  159 (1994), 265-285.

\bibitem[IK]{[IK]}K.~Ito and H.~Kanno, {\it Hamiltonian Reduction and
    Topological Conformal Algebra in $c\leq1$ Non-Critical Strings},
\MPLA9 (1994) 1377.

\bibitem[K]{[TheBook]}V.G.~Ka\v c {\sl Infinite Dimensional Lie
    Algebras\/}, Cambridge University Press, 1990.

\bibitem[KS]{[KS]}Y.~Kazama and H.~Suzuki, {\it New $N=2$
    Superconformal Field Theories and Superstring compactification\/},
  \NPB321 (1989) 232.

\bibitem[KL]{[KL]}S.V.~Ketov and O.~Lechtenfeld, {\it The String
    Measure and Spectral Flow of Critical $N=2$ Strings}, \PLB353
  (1995) 463. 

\bibitem[Kir]{[Kir]}E.B.~Kiritsis, {\it Character Formulae and
    Structure of the Representations of the $\N1$, $\N2$
    Superconformal Algebras}, \IJMPA (1988) 1871.

\bibitem[KM]{[KM]}D.~Kutasov and E.~Martinec, {\it New Principles for
    String/Membrane Unification}, \NPB477 (1996) 652;\\
  D.~Kutasov, E.~Martinec, and M.~O'Loughlin, {\it Vacua of M-theory
    and $N=2$ strings}, \NPB477 (1996) 675.

\bibitem[LVW]{[LVW]}W.~Lerche, C.~Vafa, and N.P.~Warner, {\it Chiral
    Rings In $N=2$ Superconformal Theories\/}, \NPB324 (1989) 427.

\bibitem[LZ]{[LZ]}B.H.~Lian and G.J.~Zuckerman, {\it New Selection
    Rules and Physical States in 2-D Gravity: Conformal Gauge},
  \PLB254 (1991) 417--423;\\
  {\it Semiinfinite Homology and 2-D Gravity. 1.}  Commun. Math. Phys.
  145 (1992) 561--593;


\bibitem[Mal]{[Mal]}F.~Malikov, {\it Verma Modules over Rank-2 Ka\v
    c--Moody Algebras}, Algebra i Analiz 2 No.~2 (1990) 65.

\bibitem[MFF]{[MFF]}F.G.~Malikov, B.L.~Feigin, and D.B.~Fuchs, {\it
    Singular Vectors in Verma Modules over Ka\v c--Moody Algebras},
  Funk.\ An.\ Prilozh.\ 20 N2 (1986) 25.

\bibitem[M]{[Marcus]}N.~Marcus, {\it A Tour through $N=2$ strings\/},
talk at the Rome String Theory Workshop, 1992, hep-th/9211059.

\bibitem[Mar]{[Mar]}E.~Martinec, \PLB217 (1989) 431; {\it Criticality,
    Catastrophes, and Compactifications}, in: {\sl Physics and
    Mathematics of Strings}, ed. L.~Brink, D.~Friedan, and
  A.~M.~Polyakov (World Scientific, 1990)

\bibitem[Mat]{[M]}Y.~Matsuo, {\it Character Formula of $C<1$ Unitary
    Representation of $\N2$ Superconformal ALgebra}, Prog.\ Theor.\
  Phys. 77 (1987) 793.

\bibitem[OS]{[OS]}N.~Ohta and H.~Suzuki, {\it $\N2$ Superconformal
    Models and their Free Field Realization}, \NPB332 91990) 146--168.

\bibitem[OV]{[OV23]}H.~Ooguri and C.~Vafa
  {\it Geometry of $N=2$ Strings}, \NPB361 (1991) 469--518;\\
  {\it $N=2$ Heterotic Strings}, \NPB367 (1991) 83--104.

\bibitem[RSS]{[RSS]}E.~Ragoucy, A.~Sevrin and P.~Sorba, {\it Strings
    from $N=2$ Gauged Wess--Zumino--Witten Models\/}, Commun.\ Math.\
  Phys. 181 (1996) 91--129.

\bibitem[RCW]{[RCW]}A.~Rocha-Caridi and N.R.~Wallach, {\it Highest
    Weight Modules over Graded Lie Algebras: Resolutions, Filtrations,
    and Character Formulas}, Trans. Amer. Math. Soc. 277 (1983)
  133-162.

\bibitem[SS]{[SS]}A.~Schwimmer and N.~Seiberg, {\it Comments on the
    $N=2$, $N=3$, $N=4$ Superconformal Algebras In Two-Dimensions\/},
  \PLB184 (1987) 191.

\bibitem[S]{[S-sing]}A.M.~Semikhatov, {\it The MFF singular vectors
    in topological conformal theories}, \MPLA9 (1994) 1867.

\bibitem[S2]{[S-sl21]}A.M.~Semikhatov, {\it The Non-Critical $N=2$
    String is an $\SSL21$ Theory\/}, \NPB478 (1996) 209--234.

\bibitem[S3]{[S-sl21sing]}A.M.~Semikhatov, {\it Verma Modules,
    Extremal Vectors, and Singular Vectors on the Non-Critical $\N2$
    String Worldsheet}, hep-th/9610084.

\bibitem[ST]{[ST2]}A.M.~Semikhatov and I.Yu.~Tipunin, {\it Singular
    Vectors of the Topological Conformal Algebra\/}, \IJMPA11 (1996)
  4597.

\bibitem[ST2]{[ST3]}A.M.~Semikhatov and I.Yu.~Tipunin, {\it All
    Singular Vectors of the $N=2$ Superconformal Algebra via the
    Algebraic Continuation Approach\/}, hep-th/9604176.

\bibitem[ST3]{[ST4]}A.M.~Semikhatov and I.Yu.~Tipunin, {\it The
    Structure of Verma Modules over the $N=2$ Superconformal
    Algebra\/}, hep-th/9704111, Commun.\ Math.\ Phys., to appear.

\bibitem[VW]{[VW]}C.~Vafa and N.P.~Warner, \PLB218 (1989) 51.

\bibitem[W]{[W-top]}E.~Witten, {\it Topological Sigma Models\/},
  Commun.\ Math.\ Phys.\ 118 (1988) 411.

\end{thebibliography}
\end{document}